\documentclass[Journal,letterpaper]{ascelike-new}
\usepackage[utf8]{inputenc}
\usepackage[T1]{fontenc}
\usepackage{lmodern}
\usepackage{graphicx}
\usepackage[figurename=Fig.,labelfont=bf,labelsep=period]{caption}
\usepackage{subcaption}
\usepackage{amsmath}
\usepackage{newtxtext,newtxmath}
\usepackage[colorlinks=true,citecolor=red,linkcolor=black]{hyperref}
\NameTag{Basak, \today}
\usepackage[inline]{trackchanges} 
\addeditor{LK}
\addeditor{RK}
\addeditor{JS}
\addeditor{RB}
\addeditor{MC}
\addeditor{MK}

\newcommand{\CN}{{\mbox{CN}}}
\newcommand{\FC}{{\mbox{FC}}}
\newcommand{\FP}{{\mbox{FP}}}
\newcommand{\PD}{{\mbox{PD}}}

\begin{document}

% You will need to make the title all-caps
\title{Two approaches to quantification of force networks in particulate systems}

\author[1]{Rituparna Basak}
\author[2]{C. Manuel Carlevaro}
\author[3]{Ryan Kozlowski}
\author[1]{Chao Cheng}
\author[4]{Luis A. Pugnaloni}
\author[5]{Miroslav Kram\'ar}
\author[6]{Hu Zheng}
\author[3]{Joshua E. S. Socolar}
\author[1]{Lou Kondic}

\affil[1]{Department of Mathematical Sciences, New Jersey Institute of Technology, Newark, NJ, 07102,
Email: kondic@njit.edu}
\affil[2]{Instituto de F\'isica de L\'iquidos y Sistemas Biol\'ogicos, CONICET, 59 789, 1900 La Plata, and Dpto. Ing. Mec\'anica, Universidad Tecnol\'ogica Nacional, Facultad Regional La Plata, Av. 60 Esq. 124, La Plata, 1900, Argentina}
\affil[3]{Department of Physics, Duke University, Durham, NC, 27708}
\affil[4]{Departamento de F\'isica, FCEyN, Universidad Nacional de La Pampa, CONICET,
Uruguay 151, 6300 Santa Rosa, La Pampa, Argentina}
\affil[5]{Department of Mathematics, University of Oklahoma, Norman, OK 73019}
\affil[6]{Department of Geotechnical Engineering, College of Civil Engineering, Tongji University, Shanghai 200092, China}

\maketitle

% Please include an abstract:
\begin{abstract}
The interactions between particles in particulate systems are organized in 
`force networks', mesoscale features that bridge between the particle scale and the 
scale of the system as a whole.  While such networks are known to be crucial in determining
the system wide response, extracting their properties, particularly from experimental 
systems, is difficult due to the need to measure the interparticle forces.  In this work, 
we show by analysis of the data extracted from simulations that such detailed information 
about interparticle forces may not be necessary, as long as the focus is on extracting
the most dominant features of these networks.   The main finding is that a reasonable 
understanding of the time evolution of force networks can be obtained from incomplete 
information such  as total force on the particles. To compare  the evolution of the 
networks based on the completely known particle interactions and the networks based on incomplete information (total force each grain) we use 
tools of algebraic topology. In particular we will compare simple measures defined on 
persistence diagrams that provide useful summaries of the force network features.
\end{abstract}

\section{Introduction}

In recent years, a significant amount of research has been carried on the topic of 
relating interparticle forces in static, compressed, or sheared dry or wet granular 
systems with the system-wide response; see~\cite{behringer_2018} for a recent 
review. This body of research has established clear connections between the 
particle-sale interactions, the mesoscopic structures loosely referred to as
force networks (often discussed in terms of force chains and cycles), 
and macro-scale system properties.  Therefore, in order to understand the 
properties of the system as a whole, it is crucial to understand and quantify the 
properties of the underlying force networks, by which we mean the
force field that describes the interparticle interactions. 

Force networks in particulate systems have been analyzed recently using a variety of methods.  
These include force network ensemble analysis~\cite{snoeijer_prl04,tighe_sm10,sarkar_prl13}, 
statistics-based methods~\cite{peters05,tordesillas_pre10,tordesillas_bob_pre12,makse_softmatter_2014}, 
network analysis~\cite{daniels_pre12,walker_pre12,tordesillas_pre_15,giusti_pre16}, 
and  topological data analysis (TDA), in particular persistent homology (PH)~\cite{arevalo_pre10,arevalo_pre13,ardanza_pre14,epl12,pre13,physicaD14,pre14,Pugnaloni_2016,kondic_2016}. While different methods provide complementary insights, we focus in particular on the PH approach since it allows for significant data reduction and for formulation of simple but informative  
measures describing the force networks, as well as for comparison of different networks in a
dynamic setting. Furthermore, the approach is dimension-independent, being  
easily applied in both two and three (2D and 3D) physical dimensions.   
Such an approach was used with success to discuss force networks in dry and wet 
(suspensions) systems that were 
compressed~\cite{epl12,pre14}, vibrated~\cite{Pugnaloni_2016,kondic_2016}, or sheared~\cite{gameiro_prf_2020}, as well as for analysis of yielding of 
a granular system during pullout of a buried intruder in 3D~\cite{shah_sm_2020}.  

While significant progress has been reached on quantifying properties of force networks
based on the data obtained in discrete element simulations, the progress on analysis of 
experimental data has been much slower.  The reason for this is that it is much more 
difficult to extract the information about the forces between the particles in experiments.  
Significant progress in this direction has been obtained using photoelastic systems, 
where based on photoelastic response on the single particle scale, one can extract information 
about the forces at particle-particle contacts \cite{zadeh2019enlightening}. 
Such extraction is, however, computationally expensive, and furthermore it requires high resolution input data so that the 
contact forces can be accurately extracted across a broad range of forces. Moreover, by now this method can be only applied to systems consisting of circular particles. Very often, the data does not meet the above requirements and the reconstruction cannot be carried out; instead, semiquantitative or qualitative measures must be utilized. 
%In such cases, one needs to resort to the use of the data obtained at lower resolution.  
Depending on the quality of the data, one could either use the gradient-squared (G$^2$) method \cite{zadeh2019enlightening,howell_99}
that allows for the extraction of the total force on a particle based on the photolastic signal \cite{YiqiuSingleG2_19} but amplifies noisy signals, 
or one could simply analyze the intensities of raw photoelastic images
with the hope that the available information is sufficient to extract meaningful data.  The former approach (based on
the G$^2$ data) was used  in the context of shear jamming experiments~\cite{dijksman_2018}, 
while the latter approach was used to analyze granular impact~\cite{pre18_impact}, and 
 recently stick-slip dynamics of a slider on top of a granular bed~\cite{cheng_pg_2021}.   
The analyses were carried out using the PH-based tools, leading to 
insightful information about the role of force networks in the considered systems. 
For example, in~\cite{dijksman_2018} it was found that a precise comparison between 
DEM simulations and experimental force network could be reached by adding white-noise
to the simulation data to mimic the experimental noise, and in~\cite{cheng_pg_2021} 
it was uncovered that the PH-based measures show clear correlations 
between the evolution of force networks and the stick-slip dynamics of a slider moving on top of 
granular media.  

The analysis of the force networks discussed above has proven
to be useful in developing connections between particle scale physics and macroscopic system 
response.  However, it should be emphasized that these results were obtained using 
incomplete data about the force networks: precise information about interparticle 
forces was not available.  This raises the question: how accurate is 
the analysis based on incomplete data?  Or, to formulate this question
differently, assuming that more detailed experimental information were available, 
how would the results change?  To be specific, let us consider an example of 
analysis of data in granular impact~\cite{pre18_impact}, where it was found that the loops
(cycles) in the force network play an important role in slowing down the intruder.  
This finding was obtained based on the data extracted from photoelastic images.  
If more detailed information about the structure of the force network were available, 
would such a conclusion still hold?

Answering the outlined question requires having complete information regarding
particle-particle forces, computing relevant results, and then comparing them to 
the ones obtained based on incomplete information. To 
be able to carry out such a project, two necessary conditions need to be met:
(i) one should to be able to compare the results obtained based on
different sets of data, and (ii) one should be able to obtain the information about 
the particle-particle forces.  To start from (ii), the information about particle-particle
forces is difficult and costly to obtain in experiments, as already discussed.  The simpler approach
is to consider DEM simulations, and then either use the complete information (which 
is readily available), or to use only the incomplete information about the total force on 
each particle (which can be easily computed from the information about the forces
at each contact).  
In the experiments, the total force on each particle could be estimated using
the G$^2$ method in a manner that is relatively straightforward \cite{YiqiuSingleG2_19}.  Going back to 
the item (i), it is convenient to use the PH-based approach, since the corresponding 
analysis can be carried out both using the information about the force 
networks based on particle-particle contact forces, and based on the total 
force on each particle.   

\begin{figure}
    \centering
    \includegraphics[width=0.8\columnwidth]{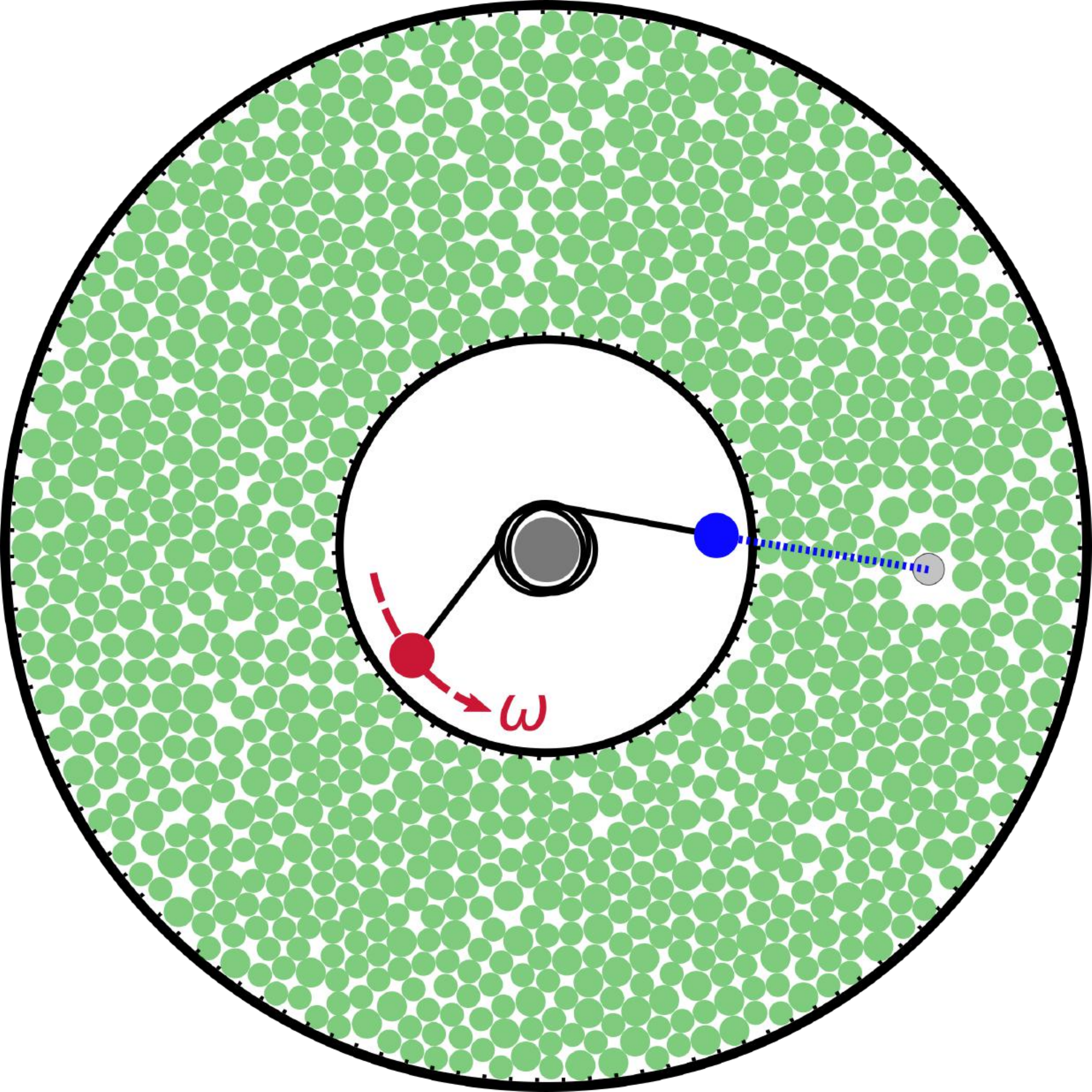}
    \caption{
    Experimental setup serving as a motivation for the present work: the intruder is 
    coupled to one end of a torque spring (blue); 
    the other end of the torque spring (red) is rotated at fixed angular velocity $\omega$. The grains fill an annular channel with rough boundaries to prevent slipping at the boundaries. 
    } 
    \label{fig:exp}
\end{figure}

\begin{figure}
    \centering
      \begin{subfigure}[b]{.4\textwidth}
    \centering
    \includegraphics[width=\linewidth]{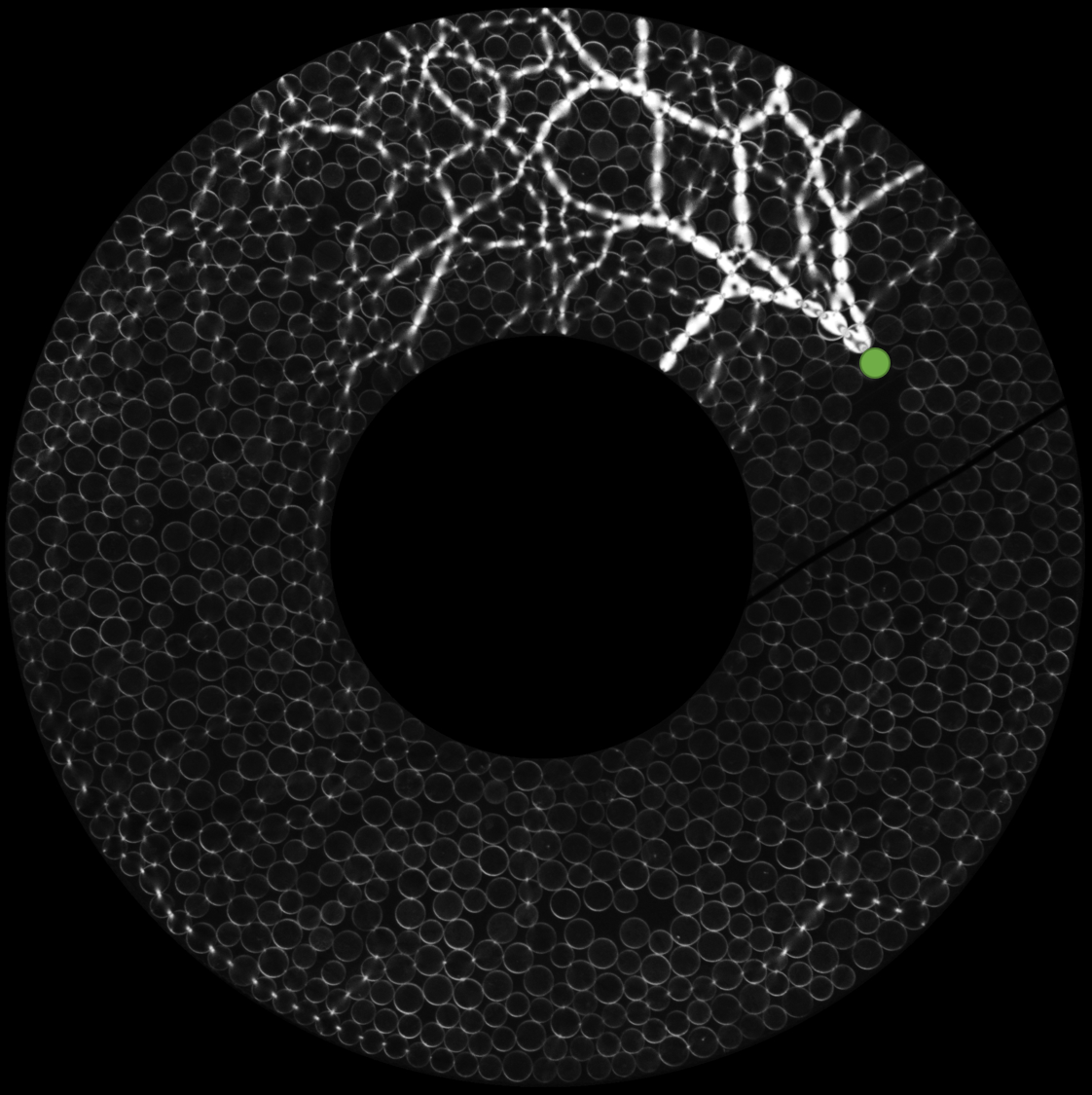}
    \caption{Disks, photoelastic image}
    \end{subfigure}
    \hfill            
    \begin{subfigure}[b]{.4\textwidth}
    \centering
    \includegraphics[width = \linewidth]{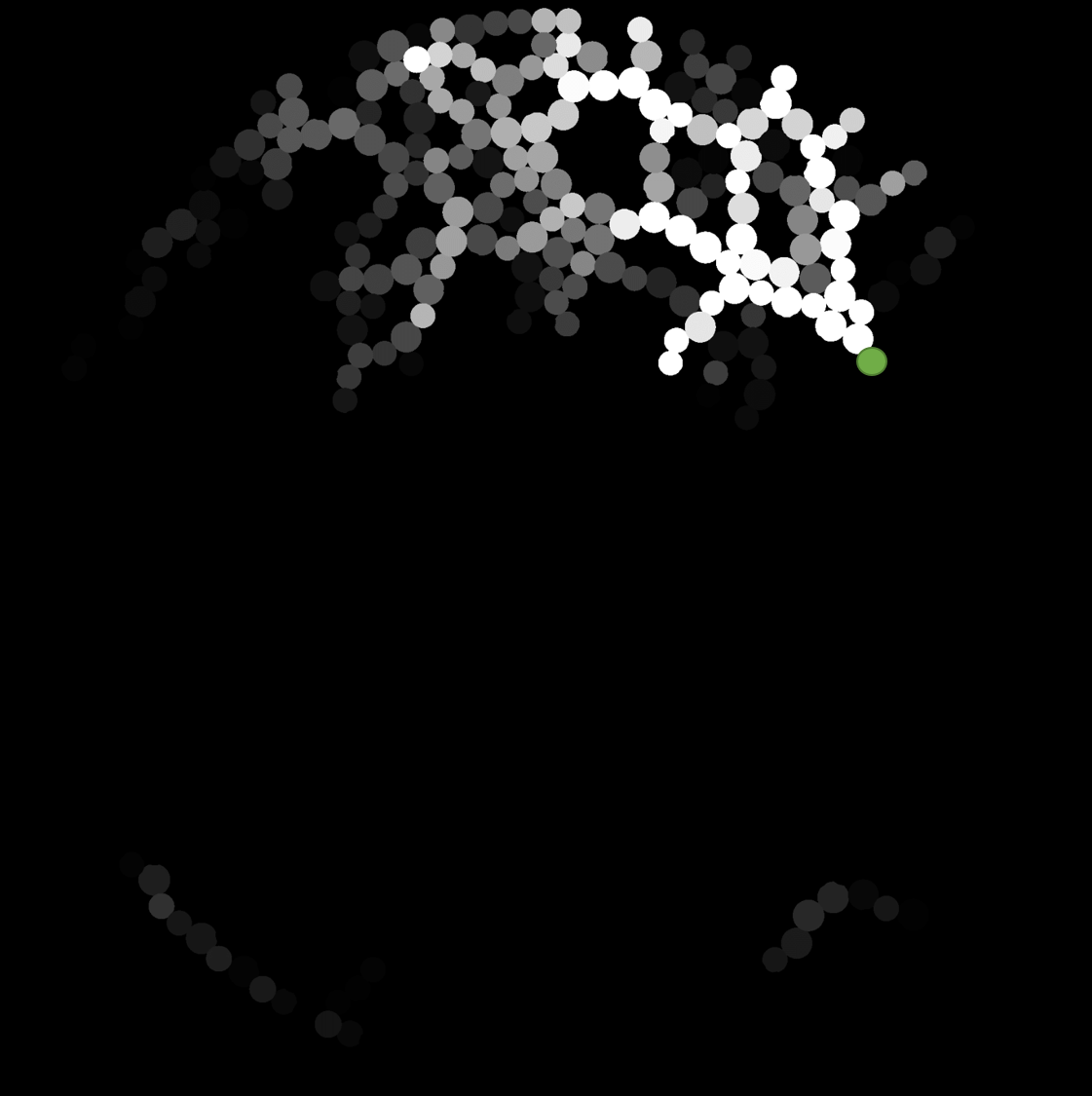}
    \caption{Disks, G$^2$.}
    \end{subfigure}
        \hfill            
    \begin{subfigure}[b]{.4\textwidth}
    \centering
    \includegraphics[width = \linewidth]{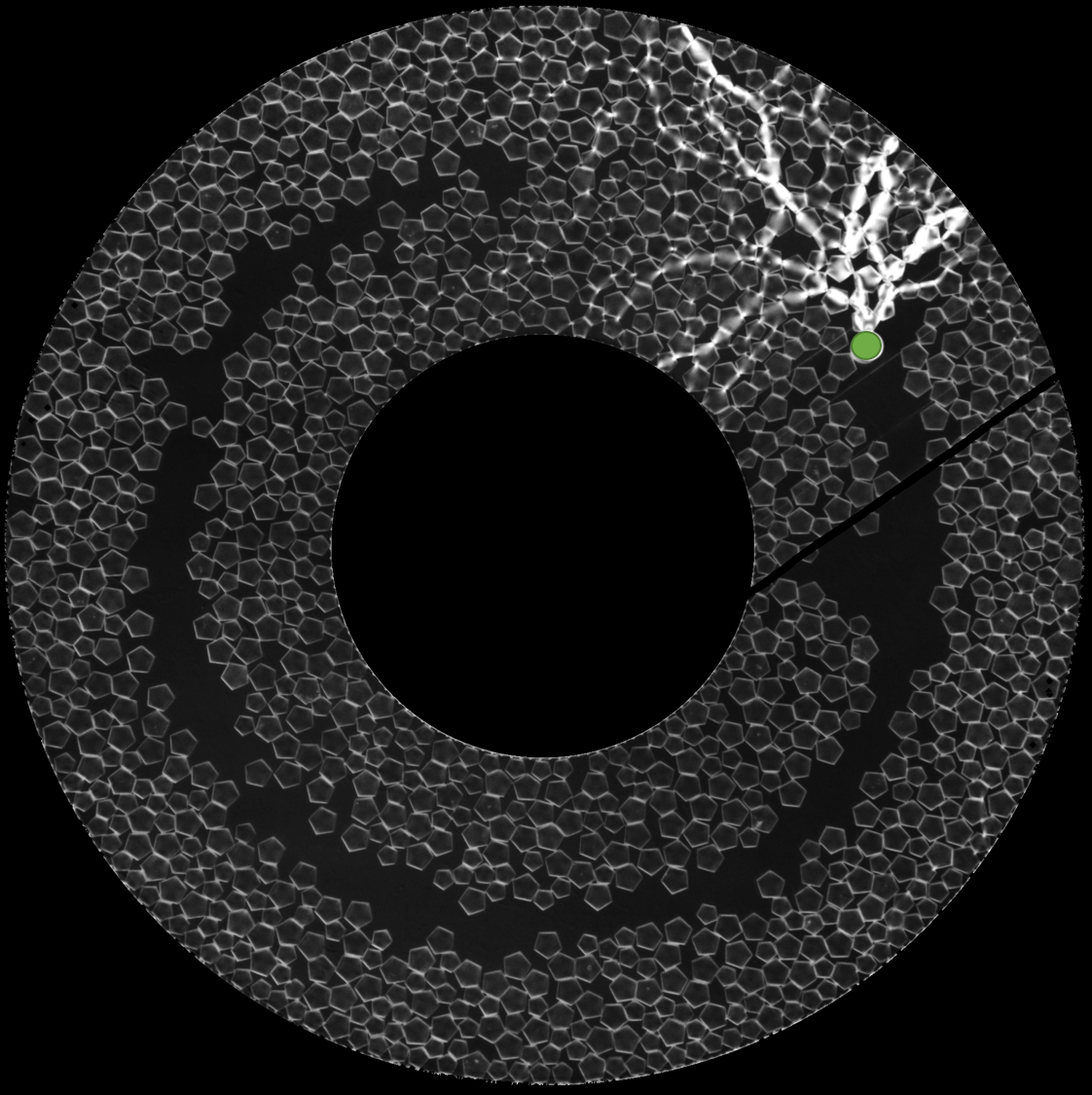}
    \caption{Pentagons, photoelastic image}
    \end{subfigure}
        \hfill            
    \begin{subfigure}[b]{.4\textwidth}
    \centering
    \includegraphics[width = \linewidth]{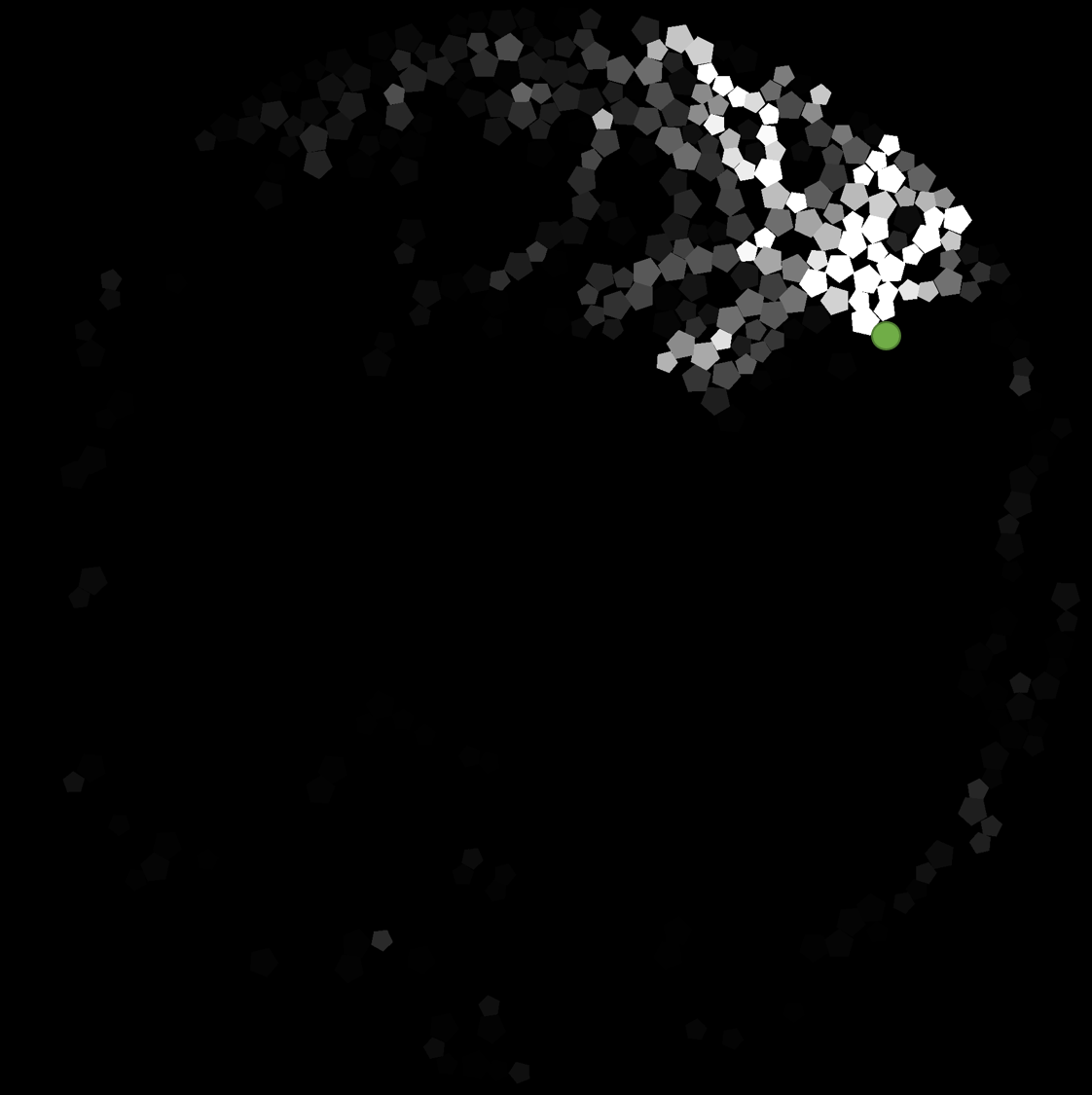}
    \caption{Pentagons, G$^2$.}
    \end{subfigure}
    \caption{Experimental photoelasic images (left) and processed images (right) of the 
    same configurations showing G$^2$ per grain. The top row corresponds to a packing of disks and the bottom row to a packing of pentagons. The intruder is shown in green.   
    } 
    \label{fig:exp_forces}
\end{figure}

In the present paper, we illustrate the outlined approach in the context of recent experiments and simulations that consider the intermittent dynamics of an intruder in an angular Couette geometry discussed in our recent works~\cite{kozlowski_pre_2019,carlevaro_pre_2020}.  Figure~\ref{fig:exp} shows the setup of the experiment motivating the system studied by simulations in this work, and Fig.~\ref{fig:exp_forces} shows the photoelastic images acquired in experiments as well as processed images of force networks extracted using the G$^2$ method. The latter are produced by measuring the average gradient-squared of the photoelastic image intensity of all pixels within each grain (tracked in a white light image of the system), and drawing all of the grains onto a black background with values corresponding to the average gradient-squared for each grain. In this system, a bidisperse monolayer of around $1000$ photoelastic disks (or pentagons) is confined to an annular region by fixed boundaries lined with ribbed rubber to prevent slipping at the boundaries. In one set of experiments described 
in~\cite{kozlowski_pre_2019}, disks were floated on a water-air interface to remove friction with the glass base, while in other experiments with disks and pentagons the particles have basal friction. An intruding disk is pushed in the counterclockwise azimuthal direction (at fixed radius) by a torque spring. One end of the torque spring is coupled to the intruder, while the other end is driven at a fixed angular rate $\omega$. By using a spring stiffness that is far smaller than the grain material stiffness, we are able to study stick-slip dynamics. %, characterized by sticking periods in which stress builds in the granular medium (and the torque spring potential energy increases) and relatively rapid, intermittent slip events in which the medium yields, the torque spring releases stored energy, and the intruder slips through the medium until a stable configuration of grains halts the intruder. 
Cameras above the system track grains and, by use of a dark-field polariscope \cite{daniels2017photoelastic}, visualize grain-scale stresses as demonstrated in Fig.~\ref{fig:exp_forces}. A detailed analysis of the insights of PH for these experimental data will be presented elsewhere; the goal of the present work is to study the same system using simulations, where exact forces are known and not limited by experimental resolution.

The rest of this paper is structured as follows. In the following section we 
discuss the simulated systems and TDA methods used in this study. This section 
also includes a brief discussion of a toy example, which illustrates how 
PH analysis of the data is carried out.  We then follow up with the Results
section, where we present the topological measures quantifying the force
networks.  We will consider both the data obtained in the stick-slip regime, 
where the intruder is essentially static for significant periods of 
time, and in the clogging regime where the intruder is rarely at rest.  The 
computations are carried out for both disks and pentagons, so that we 
can also reach some insight regarding the influence of particle shape
on the force networks.  

\section{Methods}

%\end{document}

\subsection{Simulations}\label{sec:simulations}

Our model considers the grains as rigid impenetrable 2D objects (disks or regular pentagons) that experience both normal and tangential forces when they are in contact with each other or the walls. The 2D particles slide flat on top of a frictional substrate (a ``table'') inside an angular Couette geometry. An intruder particle is dragged through the granular system in a circle (concentric with the annular cell boundaries) by pulling it via a soft torsion spring at a very low speed. The interaction with the substrate is defined by dynamic and static sliding friction coefficients, but rotational friction is set to zero. Since the model is 2D, we do not allow buckling out of the plane, and we assume the interparticle forces have no out-of-plane component.  Key parameters are chosen to match the values used in the experiments that inspired these simulations, including particle diameters and masses, dimensions of the confining annular region, driving velocity, and torque spring constant \cite{kozlowski_pre_2019,carlevaro_pre_2020}. We have found that the statistics of the intruder dynamics closely match the experimental results  \cite{carlevaro_pre_2020}.

We have carried out discrete element method (DEM) simulations of the model using the Box2D library~\cite{box}. The Box2D library uses a constraint solver to handle rigid (infinitely stiff) bodies. Before each time step, a series of iterations (typically 100) is used to resolve constraints on overlaps and on static friction between bodies through a Lagrange multiplier scheme~\cite{pytlos2015modelling,catto}. After resolving overlaps, the inelastic collision at each contact is solved and new linear and angular velocities are assigned to each body. The interaction between particles is defined by a normal restitution coefficient and a friction coefficient (dynamic and static friction coefficients are set to be equal). The equations of motion are integrated through a symplectic Euler algorithm. Solid friction between grains is also handled by means of a Lagrange multiplier scheme that implements the Coulomb criterion. The approach yields realistic dynamics for granular bodies \cite{pytlos2015modelling} with complex shapes. Box2D has been successfully used to study grains under a variety of external drivings~\cite{carlevaro_jsm11,carlevaro_pre_2020}.

Systems consisting of disks are made up of bidisperse mixtures of small disks (S, with mass $m$ and diameter $d$) and large disks (L, with mass $1.5625 m$ and diameter $1.25 d$) in a $2.75:1$ (S/L) ratio. We also simulate a bidisperse mixture of pentagons in a $1:1$ ratio, with radii $1.086 d$ and $1.20 d$ for the small and large sizes, respectively. The time step used to integrate the Newton-Euler equations of motion is $\delta t = 0.0278 \sqrt{d/g}$, with $g$ the acceleration of gravity acting in the direction perpendicular to the substrate. The restitution coefficient is set to $\epsilon = 0.05$, and the friction coefficient $\mu$ is set to $1.2$ for the grain-grain and grain-wall interactions. The static friction coefficient with the substrate is set to $0.36$ in all cases, whereas the dynamic friction is set to $0.36$ in some simulations (which we call frictional) and to $0$ when a frictionless substrate is investigated. The particles are contained in a 2D Couette cell formed by two concentric rings of radii $8.81 d$ and $22.80 d$. These rings have effectively roughened surfaces made up of immobile small equilateral triangles facing inward (toward the annular channel) to prevent the slippage of particles at the boundaries.

The intruder is a disk with $d_i = 1.25d $ and is constrained to move on a circular trajectory midway between the inner and outer rings. The intruder can interact with any other grain in the system but does not interact with the base (i.e., it has no basal friction). It is pulled by a torsion spring with a spring constant of $3591.98 mgd$/rad. One end of this spring is attached to the intruder; the other is driven at a constant angular velocity of $0.00432 \sqrt{g/d}$. This spring can only pull the intruder; no force is applied when the spring becomes shorter than its equilibrium length.

During the simulations, the intruder displays stick-slip dynamics and the particles in the system develop a force network during the sticking periods that fully rearranges after each slip event. These force networks resemble the ones observed in experiments (see Fig. \ref{fig:exp_forces}). We save the contact forces (normal and tangential components) for every single contact in the system for further analysis though persistent homology, discussed in 
what follows.  Contact forces are calculated from the impulses (normal and tangential) after resolving each contact collision. In the case of pentagonal particles, the side-to-side contacts are defined by two points and two forces (one at each point selected along the contact line). The total force at the contact is obtained as the vector sum of these two forces.

\subsection{Two networks: Contact and particle force network }

In this section, we present a toy example that clarifies definitions of the contact force and particle force networks. In the following section these networks are used to demonstrate basic properties of persistence diagrams.  Roughly speaking, both
force networks are defined by a real valued function on a contact network created by the particles.  We start by considering an ensemble of the particles $p_i,~i = 1,\ldots,~N$. The contact network $\CN$  is a network with vertices $v_i,~i= 1,\ldots,~N$ corresponding to particle centers. An edge $\langle v_i$, $v_j\rangle$ is present in the contact network if the particles $p_i$ and $p_j$ are in contact. 

\begin{figure}[!ht]
    \begin{subfigure}[b]{.25\textwidth}
    \centering
    \includegraphics[width=\linewidth]{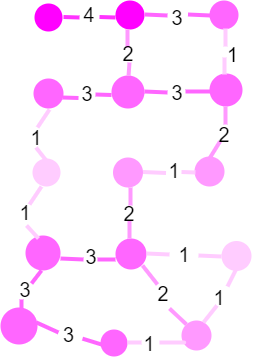}
    \caption{FC network.}
    \end{subfigure}
     \hfill           
    \begin{subfigure}[b]{.25\textwidth}
    \centering
    \includegraphics[width=\linewidth]{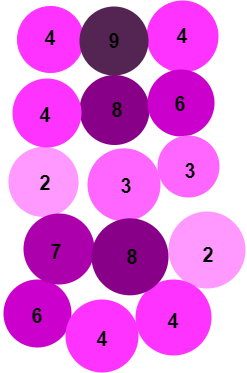}
    \caption{FP network, representation 1.}
    \end{subfigure}
     \hfill            
    \begin{subfigure}[b]{.25\textwidth}
    \centering
    \includegraphics[width=\linewidth]{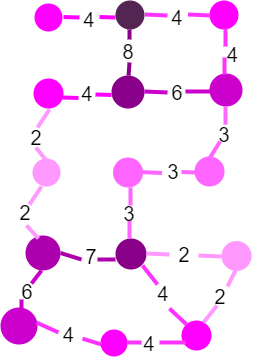}
    \caption{FP network, representation 2.}
    \end{subfigure}
   \caption{Toy example illustrating FC (contact force) and FP (particle force) networks.  
   (a) Contact force network;  the values of the forces at each age are prescribed (as shown 
   by the numbers).
   (b-c) Particle force network: (b) the number assigned to each particle shows the total 
   force on that particle (vertex), obtained by summing up the forces on the edges from (a), 
   and (c) the associated network showing the forces on the edges 
    connecting the particles in (b) as described in the text.  Clearly, the FC (a) and (FP) (c) networks are different.  Note that FP network shown in (c) does not require the information 
    from (a) as long as the total force on particles (as shown in (b)) is known.  
        }
    \label{fig:toy_network}
\end{figure}

To define the force networks we need to assign real values to both vertices and edges of the contact network $\CN$.  If we know the forces between the particles, then it is natural to define the value of a function $f_{FC}(\langle v_i, v_j\rangle)$ to be the magnitude of the force acting between the particles $p_i$ and $p_j$. For the reasons explained below we extend the definition of $f_{FC}$ to the vertices, so that the value at the vertex $v_i$ is the maximum value of $f_{FC}$ on the edges that contain $v_i$. Figure~\ref{fig:toy_network}a shows a simple example of a possible contact force (FC) network.  
In this toy example we specify the forces values by hand; for the data discussed in the Results section, these
forces are obtained from simulations. 

On the other hand, if only the total force on each particle is known, it is natural to define a particle force (FP) network by a function $f_{FP}$ with the values on the vertices $f_{FP}(v_i)$ equal to the total force experienced by the particle $p_i$.  Again for the reasons explained below, we expand the definition of $f_{FP}$ to the edges by $f_{FP}(\langle v_i, v_j\rangle = \min(f_{FP}(v_i), f_{FP}(v_j))$. Figure~\ref{fig:toy_network}b-c shows an example: 
note that the forces on the vertices are defined by the sums of 
the forces on the edges from Fig.~\ref{fig:toy_network}a (these forces are 
shown in Fig.~\ref{fig:toy_network}b), and then the forces on the edges are assigned
as a minima of the forces on the adjacent vertices, as described above.  These forces
are shown in Fig.~\ref{fig:toy_network}c.

The aim of the persistent homology is to understand the structure of the force networks for all
values $\theta$ (thresholds) of the force. For the FC network, the persistent 
homology describes how the topological structure of the super level 
sets $\FC(\theta) = \{\sigma \in \CN \colon f_{FC} (\sigma ) \geq \theta\}$ 
changes with $\theta$. Similarly, for the FP network, the level sets are 
given by $\FP(\theta) = \{\sigma \in \CN \colon f_{FP} (\sigma ) \geq \theta\}$. 
In order to use persistence homology the families of super level sets  
$\FC(\theta)$ and $\FP(\theta)$ have to satisfy the following property. 
If the edge  $\langle v_i, v_j\rangle$ belongs to a given super level set, than 
both vertices, $v_i$ and $v_j$, must belong to this set as well; this governs our 
choice for extending the functions $f_{FC}$ and $f_{FP}$. 

To relate these networks to the ones that are obtained in experiments or simulations 
of granular systems, we note that, on the one hand, the FC network requires as an 
input all the contact forces, which are difficult to obtain in experiments. 
On the other hand, the FP network requires only the total force on a particle, which 
in experiments can be estimated based on G$^2$ information only. 

Our toy example illustrates that there is no clear connection between these two 
networks, since the structure of $\FC(\theta)$ and $\FP(\theta)$ is very different.  
For example, the top two layers of particles are connected by edges with much smaller values 
in Fig.~\ref{fig:toy_network}a than in Fig.~\ref{fig:toy_network}c.  This should not come as a
surprise since the networks are defined in a very different way. The FC network describes 
structure of the force chains through which the force propagate. Thus, the edges with high 
values tend to belong to strong force chains. On the other hand, for the FP network, high 
values on the vertices indicate particles at the junctions of force chains. 

Clearly, the considered networks are different and their direct comparison is not
possible.  Instead, the natural question is whether the evolution of the features 
of one network is closely correlated to the evolution of the features of the other. 
To find the answer to this question we use persistent homology.

\subsection{Persistent Homology}

For the present purposes, one can think of persistent homology (PH) as a tool for describing 
the structure of weighted networks, such as $\FC(\theta)$ and $\FP(\theta)$, defined in the previous section. 
The reader is referred to~\cite{shah_sm_2020} for a more detailed overview of application of PH to dense granular matter 
from a physics point of view, and to~\cite{kramar_physD_2016} for a more in-depth presentation. To facilitate  understanding of the insight that can be reached by using PH, we briefly outline the main ideas and demonstrate them on the above toy example.

Given a weighted network, PH assigns to it two persistence diagrams, $\PD~\beta_0$ and 
$\PD~\beta_1$, 
which describe how the structure of the super level sets changes with the threshed $\theta$.  
The $\PD~\beta_0$ diagram encodes how distinct connected components in super level sets appear 
and then merge as the threshold is decreased. The appearance and merging of these 
components  is precisely encoded by the birth and death coordinates of the  points in 
the $\PD~\beta_0$ diagram.  To put this in context, in a granular system, high threshold 
values correspond to strong forces, the components correspond to `force chains,' and 
merging corresponds to force chains connecting at the lower force levels. 

\begin{figure}[!ht]
    \centering
             \begin{subfigure}[b]{.45\textwidth}
                 \centering
    \includegraphics[width=\linewidth]{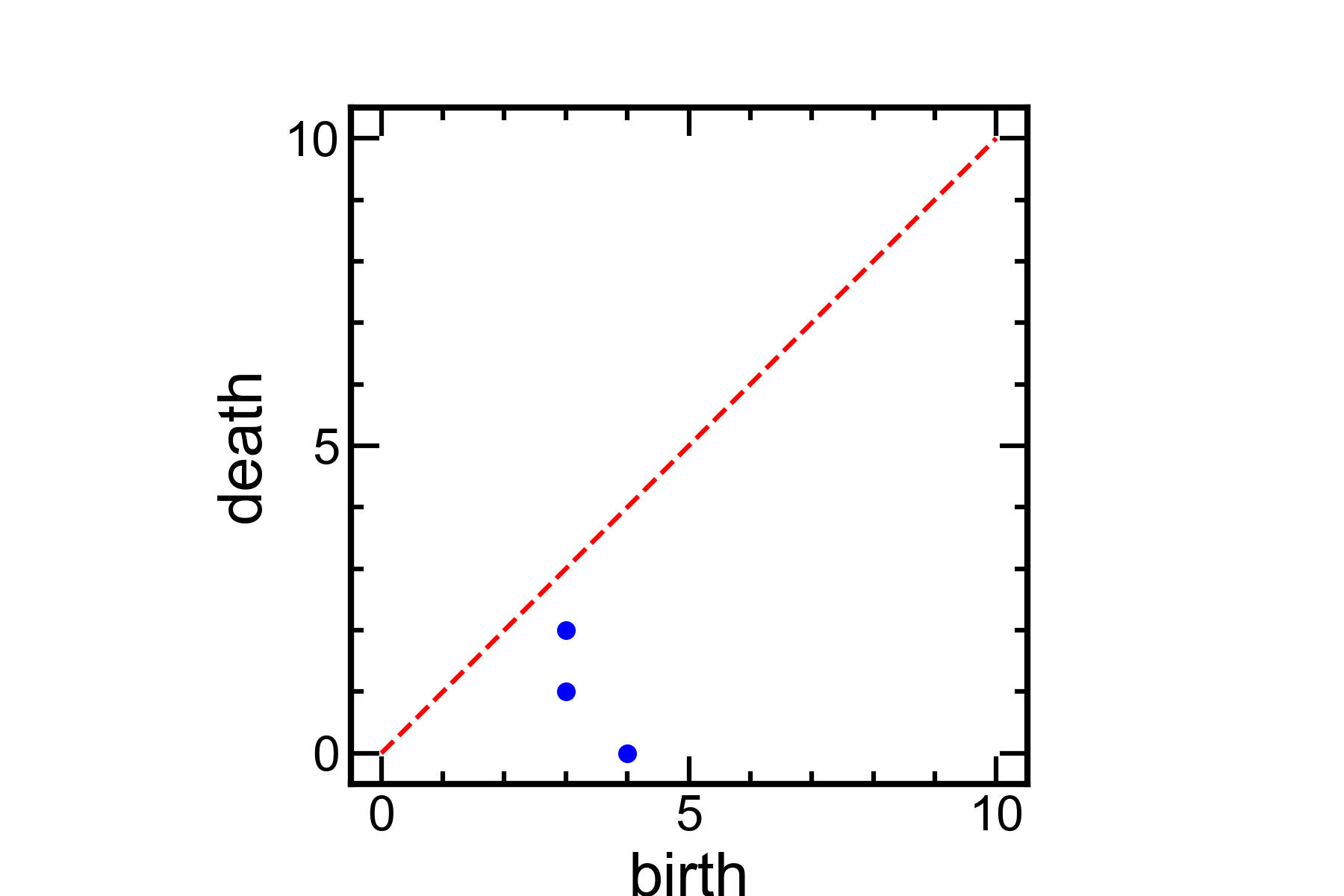}
    \caption{FC, $\beta_0$.}
    \end{subfigure}
             \hfill
           \begin{subfigure}[b]{.45\textwidth}
                 \centering
    \includegraphics[width = \linewidth]{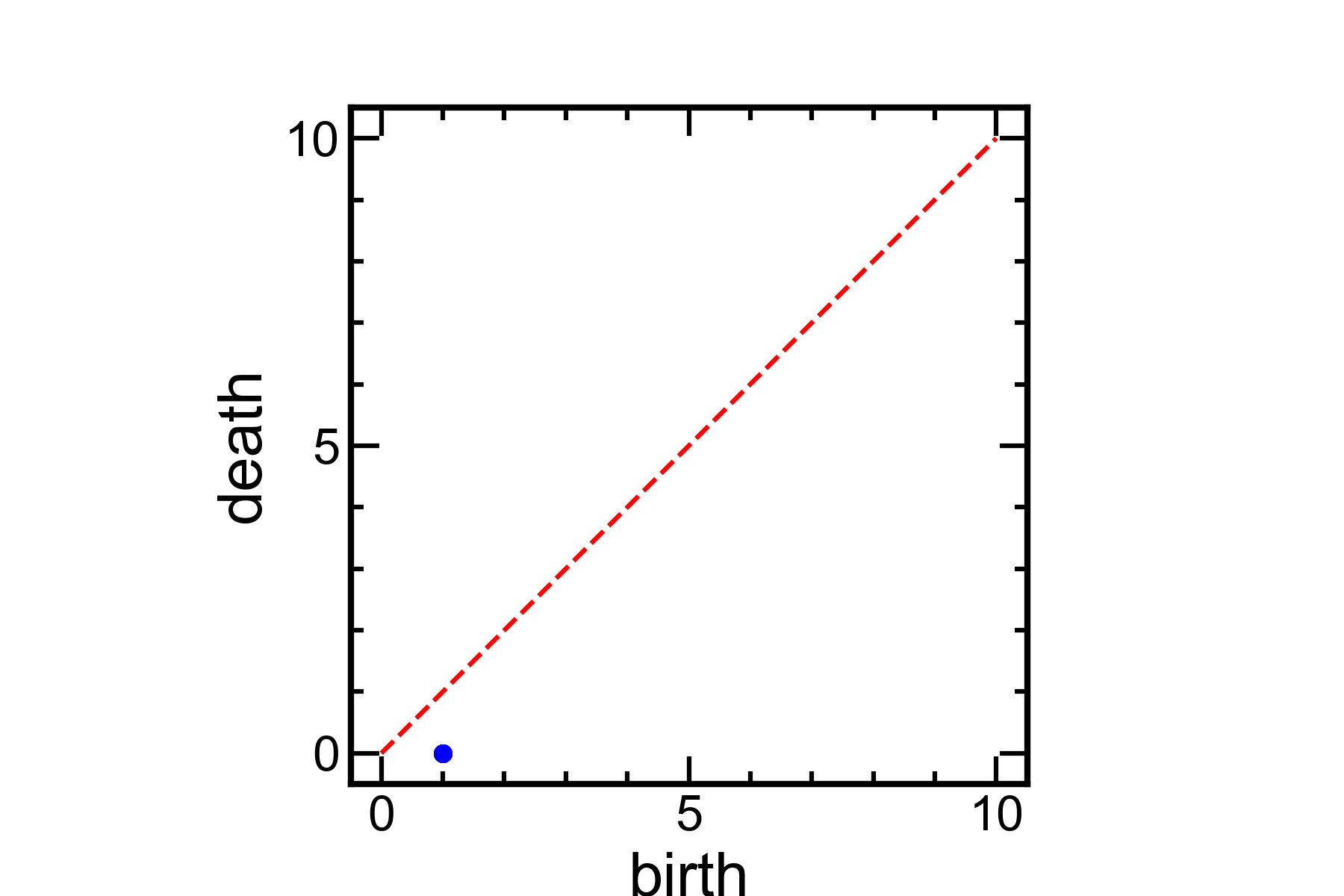}
    \caption{FC, $\beta_1$.}
    \end{subfigure}
             \hfill            \begin{subfigure}[b]{.45\textwidth}
                 \centering
    \includegraphics[width=\linewidth]{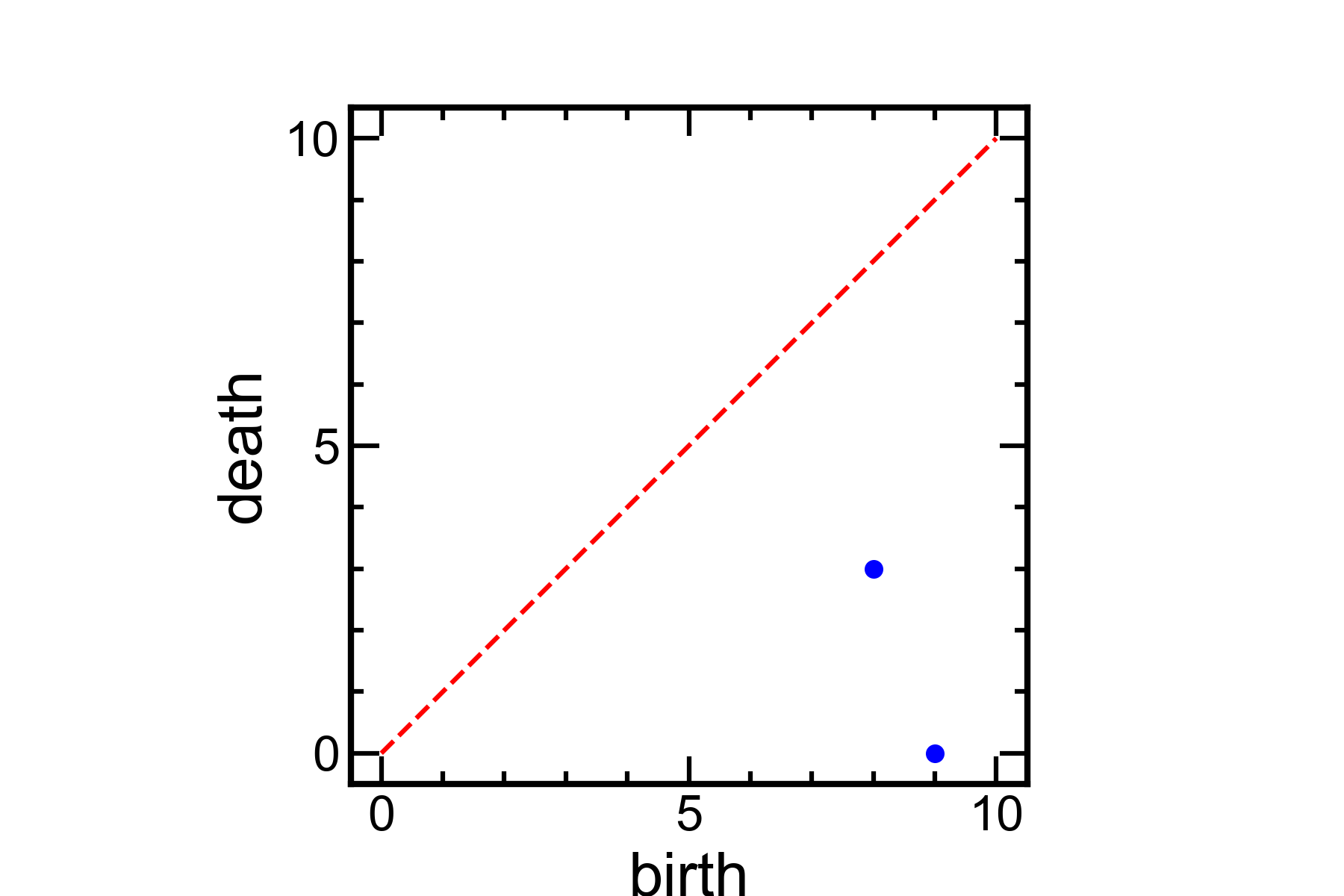}
    \caption{FP, $\beta_0$.}
    \end{subfigure}
    \hfill
             \begin{subfigure}[b]{.45\textwidth}
                 \centering
    \includegraphics[width=\linewidth]{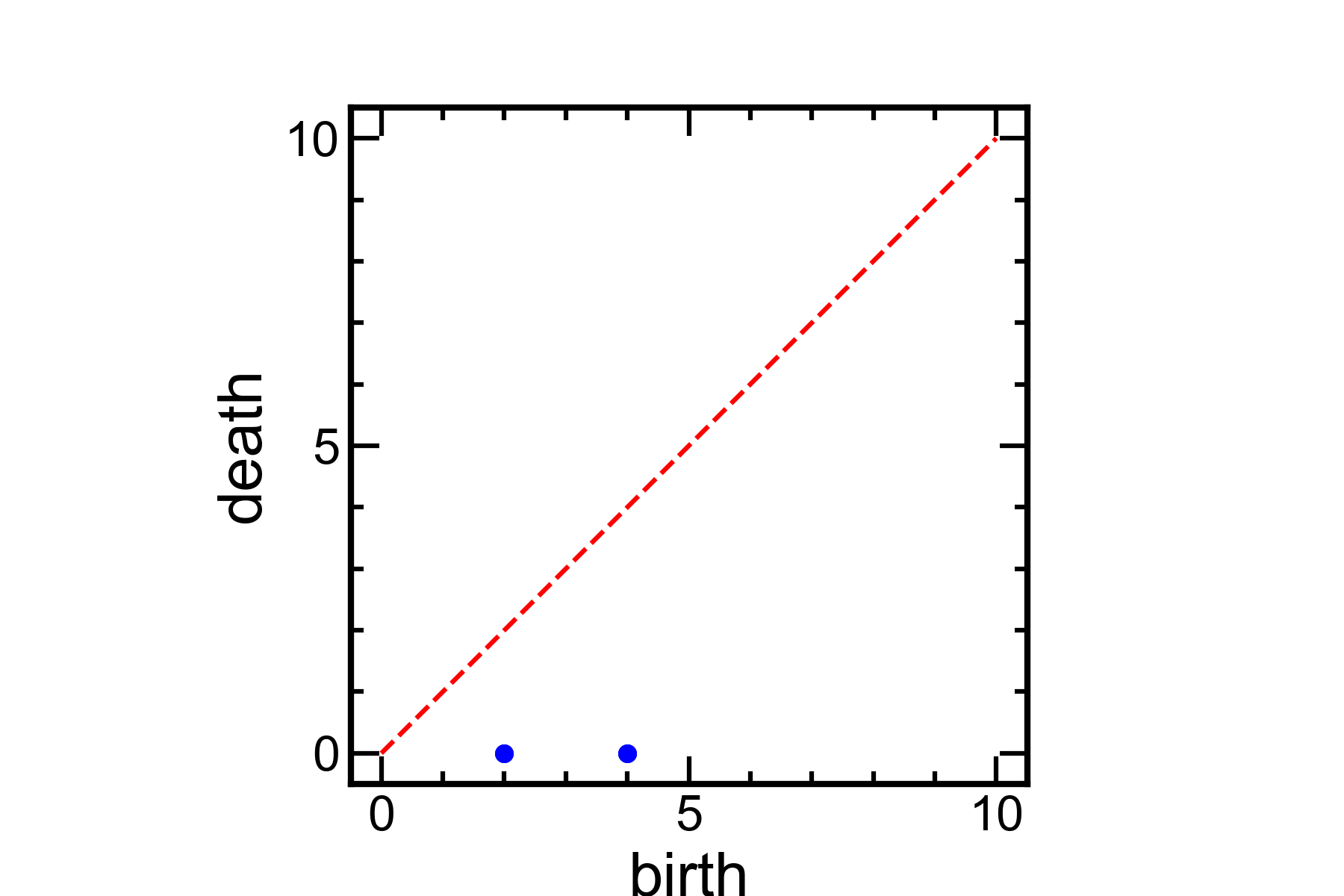}
    \caption{FP, $\beta_1$.}
    \end{subfigure}
             
    \caption{Persistence diagrams, PDs, corresponding to the FC and FP networks from  Fig.~\ref{fig:toy_network}.
     }
    \label{fig:toy_pd}
\end{figure}

Figure~\ref{fig:toy_pd}a illustrates this process using 
$\FC(\theta)$ networks from Fig.~\ref{fig:toy_network}a. 
In this network, the first connected component appears for $\theta = 4$ 
and is represented by the  point in $\PD~\beta_0$ with the birth coordinate value $4$.  
There are two more points in this diagram with birth coordinates 
at $3$, and they represent two distinct components that appear at 
this threshold, one at the bottom of the network and the other 
consist of the edges in the second layer from the top. The latter 
component merges with the top layer at $\theta = 2$, and we say that 
it disappears (dies) at this value so that the lifetime of this component 
is described by the point $(3,2) \in \PD~\beta_0$. The bottom component 
merges with the rest of the network at $\theta = 1$ and is represented by  
$(3,1) \in \PD~\beta_0$. Finally, the first component that appeared 
for $\theta = 4$ is present for all values of $\theta$ and is identified 
by the point $(4,0) \in \PD~\beta_0$. Again to put this in context, we note 
that the points in the diagram can be related to the common (even if not always 
well defined) concept of force chains. The birth and death coordinate indicate the 
force levels at which different force chains form and merge.

The persistence diagram $\PD~\beta_0$, shown in Fig.~\ref{fig:toy_pd}c for the
FP network, describes the appearance and  disappearance of the connected components 
in the network $\FP(\theta)$ depicted in Fig.~\ref{fig:toy_network}c.  Clearly, 
this diagram is different from the one for $\FC(\theta)$, since the networks are 
different. As we already mentioned, the high values in FP network are attained at 
the places where 'force chains' come close to each other or intersect. As indicated 
by the presence of two points in $\PD_0$, there are two distinct place in the FP 
network where this happens, as visible in Fig.~\ref{fig:toy_network}b-c.

In a similar manner, the $\PD~\beta_1$ diagram describes the appearance of 
the loops (cycles).  Note that if a loop appears in the super level set 
at a given threshold $\theta_1$, then it is present for all 
values $\theta \leq \theta_1$, and is represented by the point $(\theta_1, 0)\in \PD~\beta_1$.
For the FC network shown in Fig.~\ref{fig:toy_network}c, there are 
four loops that appear in the $\PD~\beta_1$ at $\theta_1 = 1$, as shown in 
Fig.~\ref{fig:toy_pd}b and they are represented by four copies of the point $(1,0)$ in $\PD~\beta_1$. For the FP network, there are four loops. Two appear at 
the top right and the bottom left 
of the network at $\theta_1 =4$, and are represented in $\PD~\beta_1$ by two copies of the point $(4,0)$. The other two 
are born at $\theta_1 = 2$ and are described by two copies of $(2,0)$ in $\PD~\beta_1$. Similarly 
as we discussed in the context of $\PD~\beta_0$, the $\PD~\beta_1$ are different
for the FC and FP networks. 

An important aspect of PH is that it provides  information about the force 
networks at all force levels. So, unlike other measures, it does not require 
separation of a force network into a `strong' or a `weak' network, although
it allows for such classification, as we will discuss also in the Results 
section. We note that each feature of the network can be described by a point $(b,d)$ 
(where $b$ stands for birth and $d$ for death) 
in one of the persistence diagrams. Moreover, the prominence of a feature
is encoded by its lifespan defined as $b-d$. 

The description of a weighted force network in terms of $\PD$s provides a compact 
but meaningful description of the features of the underlying network. As demonstrated by
Fig.~\ref{fig:toy_network}, the $\PD$s clearly describe the differences between the FC
and FP networks.  It should be noted, however, that the space of $\PD$s is 
a nonlinear complete metric space~\cite{mileyko2011probability}, and there is no readily available method for correlating the diagrams.  
Hence, in the rest of this paper, we will consider several different metrics that 
can be defined for $\PD$s. Introducing these metrics leads to a further data reduction.  
Still, as we will  demonstrate, such metrics provide relevant summaries of the 
properties of the considered networks.  One considered metric is the number 
of points (generators) in a diagram.  Another option is the lifespan, already
introduced above, which describes for how long (that is, for how many 
thresholds levels) a point persists. Using a landscape (mountains and valleys) 
as an analogy, the number of points, N$_{\rm G}$ in  $\PD~\beta_0$ corresponds
to the number of mountain peaks, and the lifespan corresponds to the difference
in altitude between a peak and a valley. Lifespans of all the points in a persistence diagram can be  aggregated into a single number by
defining the total persistence, TP, as a sum of all lifespans.   We will be using 
both N$_{\rm G}$ and TP in discussing some properties of the force networks in 
the considered system. 
   
In our calculations, we define FC and FP networks based on the normal force
between the particles, suitably normalized.  The PH calculations leading to 
persistence diagrams are carried out to compute PDs for both FC and FP networks using the 
software package Gudhi~\cite{gudhi}.  We focus on the interparticle 
interactions only, and do not consider particle - wall forces in the 
present work. 

\section{Results}

In this section we first discuss the general features of the results for the considered
networks, and then focus our discussion on the main topic of the paper: the differences 
between the contact force (FC) and particle force (FP) networks. 

\subsection{Contact force and Particle force networks: General features}

\begin{figure}[!ht]
    \centering
             \begin{subfigure}[b]{.4\textwidth}
                 \centering
    \includegraphics[width=\linewidth]{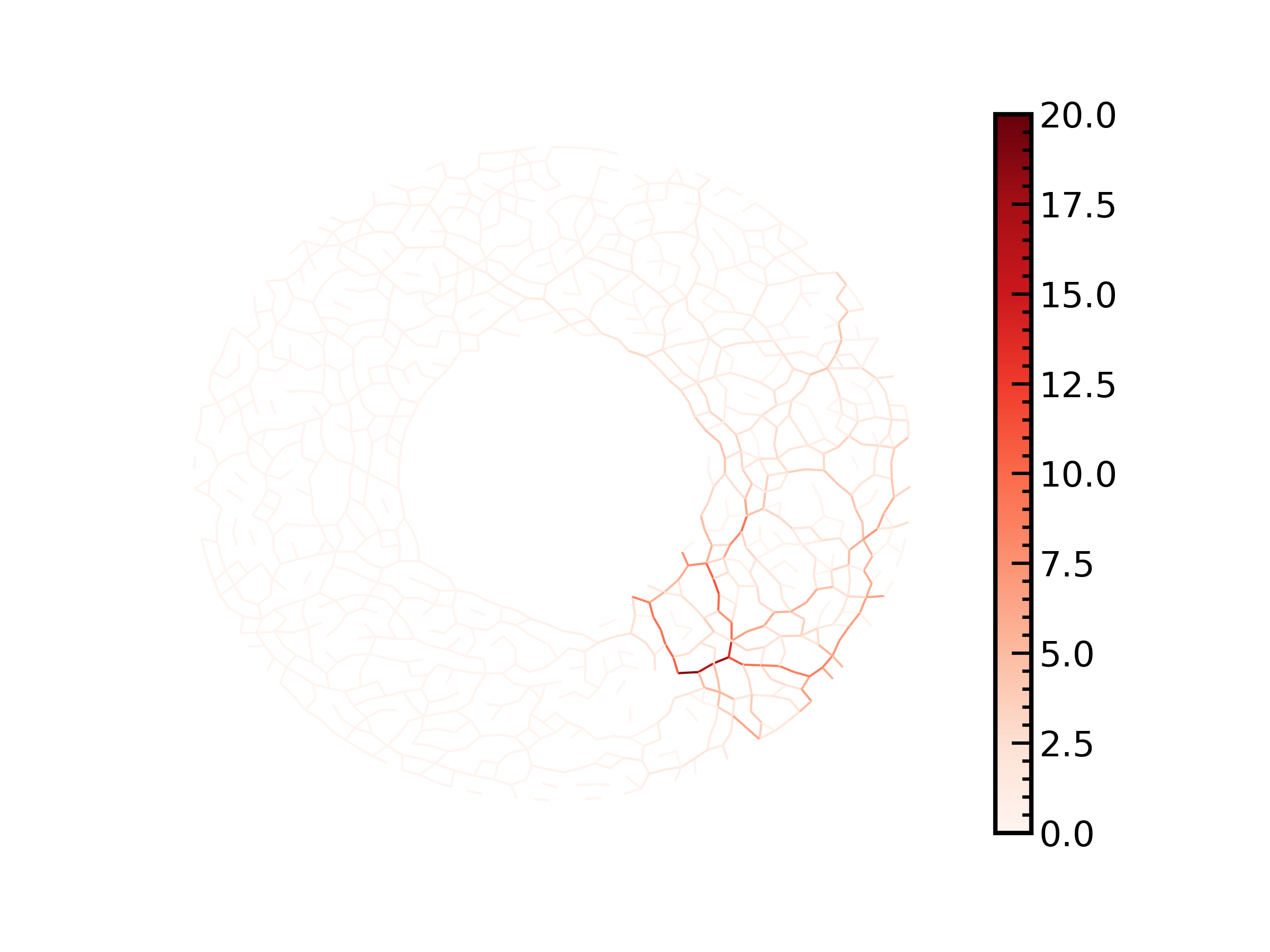}
    \caption{Disks, FC.}
    \end{subfigure}
             \hfill
             \begin{subfigure}[b]{.4\textwidth}
                 \centering
    \includegraphics[width=\linewidth]{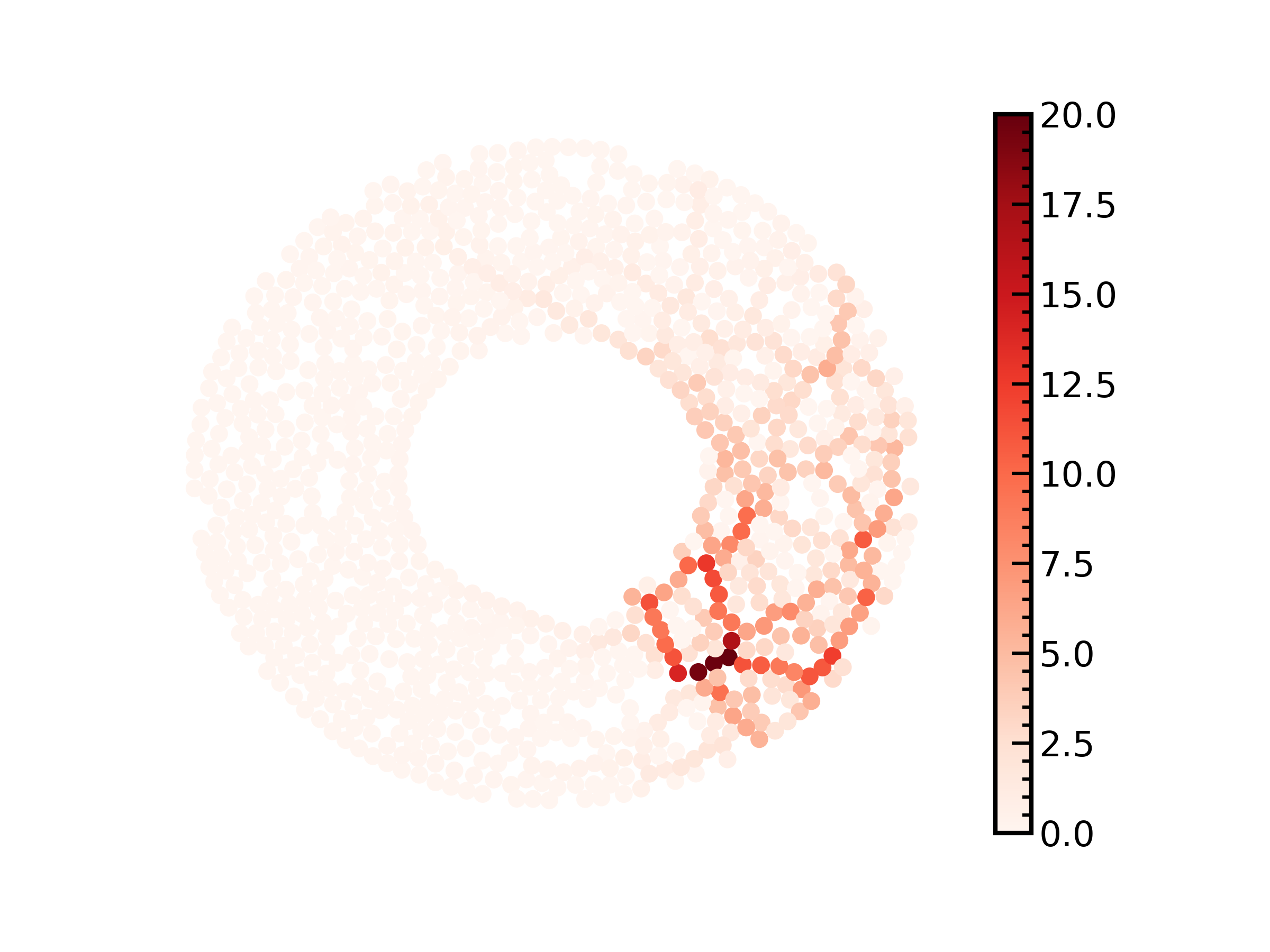}
    \caption{Disks, FP.}
    \end{subfigure}
    \centering
             \begin{subfigure}[b]{.4\textwidth}
                 \centering
    \includegraphics[width=\linewidth]{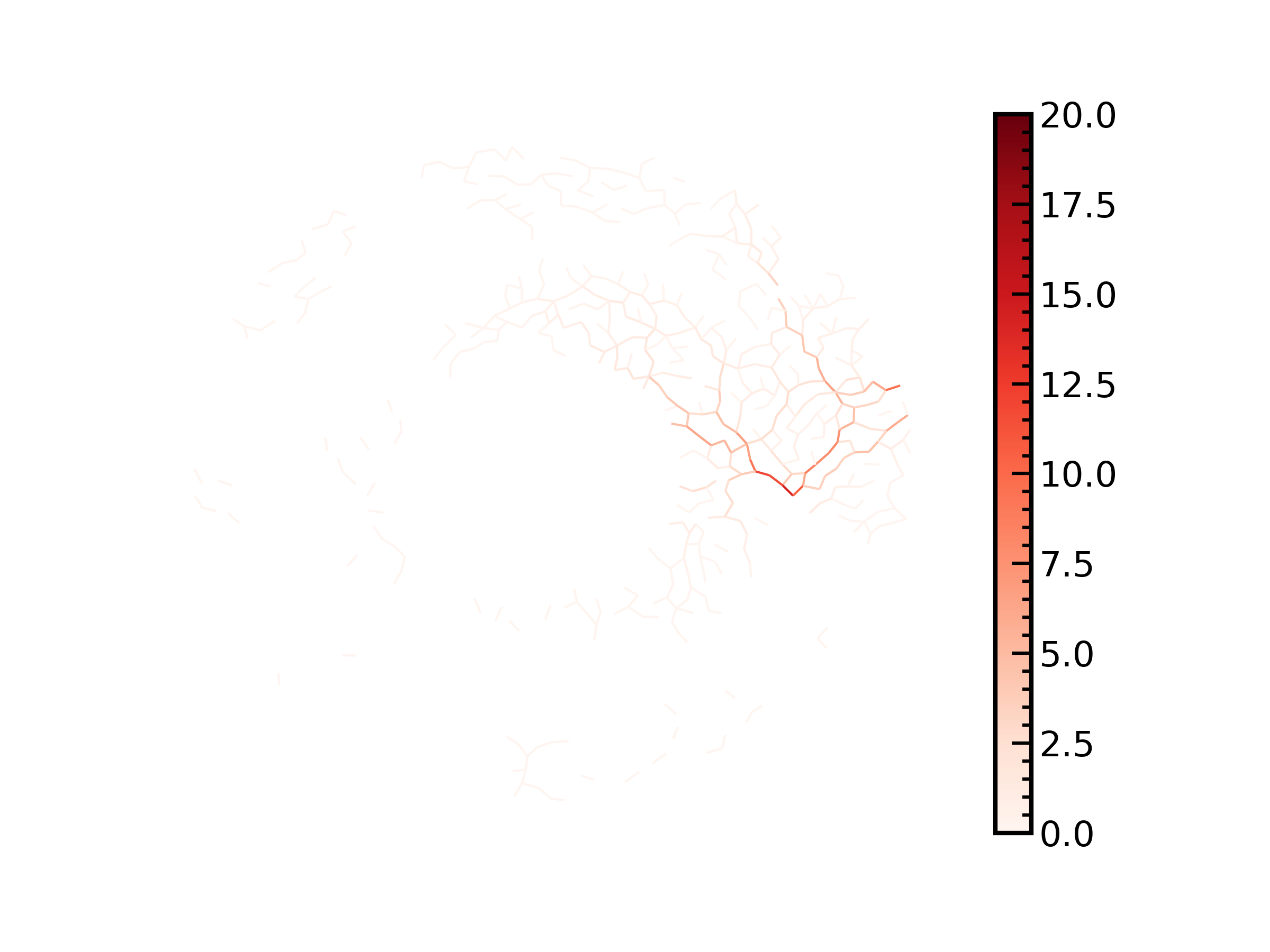}
    \caption{Pentagons, FC.}
    \end{subfigure}
             \hfill
             \begin{subfigure}[b]{.4\textwidth}
                 \centering
    \includegraphics[width=\linewidth]{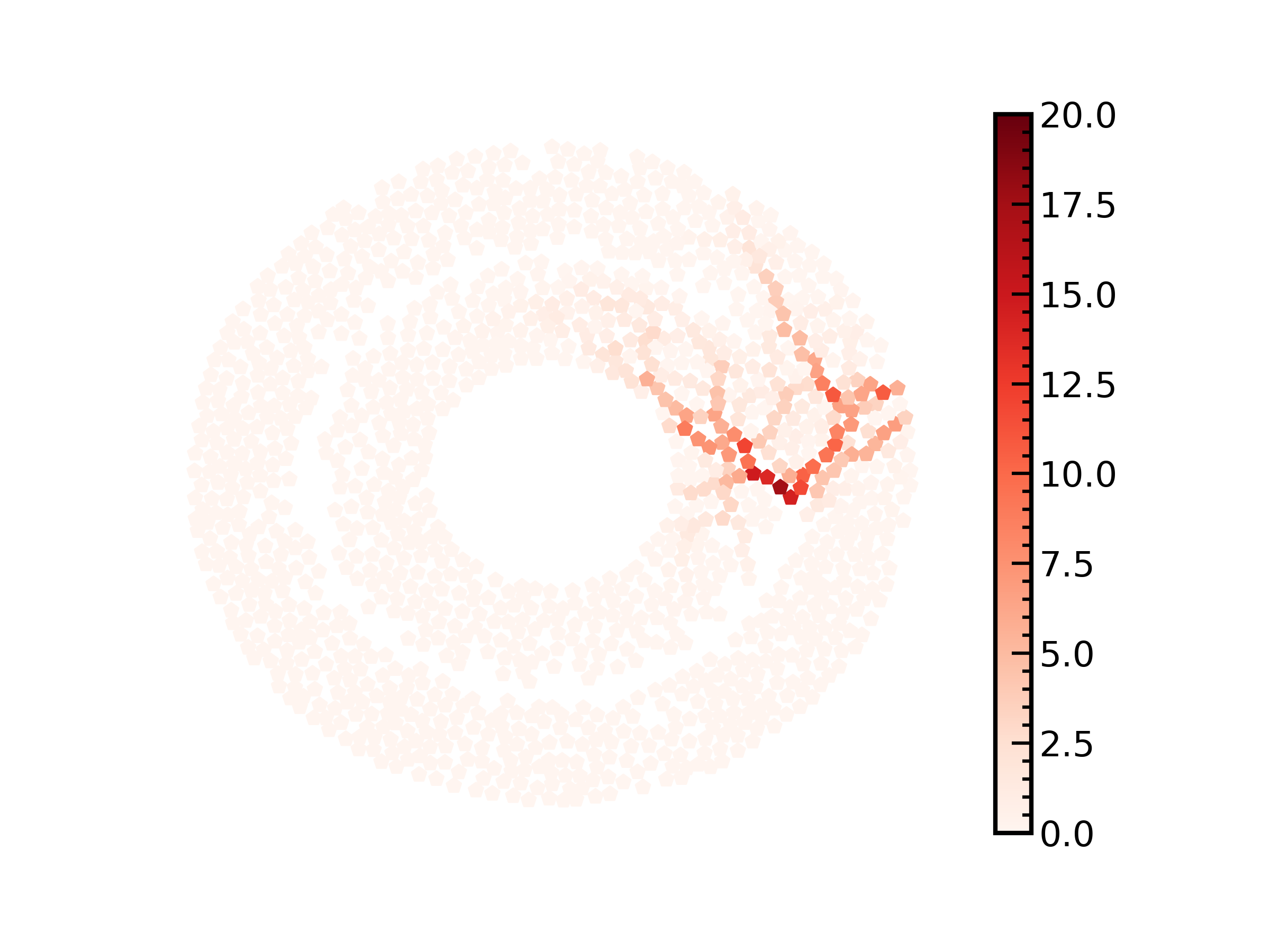}
    \caption{Pentagons, FP.}
    \end{subfigure}
    \caption{Snapshot of force networks obtained from simulation results, for disks (a-b) at 
    packing fraction $\phi = 0.78$, 
    and pentagons (c-d) at $\phi = 0.62$.  The information obtained from 
    simulations is the same in (a, b) and (c, d), but in (a, c) we use the 
    force contact (FC) information, while in (b, d) we use the force on a particle (FP) information only.  
    The color bars represent the 
    normalized forces $\hat f_{i,j}$ and $\hat f_{i}$ for the FC and FP networks, respectively, 
    as discussed in the text.  All results are obtained in the simulations that include
    basal friction.  Animations of the networks are available as Supplementary Materials, see FN-disk and FP-pent.  
        }
    \label{fig:sim}
\end{figure}

Figure~\ref{fig:sim} (see also associated animations FN-disk and FN-pent) shows the 
FC and FP networks obtained by 
simulations based on the same setup as the experiments, depicted in Fig.~\ref{fig:exp_forces}.  
As we have already discussed, these two networks exhibit different features and cannot be 
directly related.  Instead, we show that the time evolution of the features 
exhibited by these networks is correlated. We will demonstrate this by first extracting the 
topological measures introduced in the previous section for a large number of both networks 
and then cross-correlating them. 

\begin{figure}[!ht]
    \centering
             \begin{subfigure}[b]{.35\textwidth}
                 \centering
    \includegraphics[width=.8\linewidth]{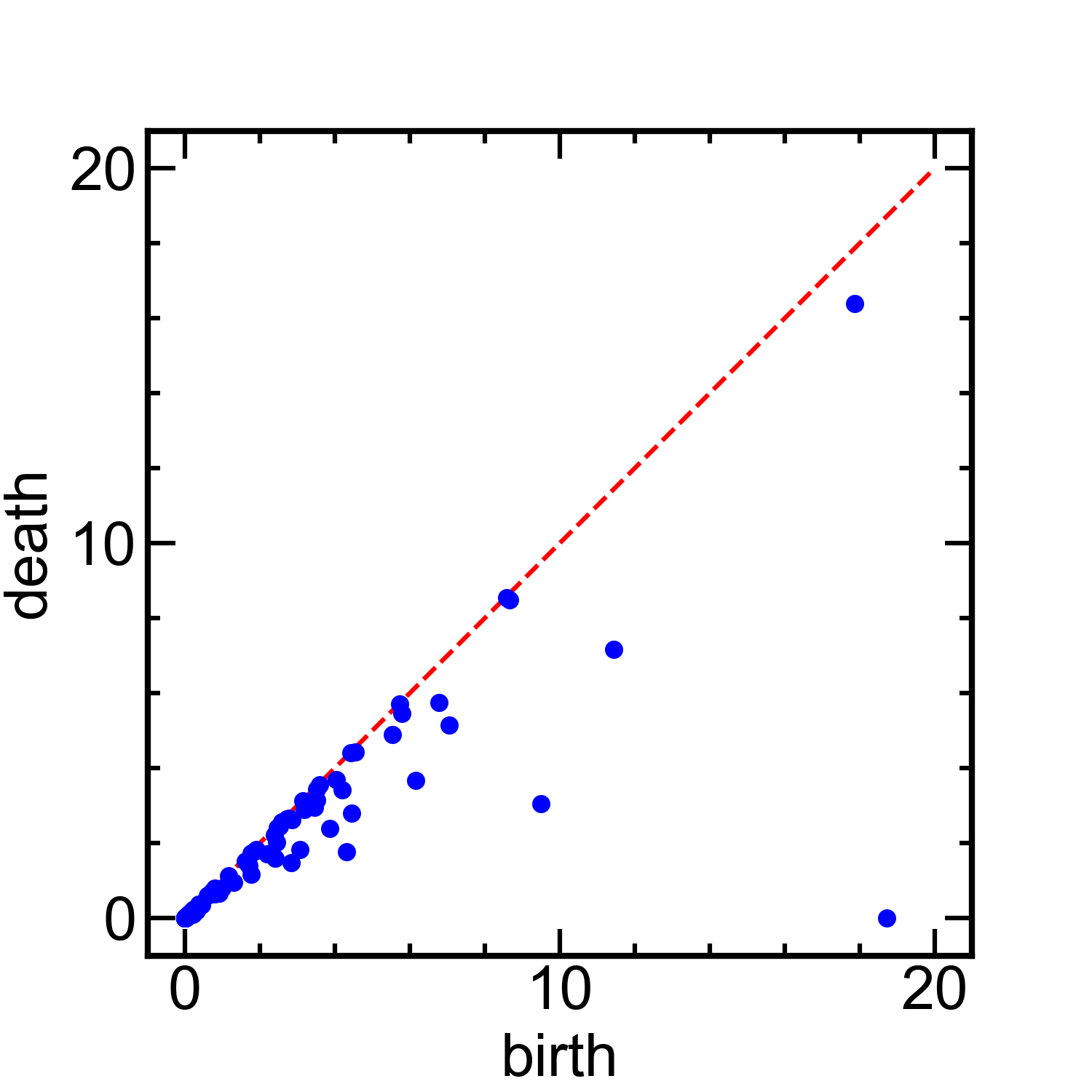}
    \caption{Disks, FC, $\beta_0$.}
    \end{subfigure}
    \hfill
             \begin{subfigure}[b]{.35\textwidth}
                 \centering
    \includegraphics[width=.8\linewidth]{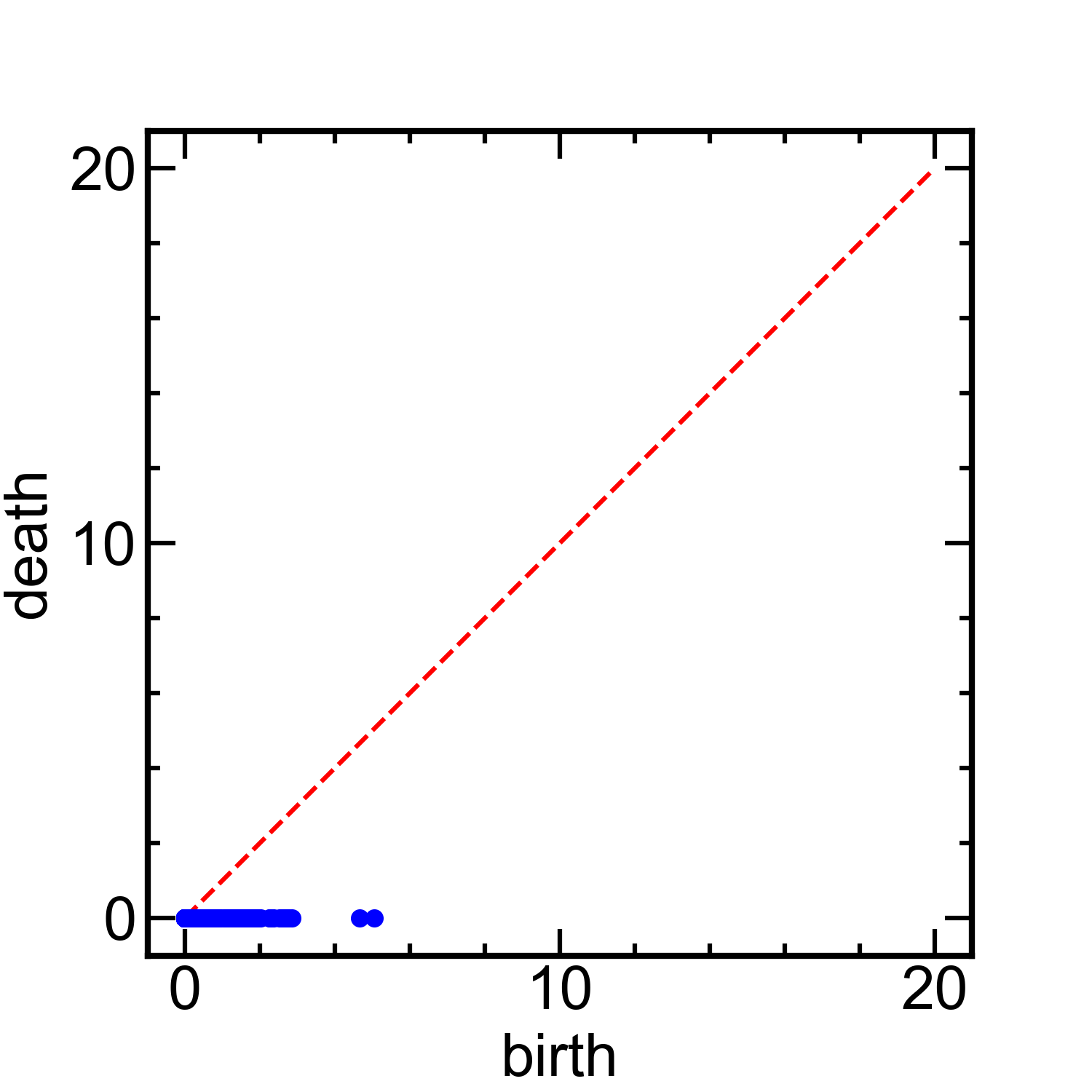}
    \caption{Disks, FC, $\beta_1$.}
    \end{subfigure}
             \hfill
             \begin{subfigure}[b]{.35\textwidth}
                 \centering
    \includegraphics[width=.8\linewidth]{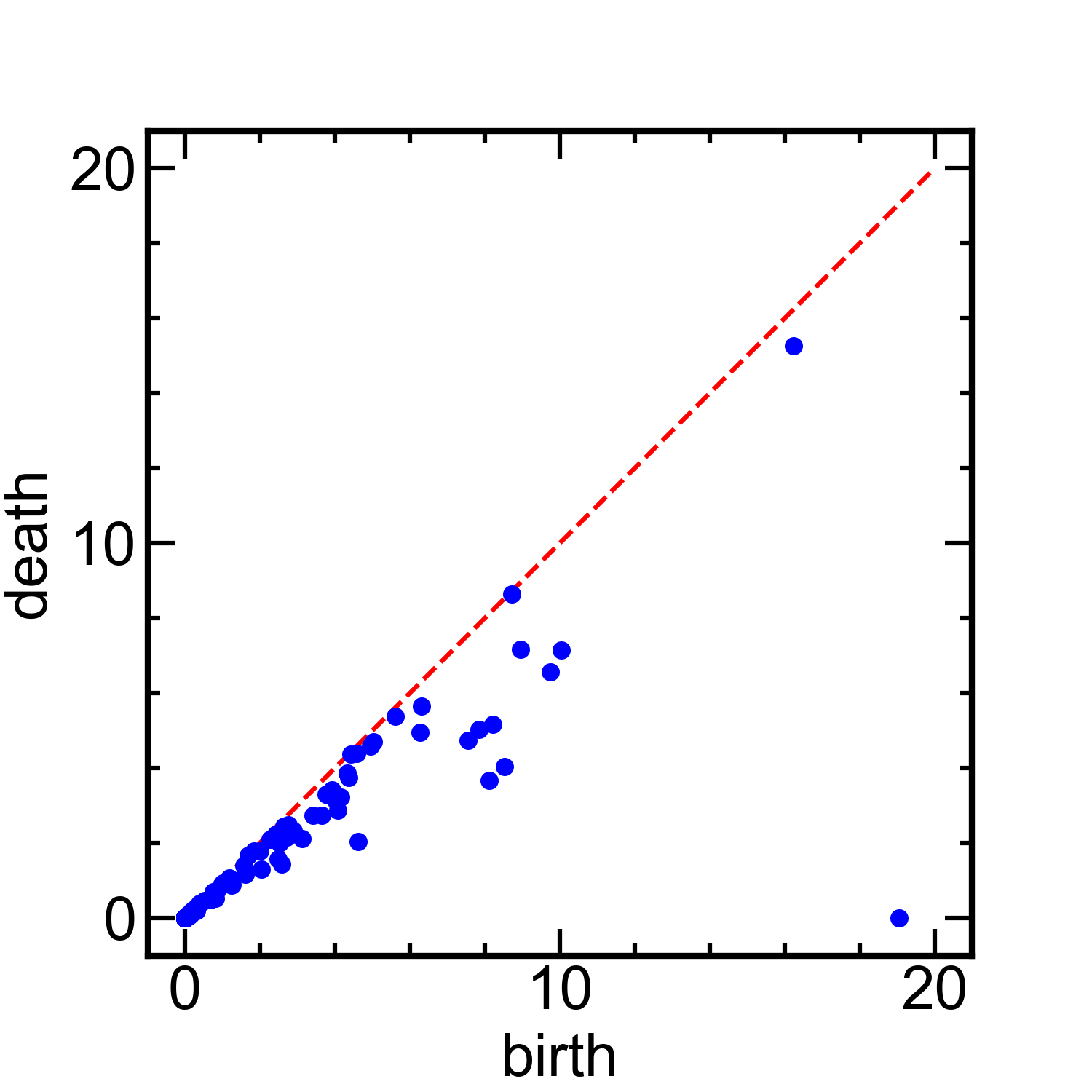}
    \caption{Disks, FP, $\beta_0$.}
    \end{subfigure}
             \hfill
             \begin{subfigure}[b]{.35\textwidth}
                 \centering
    \includegraphics[width=.8\linewidth]{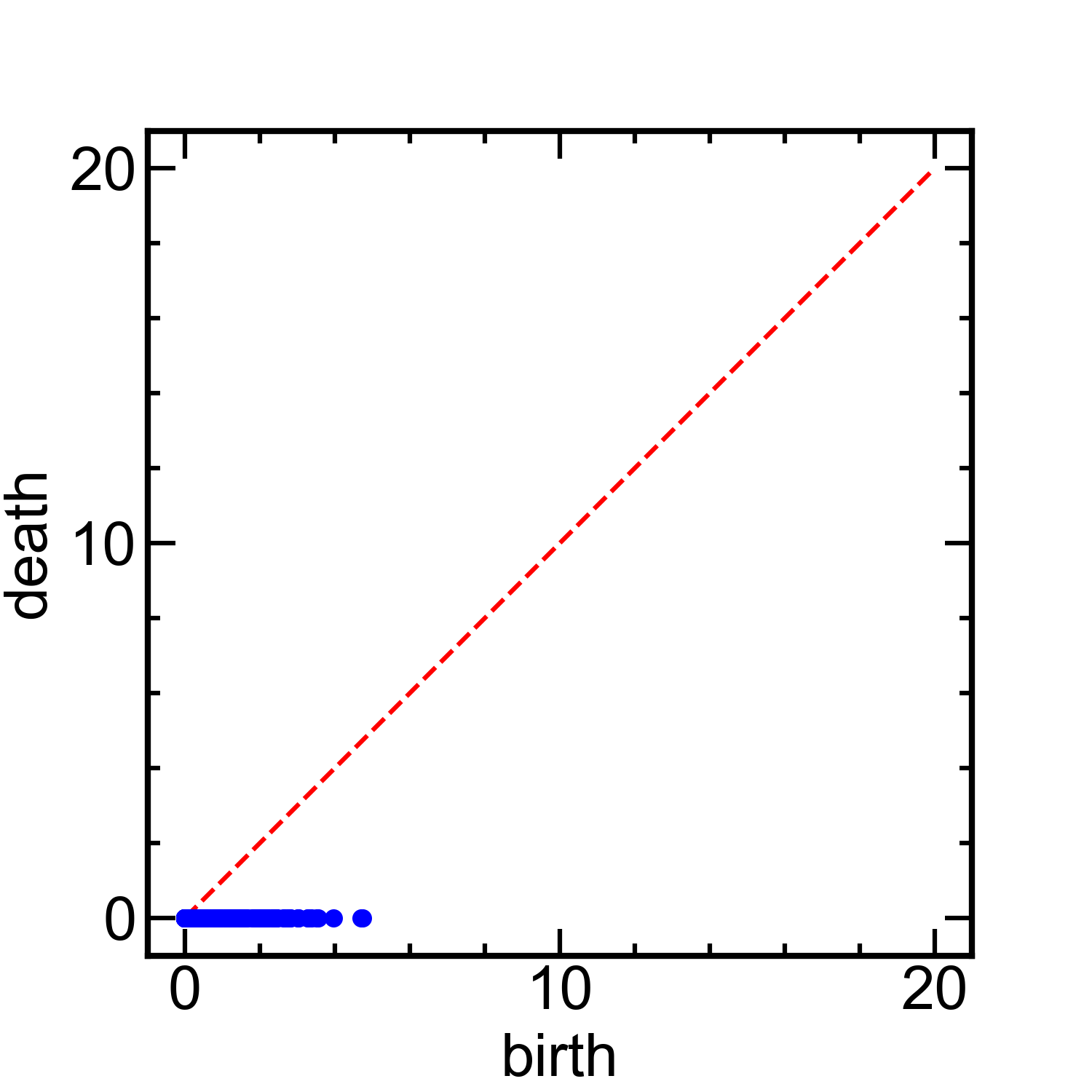}
    \caption{Disks, FP, $\beta_1$.}
    \end{subfigure}
    
    \centering
             \begin{subfigure}[b]{.35\textwidth}
                 \centering
    \includegraphics[width=.8\linewidth]{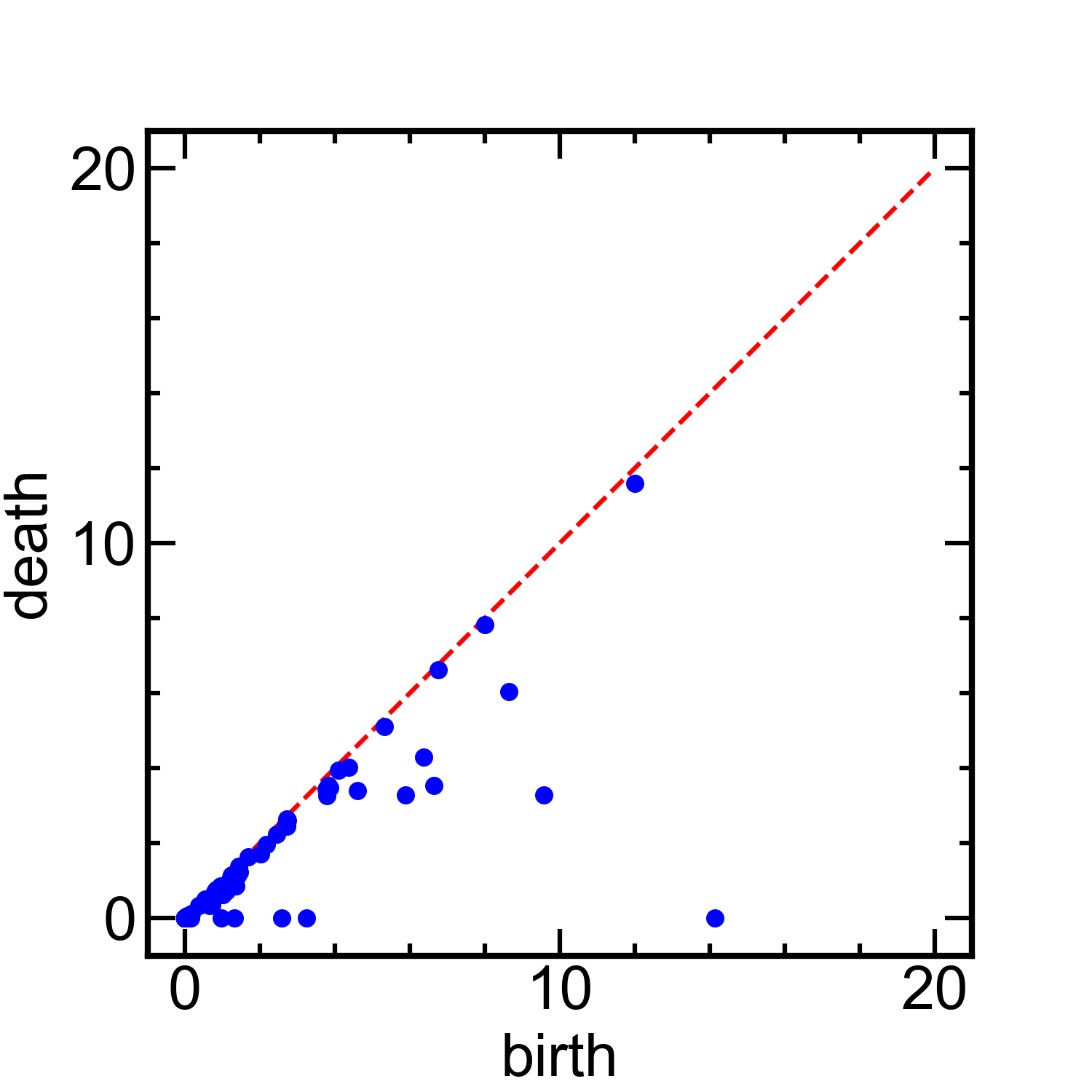}
    \caption{Pentagons, FC, $\beta_0$.}
    \end{subfigure}
    \hfill
             \begin{subfigure}[b]{.35\textwidth}
                 \centering
    \includegraphics[width=.8\linewidth]{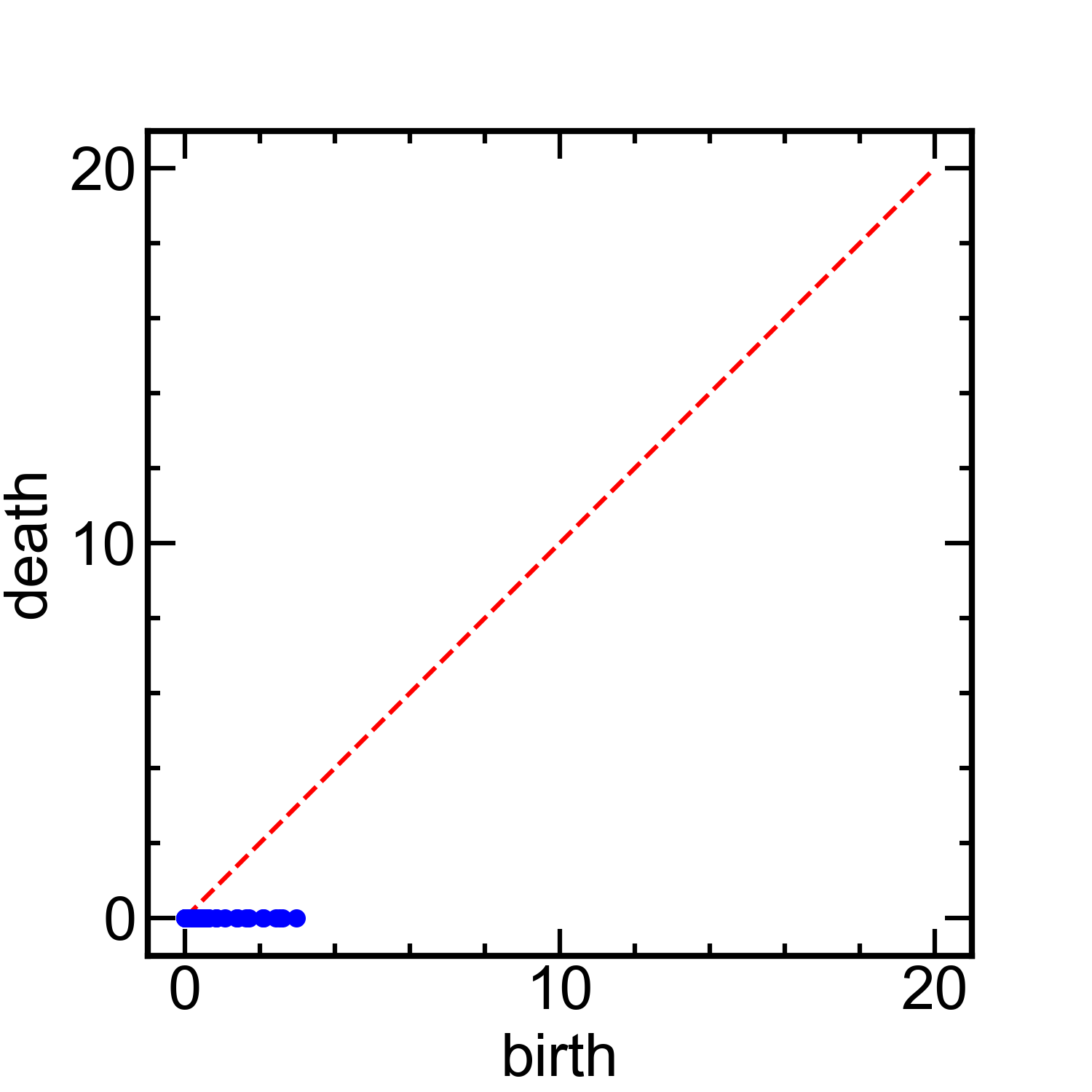}
    \caption{Pentagons, FC, $\beta_1$.}
    \end{subfigure}
             \hfill
             \begin{subfigure}[b]{.35\textwidth}
                 \centering
    \includegraphics[width=.8\linewidth]{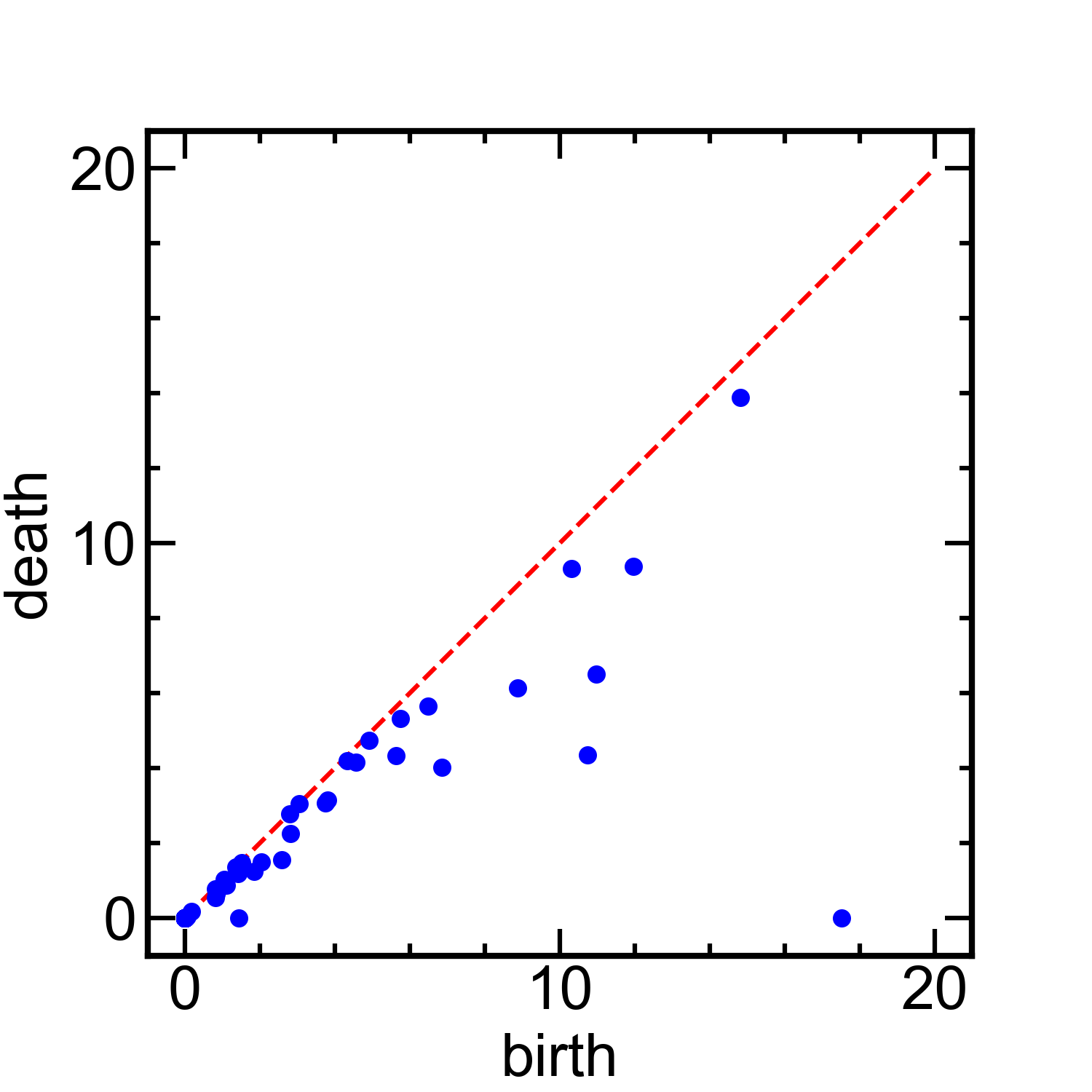}
    \caption{Pentagons, FP, $\beta_0$.}
    \end{subfigure}
             \hfill
             \begin{subfigure}[b]{.35\textwidth}
                 \centering
    \includegraphics[width=.8\linewidth]{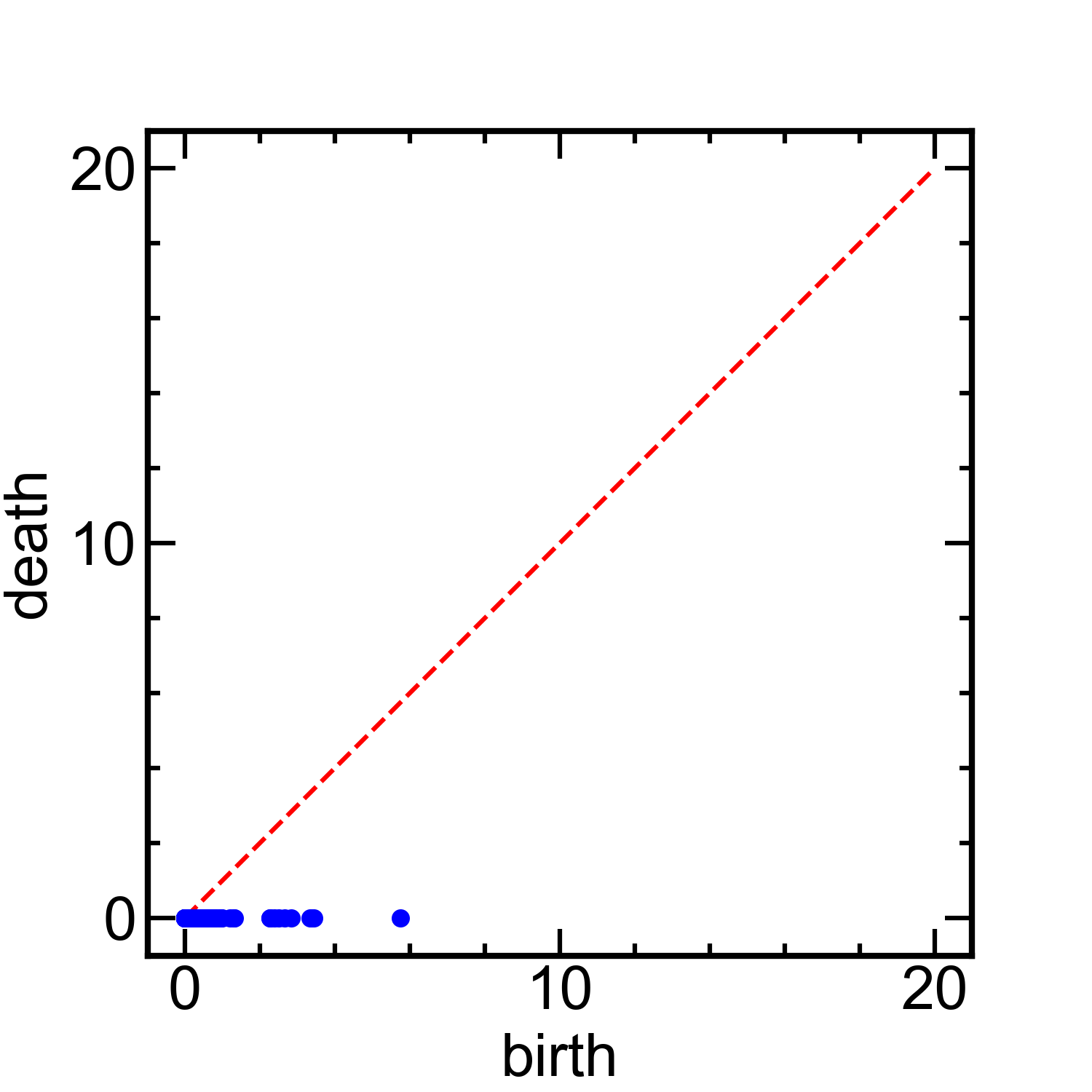}
    \caption{Pentagons, FP, $\beta_1$. }
    \end{subfigure}
    \caption{Persistence diagrams (PDs) corresponding to the networks shown in Fig.~\ref{fig:sim}.
    Animations of the $\beta_0$ PDs are available, see pd-disk and pd-pent.   
        }
    \label{fig:sim_pd}
\end{figure}

The functions $f_{FC}$ and $f_{FP}$ describing the FC and FP networks are 
defined as in the previous section using the normal forces between the particles.  
In the considered system, the average force between 
the particles fluctuates significantly during the evolution, so we normalize the 
computed forces by their (time-dependent) average first.  
The function  $f_{FC}$ is normalized  by the sum of its original values $f_{FC}(e)$ 
over the  edges $e \in CN$ divided by the number of edges (only the edges 
characterized by non-zero forces are considered).  
The function $f_{FP}$ is normalized by the sum of its original values 
$f_{FC}(v)$ over the vertices (particles) $v \in \CN$ divided by the number of 
vertices, again considering only the vertices (particles) characterized by 
a nonzero force.  Figure~\ref{fig:sim} shows these normalized functions.
For visualization purposes, each vertex is plotted as the approximate size of the 
particle at whose center it resides.

The persistence diagrams corresponding to Fig.~\ref{fig:sim}, are computed from
the $\FC(\theta)$  and $\FP(\theta)$ networks, respectively.  
Figure~\ref{fig:sim_pd} shows the corresponding diagrams (see also associated
animations, pd-disk and pd-pent).  
We will analyze the properties of a large number of such diagrams in the next section, focusing 
mostly on two measures: the total persistence (TP) and the number of generators, N$_{\rm G}$.  
In interpreting the results, it is useful to remember that the generators 
that are close to the diagonal represent the features that persist just for a small range 
of thresholds and therefore are not significant. The significant features are the ones that 
are far away from the diagonal. We will see later in the paper that excluding these insignificant 
features may help considerably in relating the FC and FP networks.

\subsection{Comparison of contact force and particle force networks: Disks with basal friction}  

After specifying how the PDs for the two types of networks are computed, we now proceed with 
the comparison of the large number of diagrams, extracted from time-dependent simulation data.
In our analysis, we consider four systems, with their 
choice motivated by our previously reported results discussing intruder dynamics~\cite{carlevaro_pre_2020}.  
First, we discuss the results obtained using disks in simulations that include the basal friction 
(friction with the substrate) for packing fraction $\phi = 0.78$; then we proceed with pentagons
for $\phi = 0.62$.  In such systems, for the simulation parameters that we use, the intruder exhibits stick-slip dynamics.  Then we proceed with briefly considering the same particles shapes
and $\phi$'s, but without basal friction.  Such systems experience clogging type of dynamics.   
These four considered systems therefore differ by both particle shapes and the type of 
dynamics.

We start by considering disks with basal friction, measuring the total persistence (TP) 
and the number of generators (N$_{\rm G}$) in the PDs, as discussed in the Methods section.  Both measures
are considered for all force thresholds, and also separately for the 
forces with the birth coordinate above (TP$_{\rm above}$) and below (TP$_{\rm below}$) the mean 
force (similarly for N$_{\rm G}$). All the 
measures are considered for both FC and FP networks, and both for the components ($\beta_0$) and for the  
loops ($\beta_1$). On each figure we also plot the magnitude of the intruder's velocity, so to facilitate
the comparison between the PH-derived measures and the intruder's dynamics.

\begin{figure}[!ht]
    \centering
             \begin{subfigure}[b]{.4\textwidth}
                 \centering
    \includegraphics[width=\linewidth]{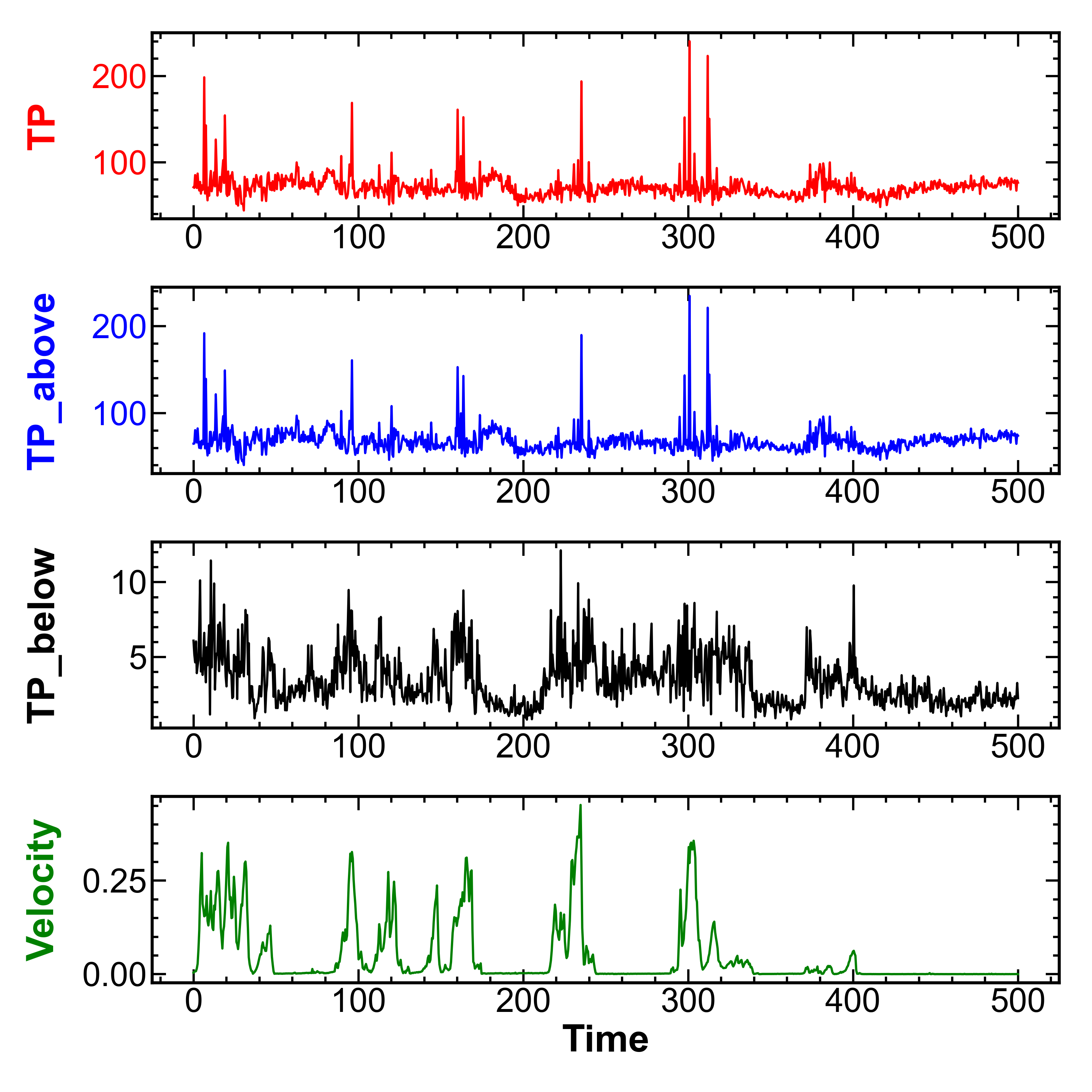}
    \caption{FC, TP.}
    \end{subfigure}
             \hfill
             \begin{subfigure}[b]{.4\textwidth}
                 \centering
    \includegraphics[width = \linewidth]{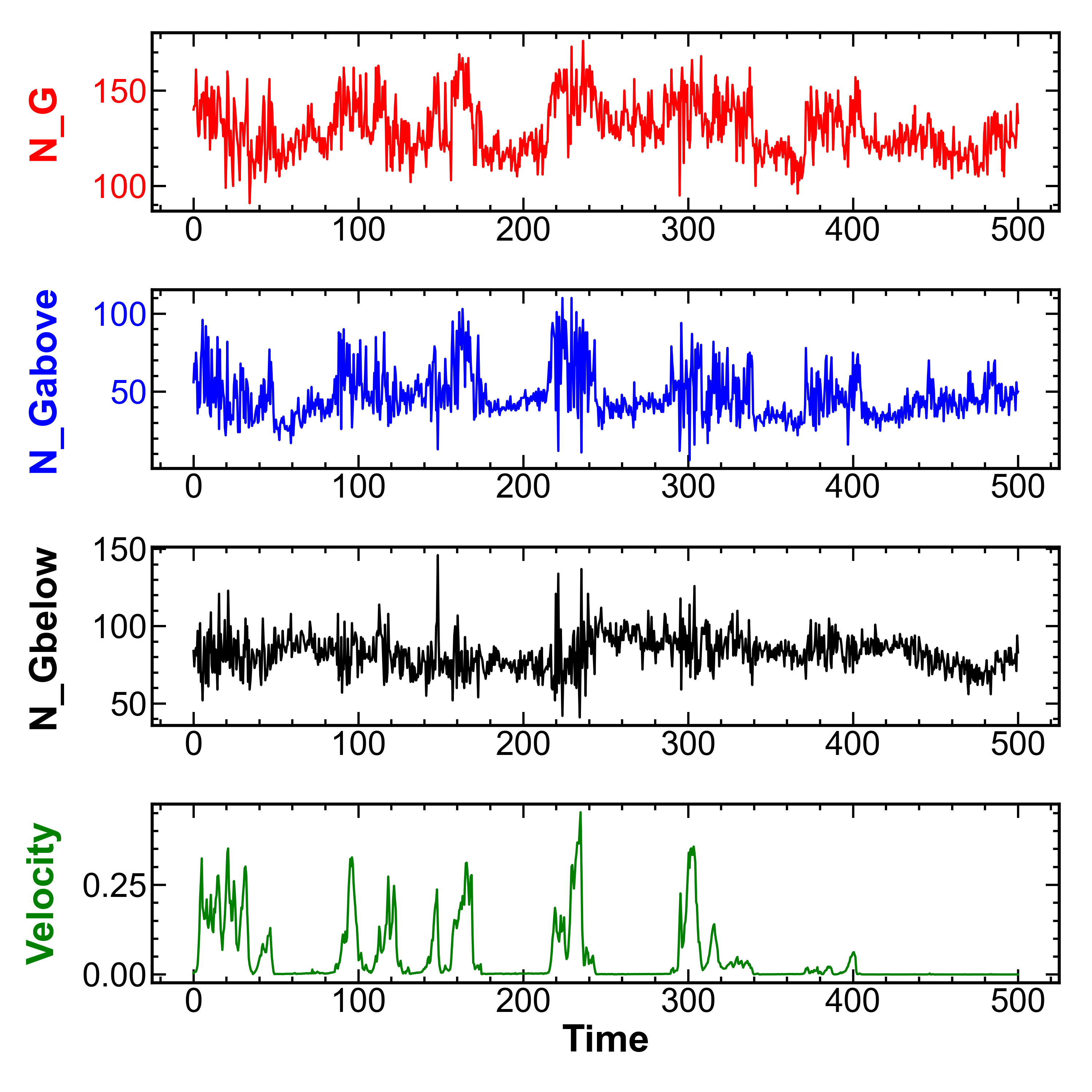}
    \caption{FC, N$_{\rm G}$.}
    \end{subfigure}
             \hfill
             \begin{subfigure}[b]{.4\textwidth}
                 \centering
    \includegraphics[width=\linewidth]{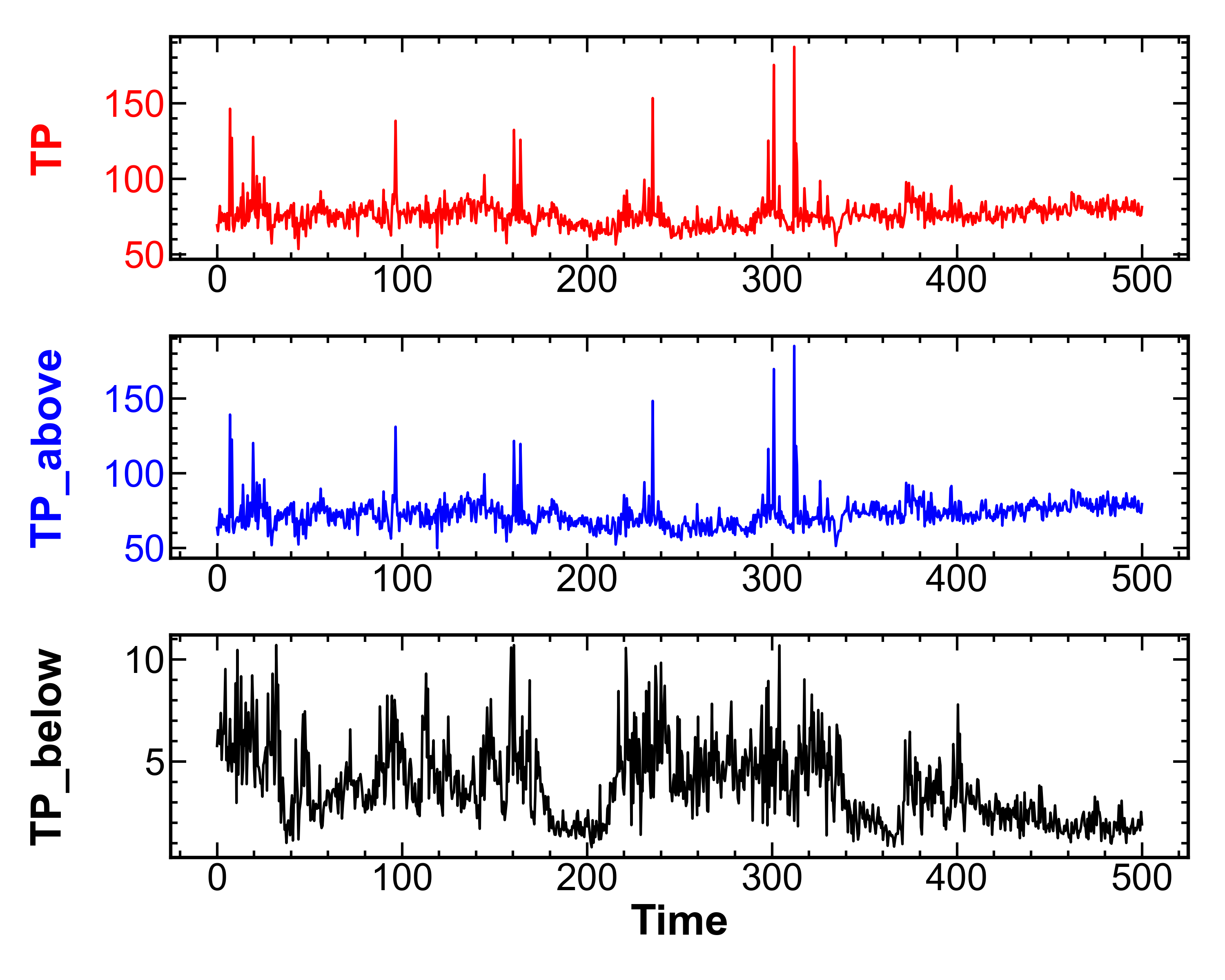}
    \caption{FP, TP.}
    \end{subfigure}
             \hfill
             \begin{subfigure}[b]{.4\textwidth}
                 \centering
                 \includegraphics[width = \linewidth]{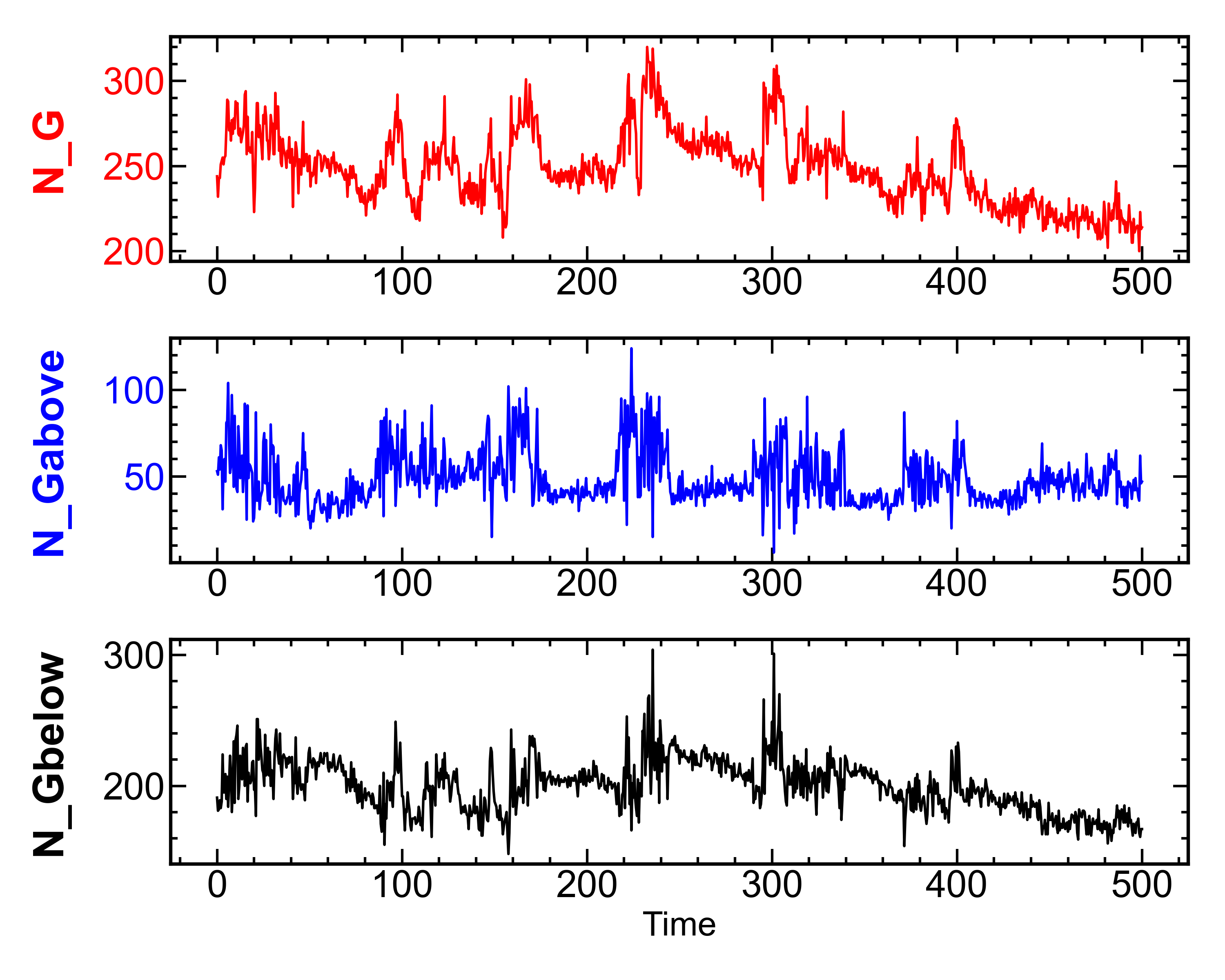}
                 \caption{FP, N$_{\rm G}$.}
             \end{subfigure}
             
    \caption{Disks with basal friction, $\beta_0$ (components); total persistence (TP), and 
    number of generators,  N$_{\rm G}$, for the force contact network (FC) and the force particle network (FP).
    The bottom plot in (a) and (b) shows the magnitude of the intruder's velocity (the velocity plots in 
    (a) and (b) are identical, and are replotted for the ease of comparison with the force network 
    results). One unit of time in this and the following figures correspond to $1000 \delta t$.
        }
    \label{fig:TP0_disk_withbf}
\end{figure}

\begin{figure}[!ht]
    \centering
             \begin{subfigure}[b]{.4\textwidth}
                 \centering
    \includegraphics[width=\linewidth]{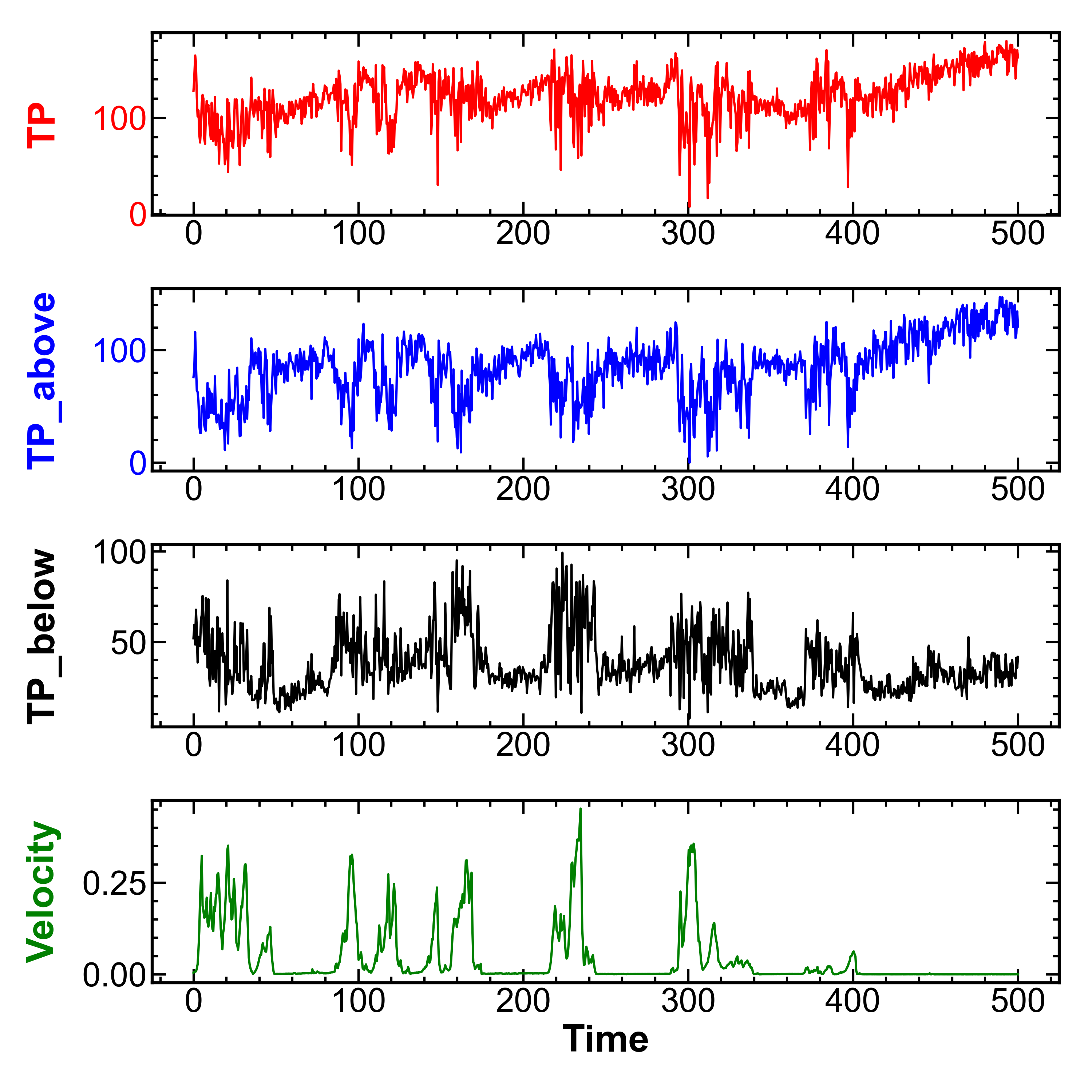}
    \caption{FC, TP.}
    \end{subfigure}
             \hfill
             \begin{subfigure}[b]{.4\textwidth}
                 \centering
    \includegraphics[width = \linewidth]{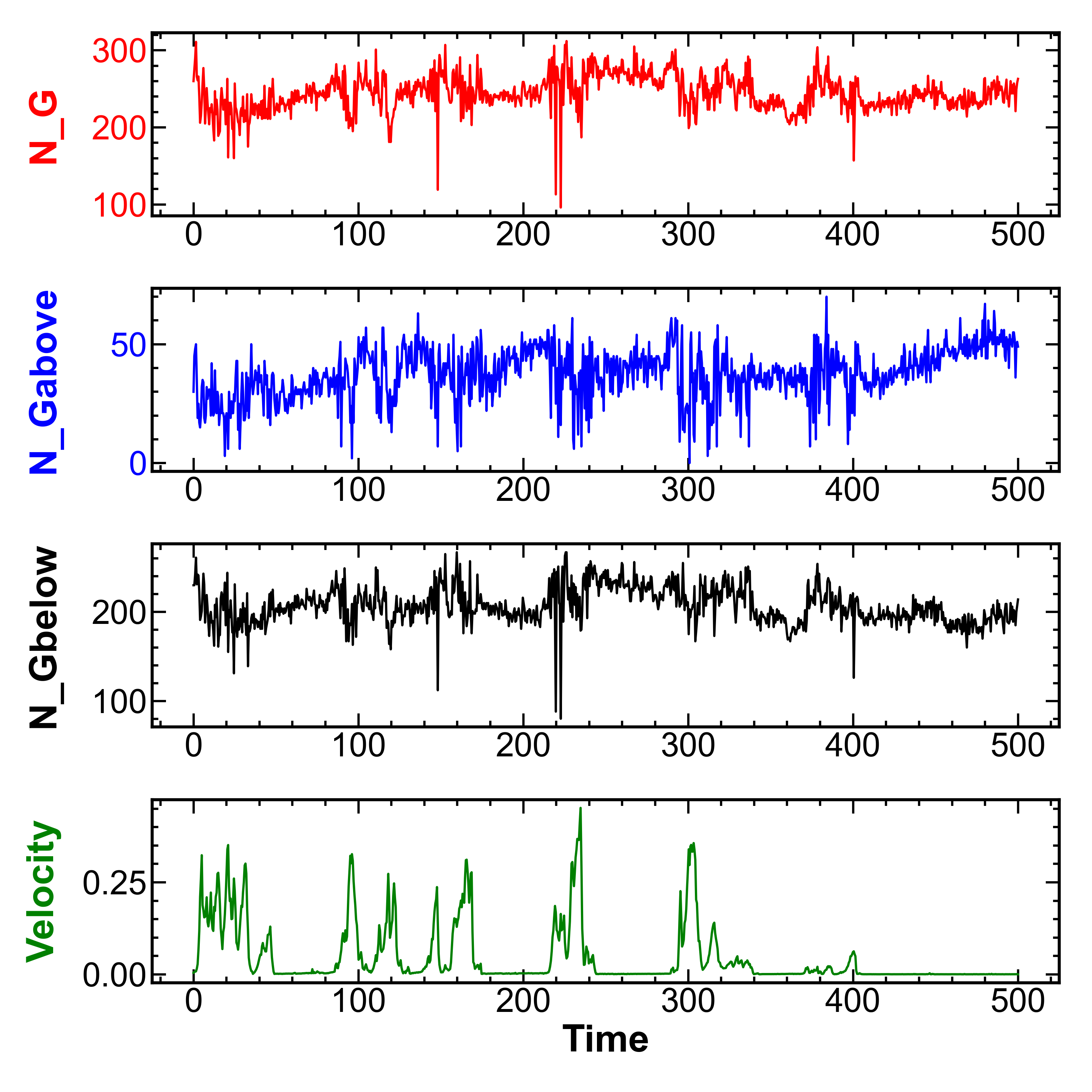}
    \caption{FC, N$_{\rm G}$.}
    \end{subfigure}
             \hfill
             \begin{subfigure}[b]{.4\textwidth}
                 \centering
    \includegraphics[width=\linewidth]{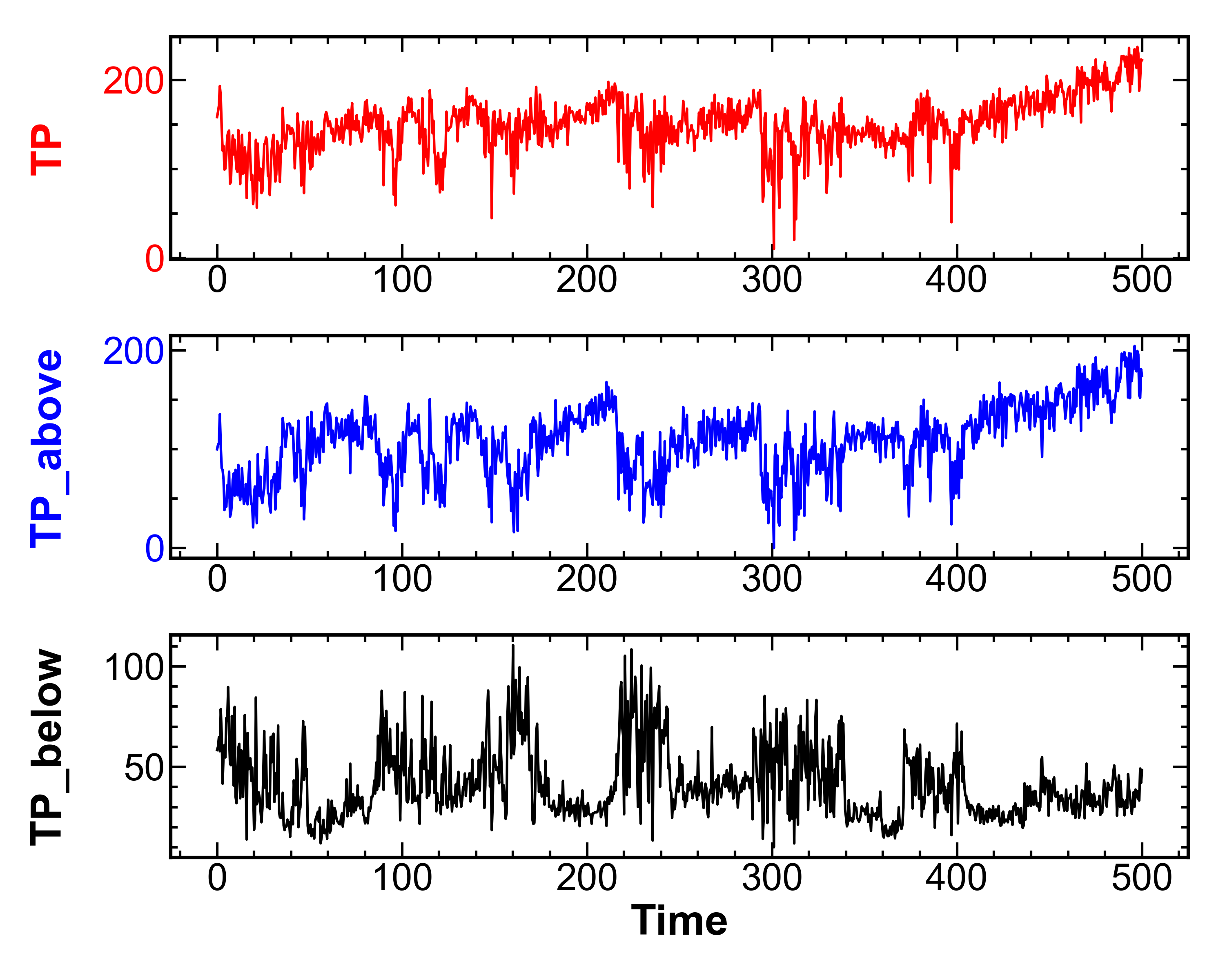}
    \caption{FP, TP.}
    \end{subfigure}
             \hfill
             \begin{subfigure}[b]{.4\textwidth}
                 \centering
                 \includegraphics[width = \linewidth]{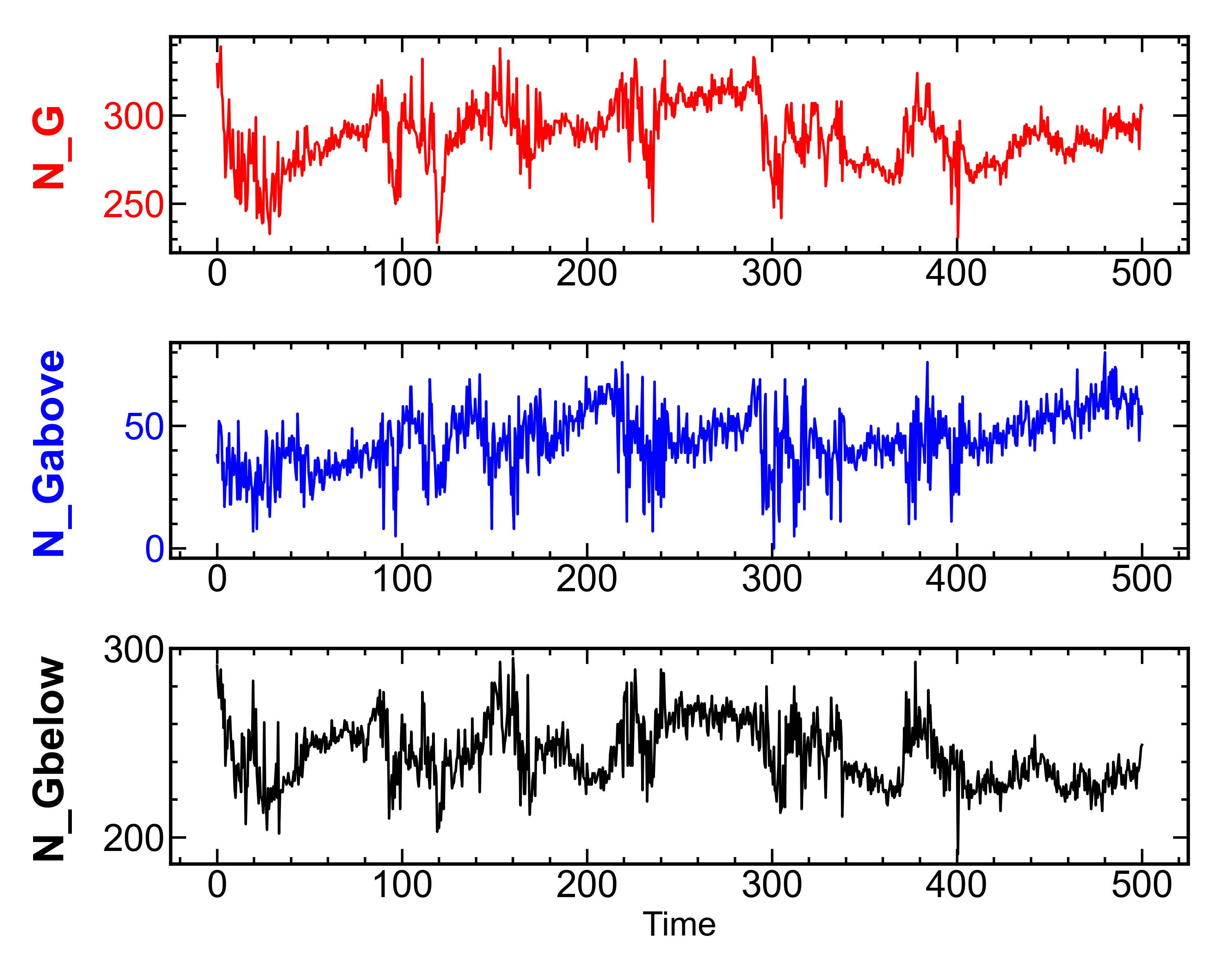}
                 \caption{FP, N$_{\rm G}$.}
             \end{subfigure}
             
    \caption{Disks with basal friction, $\beta_1$ (loops).}
    \label{fig:TP1_disk_withbf}
\end{figure}

\begin{figure}[!ht]
    \centering
             \begin{subfigure}[b]{.4\textwidth}
                 \centering
    \includegraphics[width=\linewidth]{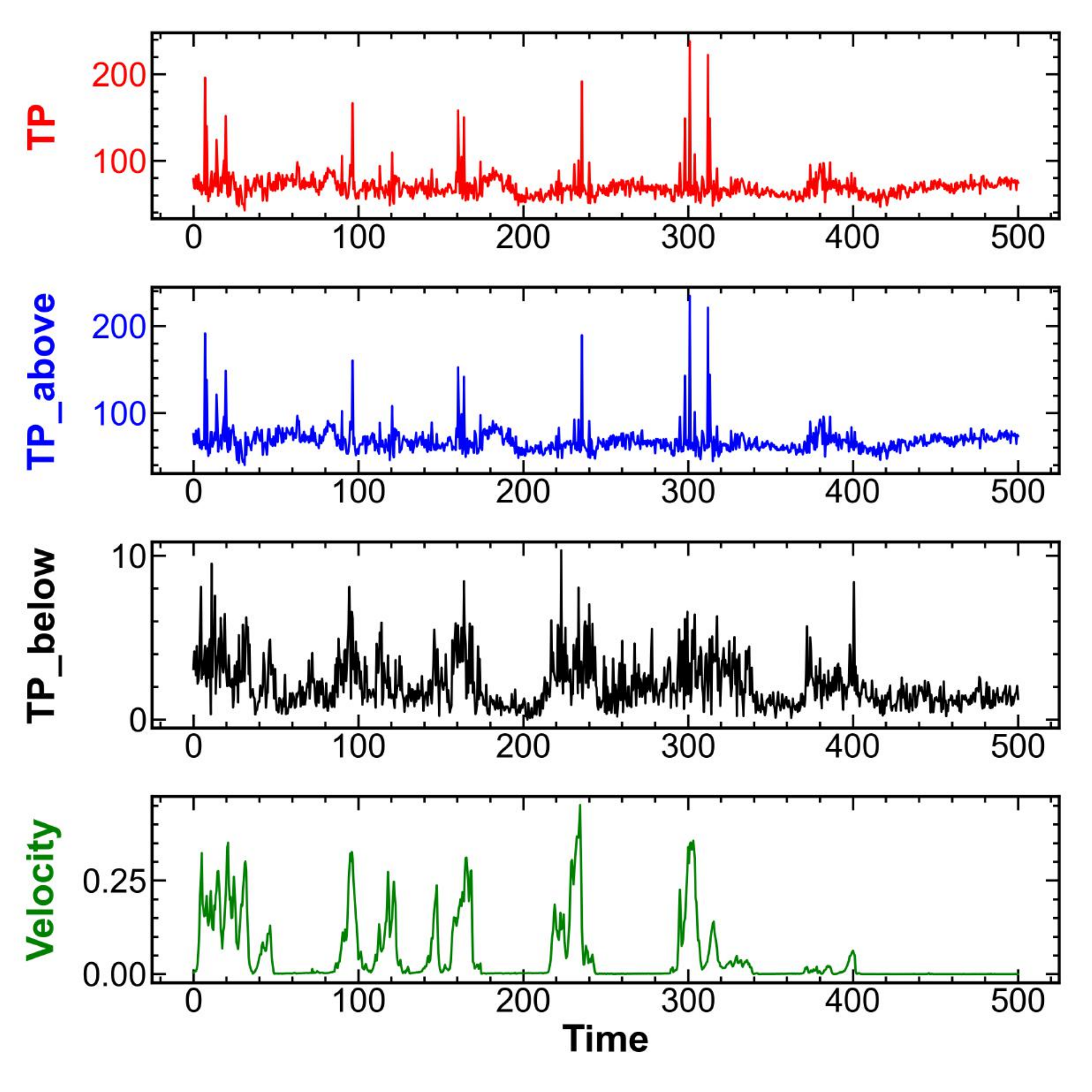}
    \caption{FC, TP.}
    \end{subfigure}
             \hfill
             \begin{subfigure}[b]{.4\textwidth}
                 \centering
    \includegraphics[width = \linewidth]{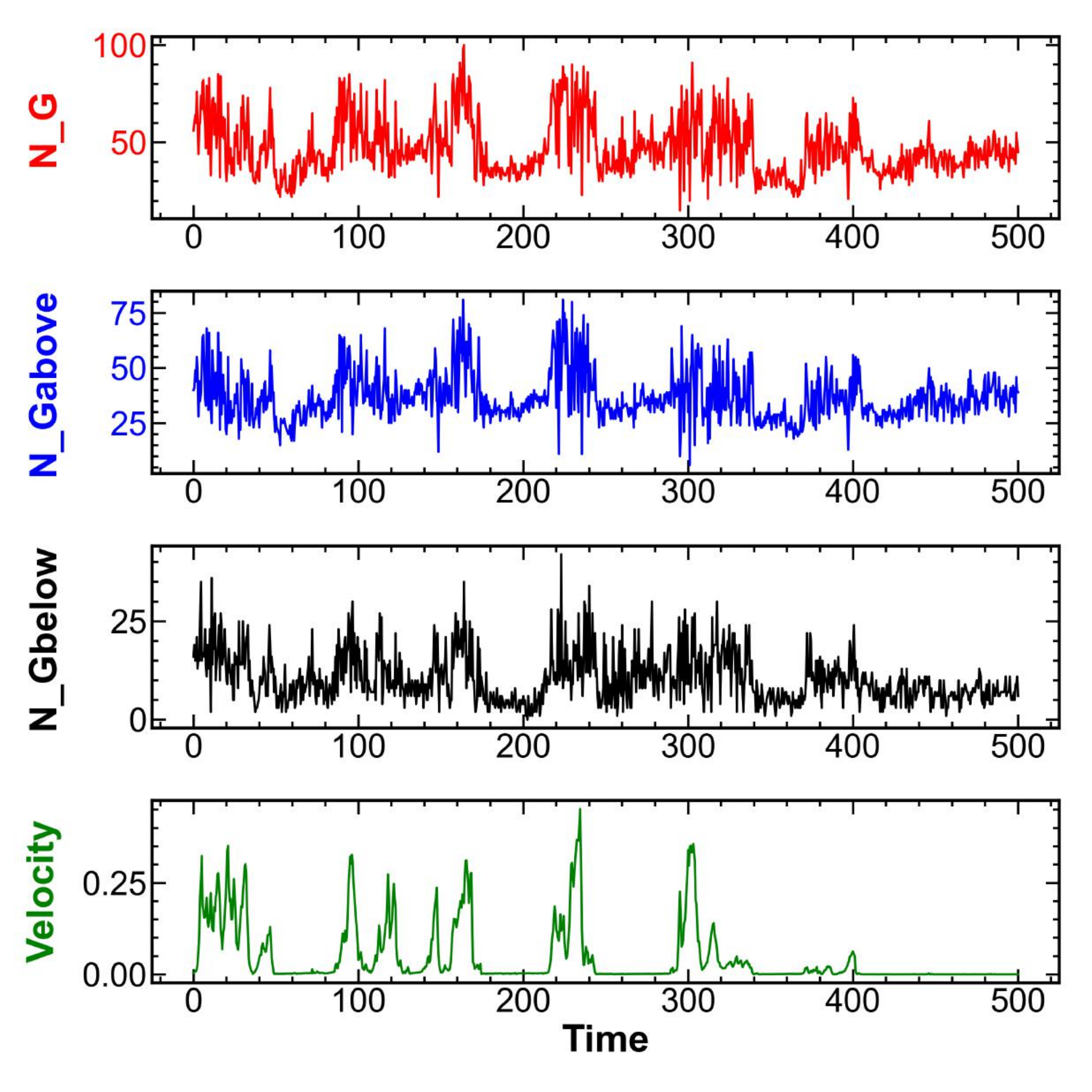}
    \caption{FC, N$_{\rm G}$ }
    \end{subfigure}
             \hfill
             \begin{subfigure}[b]{.4\textwidth}
                 \centering
    \includegraphics[width=\linewidth]{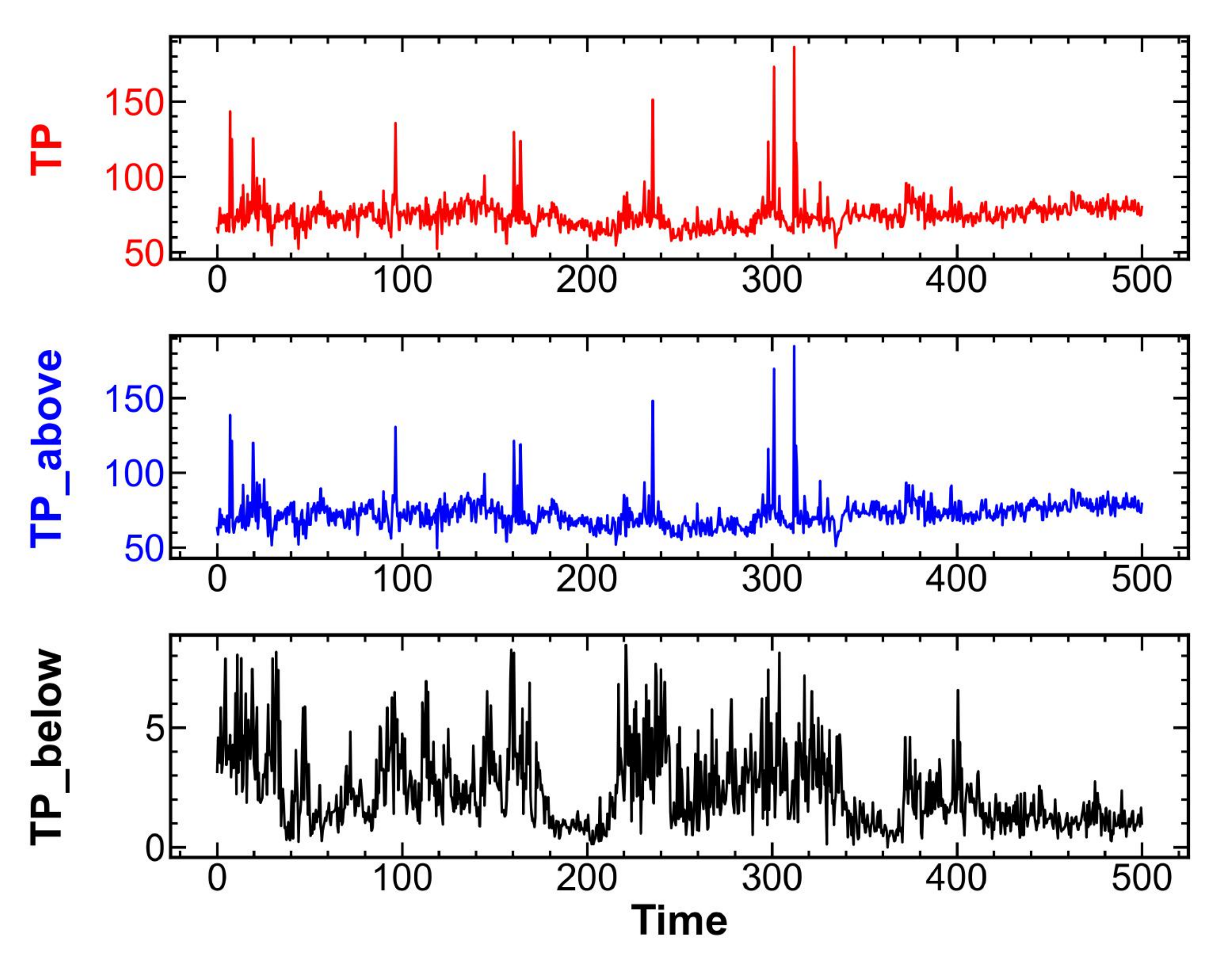}
    \caption{FP, TP.}
    \end{subfigure}
             \hfill
             \begin{subfigure}[b]{.4\textwidth}
                 \centering
                 \includegraphics[width = \linewidth]{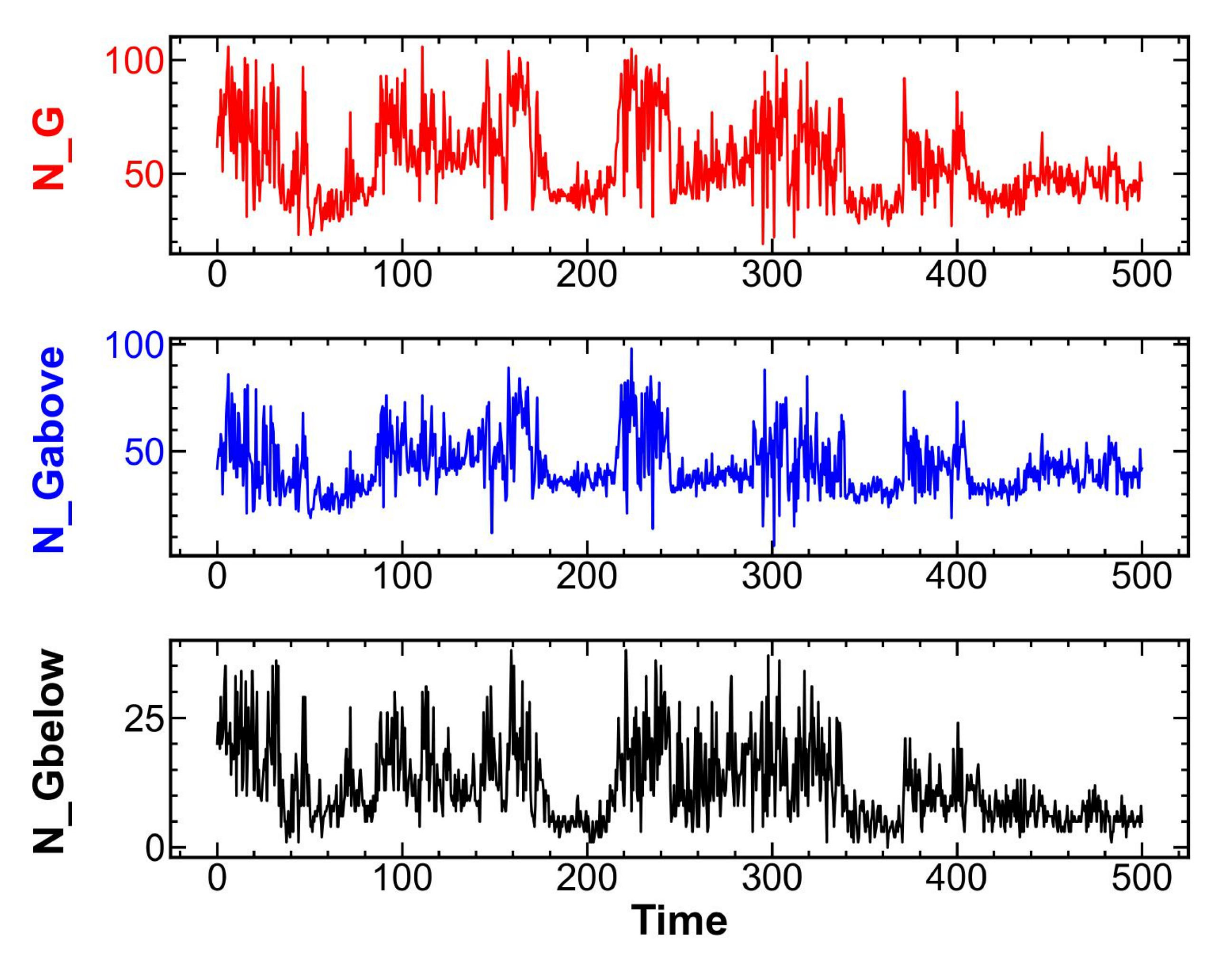}
                 \caption{FP, N$_{\rm G}$.}
             \end{subfigure}
            
    \caption{Disks with basal friction, $\beta_0$ (components) after removing the band next to 
    the diagonal; viz. Fig~\ref{fig:TP0_disk_withbf}.}
    \label{fig:TP0_disk_withbf_removeband}
\end{figure}

\begin{figure}[!ht]
    \centering
             \begin{subfigure}[b]{.4\textwidth}
                 \centering
    \includegraphics[width=\linewidth]{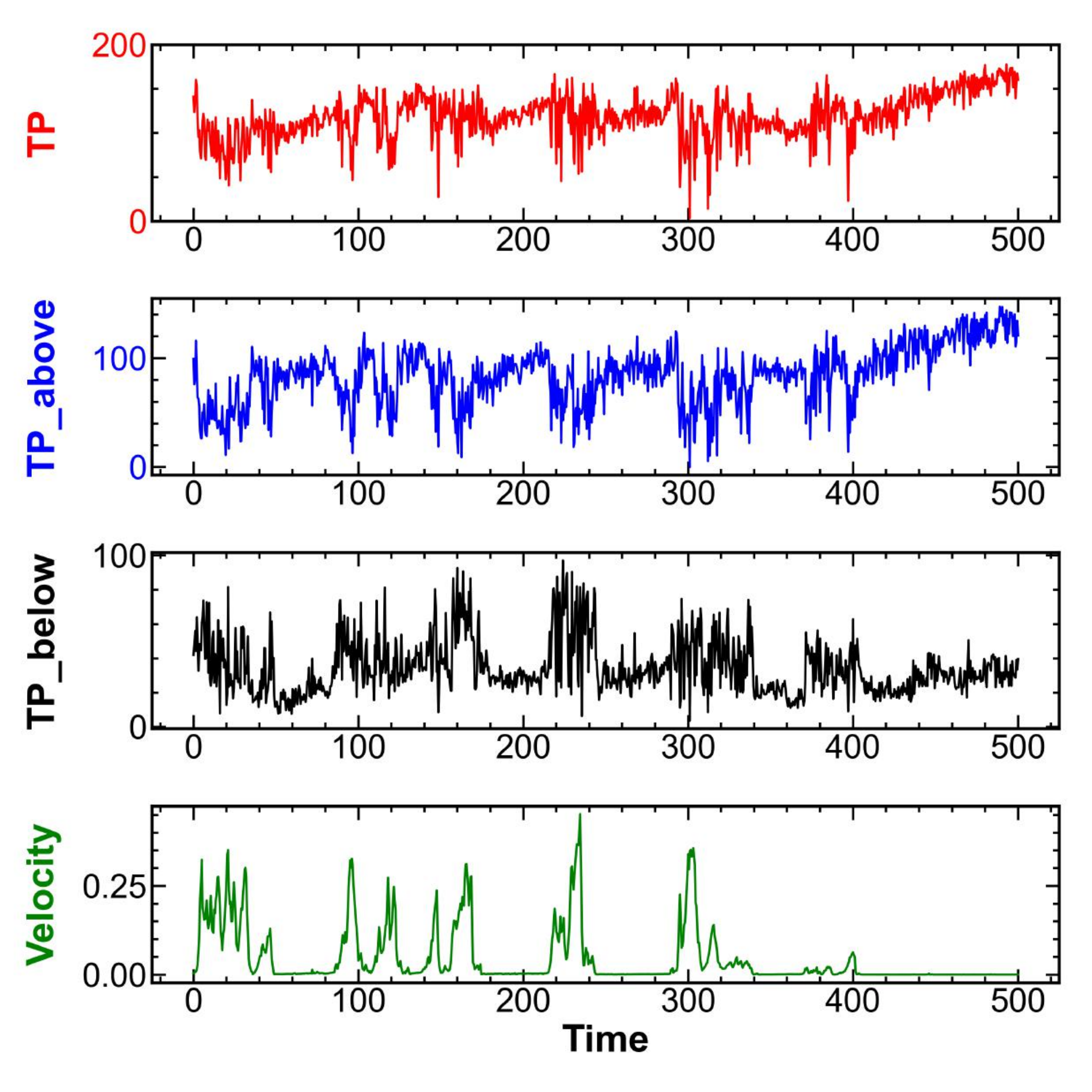}
    \caption{FC, TP.}
    \end{subfigure}
             \hfill
             \begin{subfigure}[b]{.4\textwidth}
                 \centering
    \includegraphics[width = \linewidth]{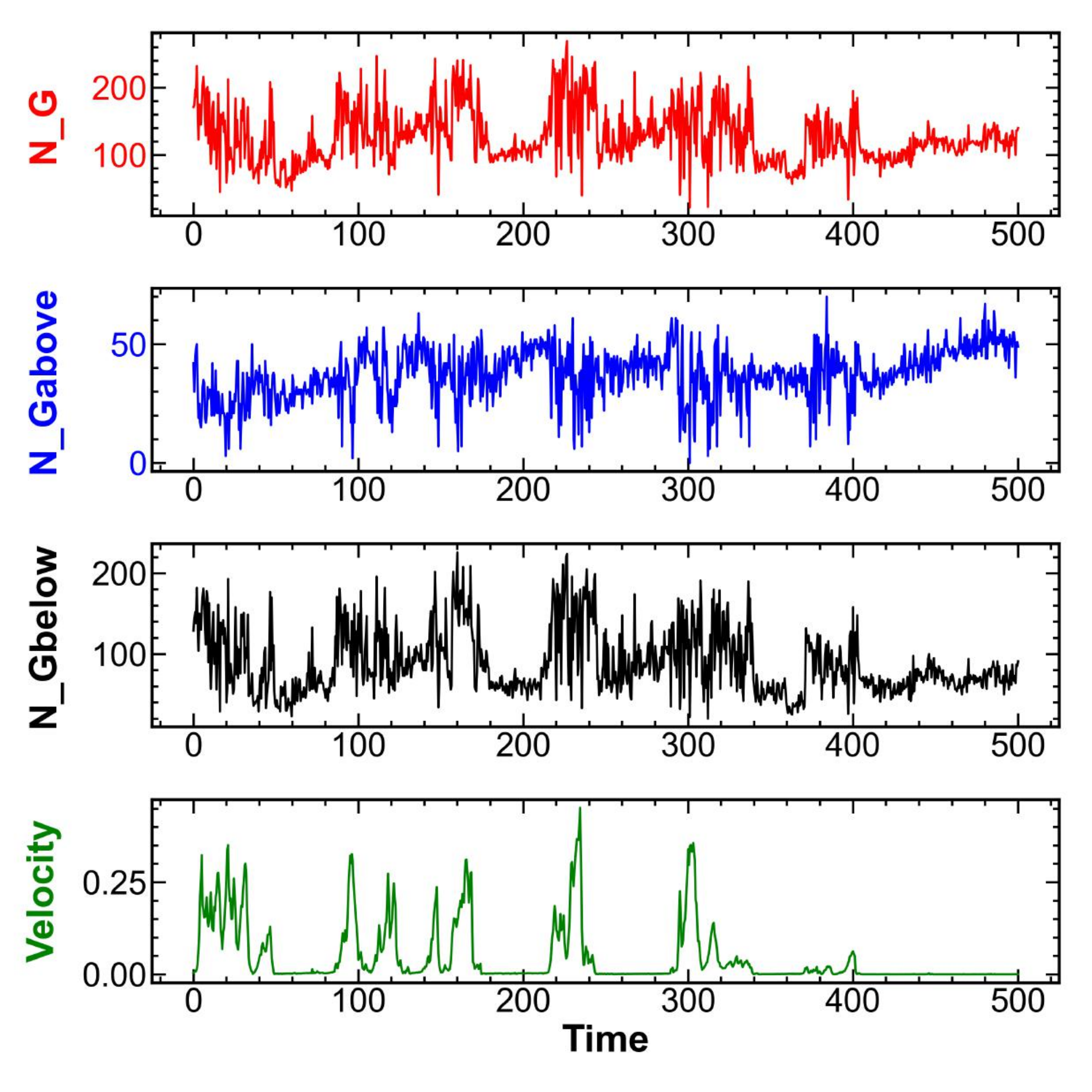}
    \caption{FC, N$_{\rm G}$.}
    \end{subfigure}
             \hfill
             \begin{subfigure}[b]{.4\textwidth}
                 \centering
    \includegraphics[width=\linewidth]{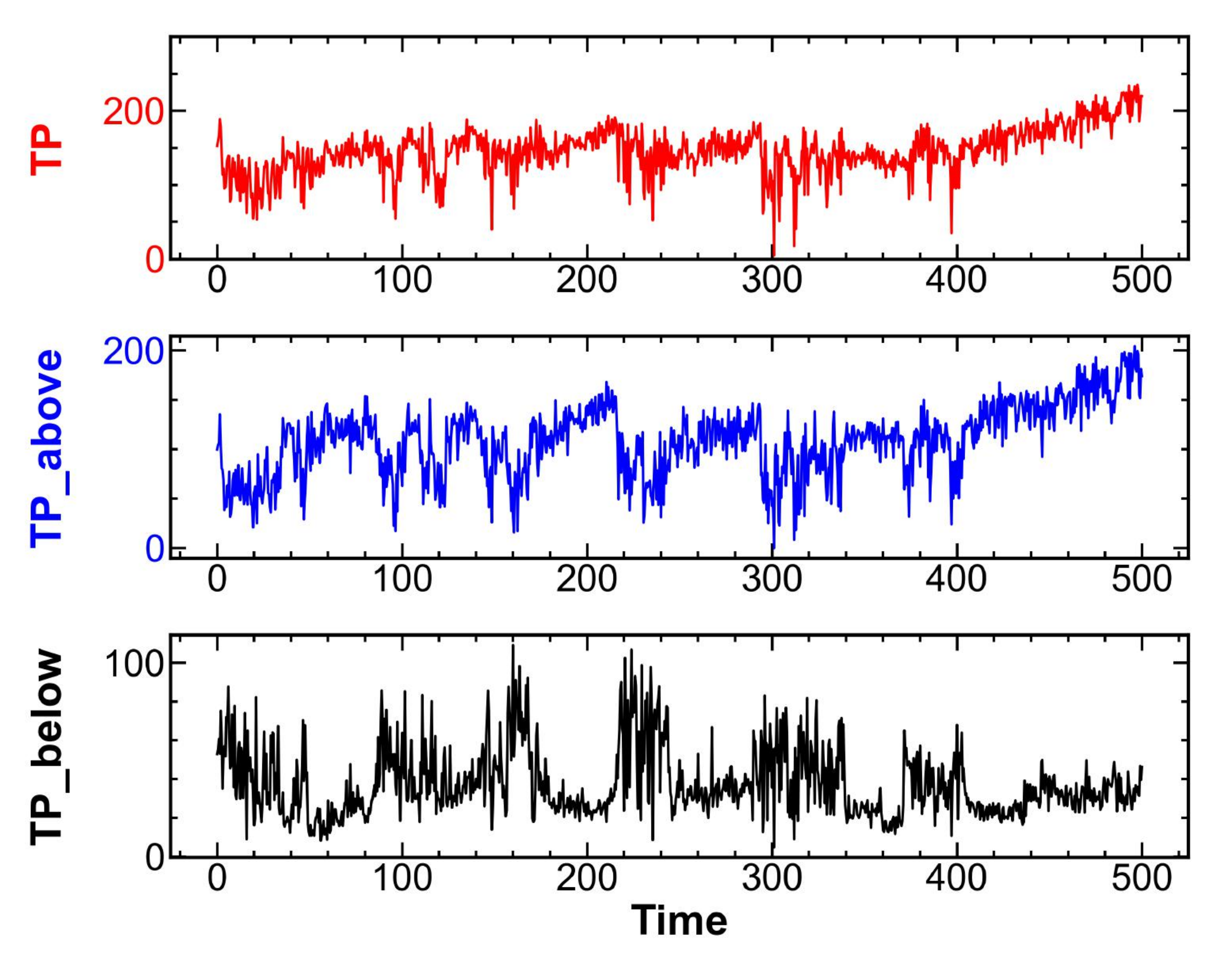}
    \caption{FP, TP.}
    \end{subfigure}
             \hfill
             \begin{subfigure}[b]{.4\textwidth}
                 \centering
                 \includegraphics[width = \linewidth]{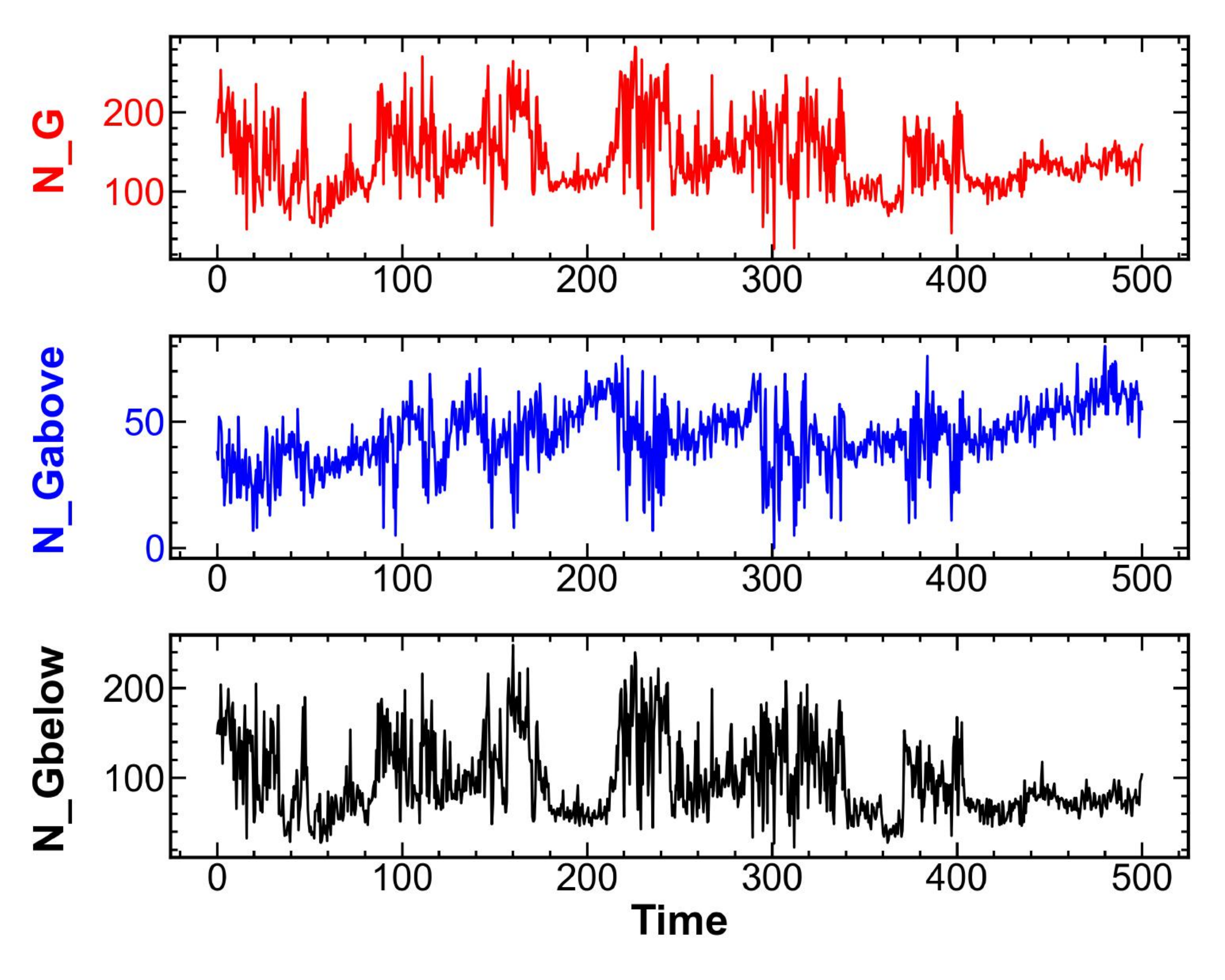}
                 \caption{FP, N$_{\rm G}$.}
             \end{subfigure}
             
    \caption{Disks with basal friction, $\beta_1$ (loops) after removing the band next to 
    the diagonal; viz. Fig~\ref{fig:TP1_disk_withbf}.}
    \label{fig:TP1_disk_withbf_removeband}
\end{figure}

Figure \ref{fig:TP0_disk_withbf} shows TP and number of generators (N$_{\rm G}$) for the components in the 
FC and FP networks.  First, we note that motion of the intruder always leads to significant 
changes for both TP and N$_{\rm G}$.  The TP results, shown in Fig.~\ref{fig:TP0_disk_withbf}(a, c) appear to show similar behavior for the FC and FP networks; however, the generators appear to be 
different.  A similar conclusion is obtained when we consider the results for loops, shown in 
Fig.~\ref{fig:TP1_disk_withbf}. Furthermore, we notice that N$_{\rm G}$ is significantly 
larger than what one would expect from the PDs for the force network snapshots, viz.~Figs.~\ref{fig:sim_pd}. Detailed inspection 
uncovers that a significant number of generators is located at rather small forces, and very 
close to the diagonal.  These generators correspond to the features which are insignificant, 
and may be due to the small variations of the forces between the particles; in experiments, 
such features may be very difficult to detect.  These generators do not have strong influence
on TP, since they are characterized by very small lifespans. To analyze the significant
features characterized by the generators that are further away from the diagonal, we consider next
the results obtained by removing a narrow band of the generators that are 
very close to the diagonal.  We choose the thickness of this narrow band to be 0.1 (so, 
10\% of the mean force).  Figures~\ref{fig:TP0_disk_withbf_removeband} 
and~\ref{fig:TP1_disk_withbf_removeband} show the corresponding results: as expected, we find that 
 the TP results are similar as if the band of generators were not removed, 
while the number of generators is significantly smaller, in particular for loops and for small forces.   

%\end{document}

While the visual comparison of the figures presented so far appears promising, in the sense 
that the two considered networks appear similar, one wishes to provide more precise 
comparisons between the two network types.  For this purpose, we compute the correlation 
coefficient, $C$, between the time series defined by the TP and N$_{\rm G}$ data, after subtracting 
their respective means.  This calculation is carried out for 8 sets of simulations 
such as ones shown in Figs.~\ref{fig:TP0_disk_withbf}~-~\ref{fig:TP1_disk_withbf_removeband}.  To help interpret the results discussed next, we note that  $C=1$ means
perfect correlations, $C=0$ means no correlation, and $C=-1$ means anticorrelation.  

Table~\ref{table:disk_bf} shows the results for disks with basal friction.  While the correlations
are not particularly high when all generators are included, the correlation after removing 
the band next to diagonal is very good, both for the components and for the loops.   This result
suggests that once the insignificant features are removed, the correlation of significant, 
dominant features between the two networks is excellent.  

\begin{table}
\caption{Correlation between the topological measures for disks with basal friction shown in Fig. \ref{fig:TP0_disk_withbf} and Fig. \ref{fig:TP1_disk_withbf}.}
\label{table:disk_bf}
\centering
\small
\renewcommand{\arraystretch}{1.25}
\begin{tabular}{lllll}
\hline\hline
\multicolumn{1}{c}{} &
\multicolumn{1}{c}{$\beta_0$ (complete)} &
\multicolumn{1}{c}{$\beta_0$ (band removed)} &
\multicolumn{1}{c}{$\beta_1$ (complete)} &
\multicolumn{1}{c}{$\beta_1$ (band removed)} \\
\hline
TP& 0.86 & 0.92 & 0.95 & 0.94\\
TP$_{\rm above}$ & 0.85 & 0.92 & 0.93 & 0.95 \\
TP$_{\rm below}$ & 0.91 & 0.87 & 0.97 & 0.97 \\
N$_{\rm G}$ &  0.42 & 0.92 & 0.80 & 0.96 \\
N$_{\rm Gabove}$ &  0.65 & 0.90 & 0.76 & 0.95 \\
N$_{\rm Gbelow}$ & 0.50 & 0.80 & 0.78 & 0.98 \\
\hline\hline
\end{tabular}
\normalsize
\end{table}

\subsection{Comparison of contact force and particle force networks: Other systems}

We proceed by discussing briefly the other three considered configurations: the disks
without basal friction and pentagons both with and without basal friction.  As specified
previously for disks we consider always $\phi = 0.78$ and for pentagons $\phi = 0.62$.  
With basal friction both systems show stick-slip type of dynamics both in simulations 
and in experiments; without basal friction the systems follow clogging type  
dynamics~\cite{carlevaro_pre_2020},~\cite{kozlowski_pre_2019}.  Therefore, by considering 
the four outlined configurations, we are in the position to discuss both the influence
of particle shape and of particle dynamics on the force networks, and in particular, 
on the degree of agreement between the FC and FP networks. 
Motivated by the high degree of correlation found for disks with basal friction when 
a narrow band of generators close to the diagonal is removed, we report only such 
results in what follows; furthermore, for brevity we show time series of the results
for TP and N$_{\rm G}$ for pentagons with basal friction only.   

\begin{figure}[!ht]
    \centering
             \begin{subfigure}[b]{.4\textwidth}
                 \centering
    \includegraphics[width=\linewidth]{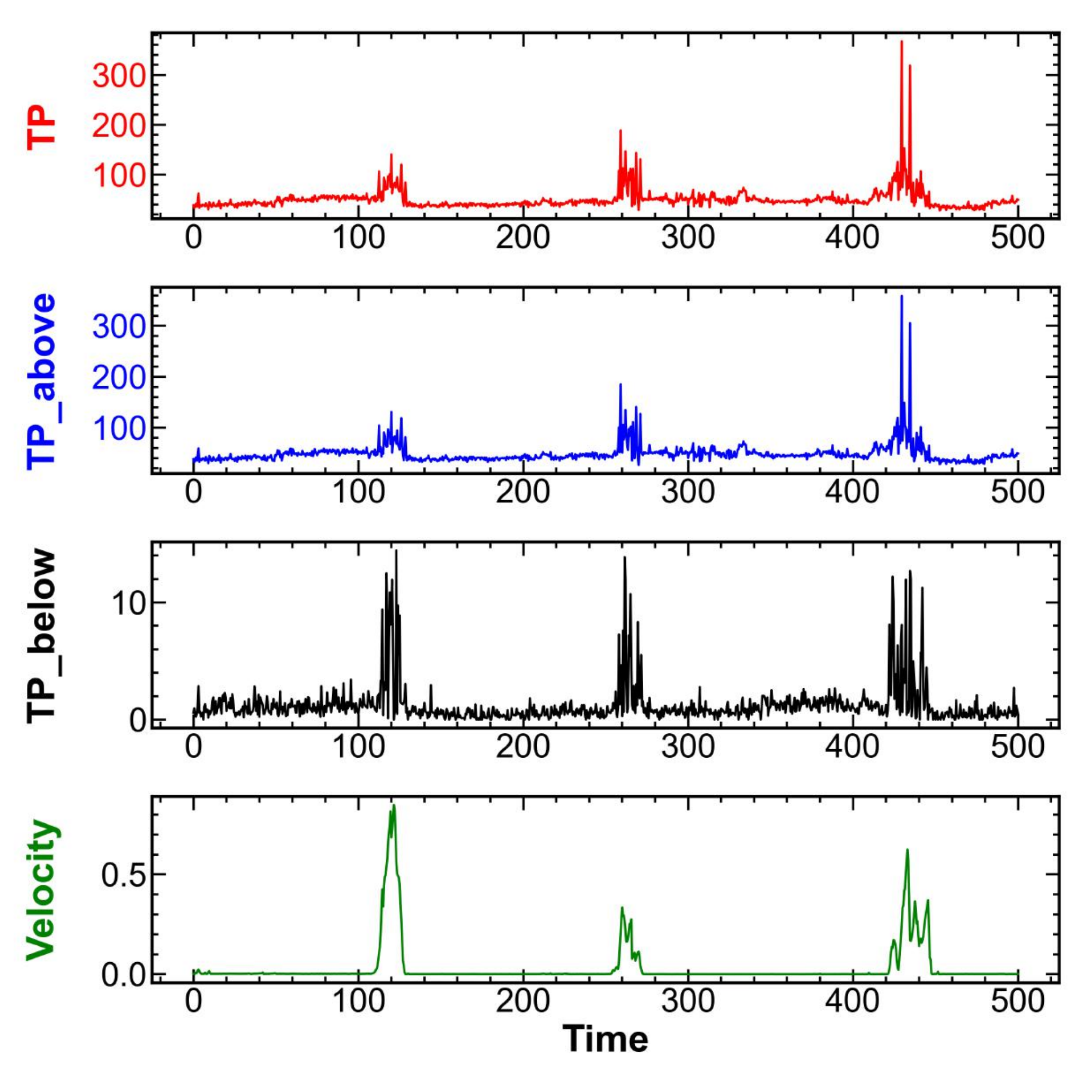}
    \caption{FC, TP.}
    \end{subfigure}
             \hfill
             \begin{subfigure}[b]{.4\textwidth}
                 \centering
    \includegraphics[width = \linewidth]{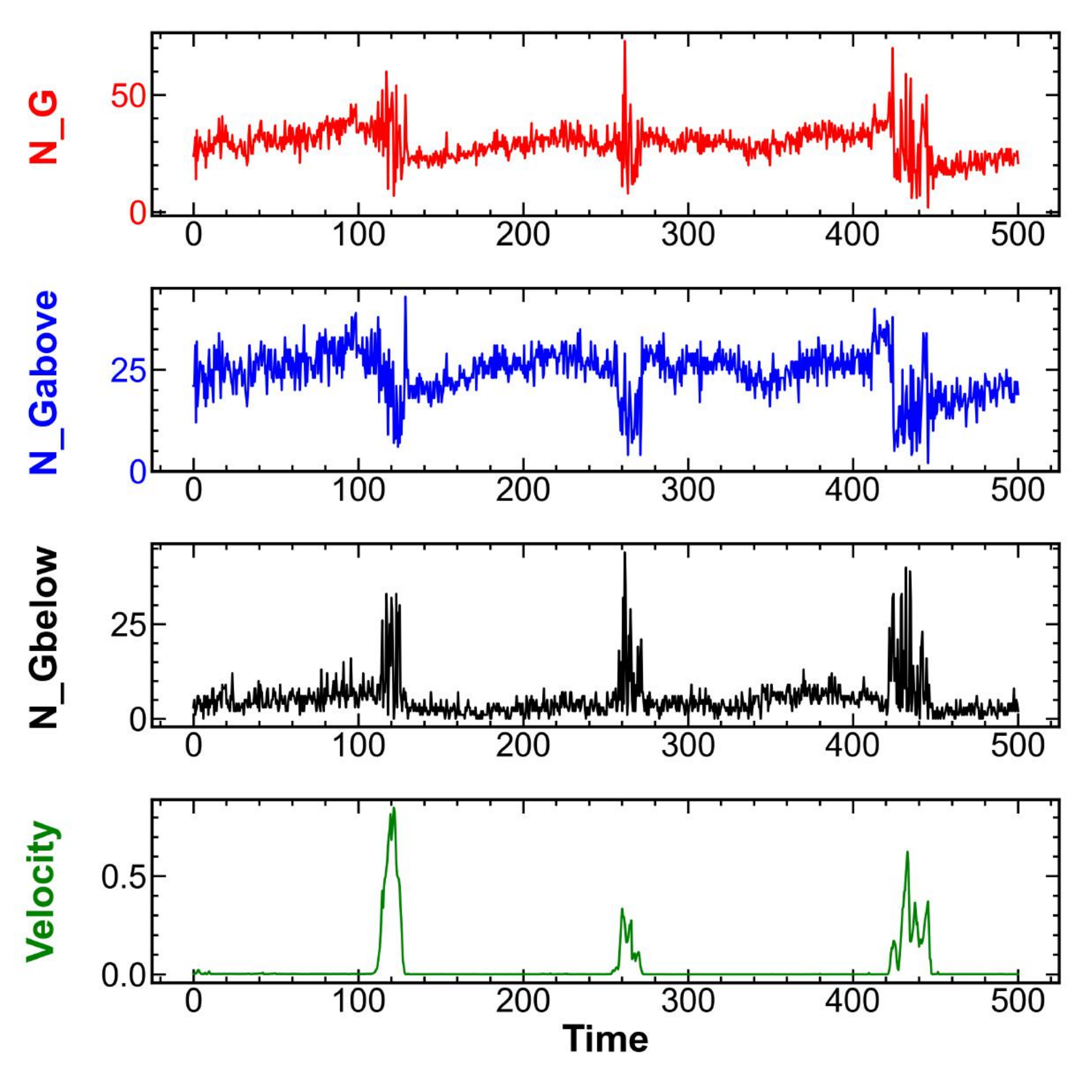}
    \caption{FC, N$_{\rm G}$.}
    \end{subfigure}
             \hfill
             \begin{subfigure}[b]{.4\textwidth}
                 \centering
    \includegraphics[width=\linewidth]{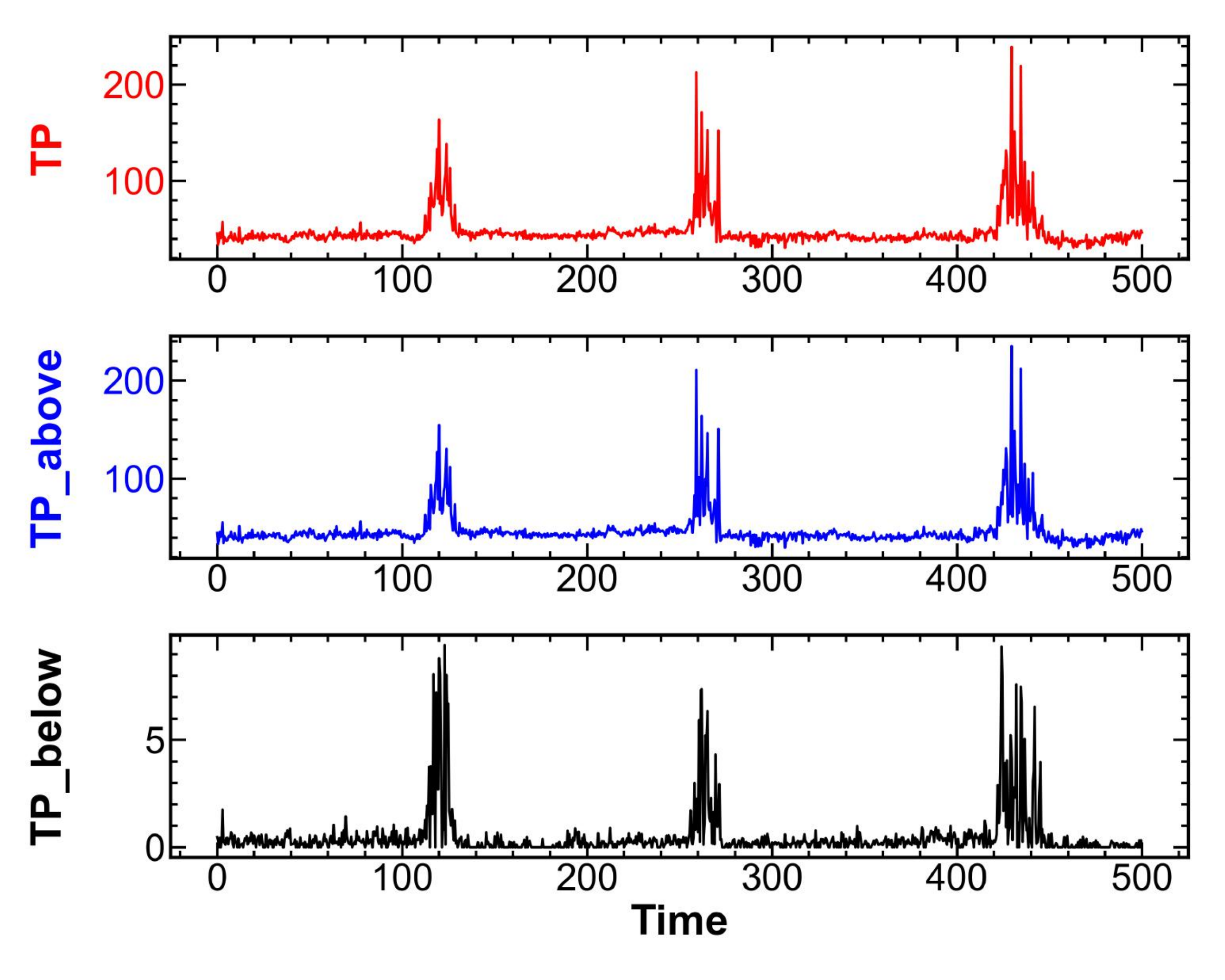}
    \caption{FP, TP.}
    \end{subfigure}
             \hfill
             \begin{subfigure}[b]{.4\textwidth}
                 \centering
                 \includegraphics[width = \linewidth]{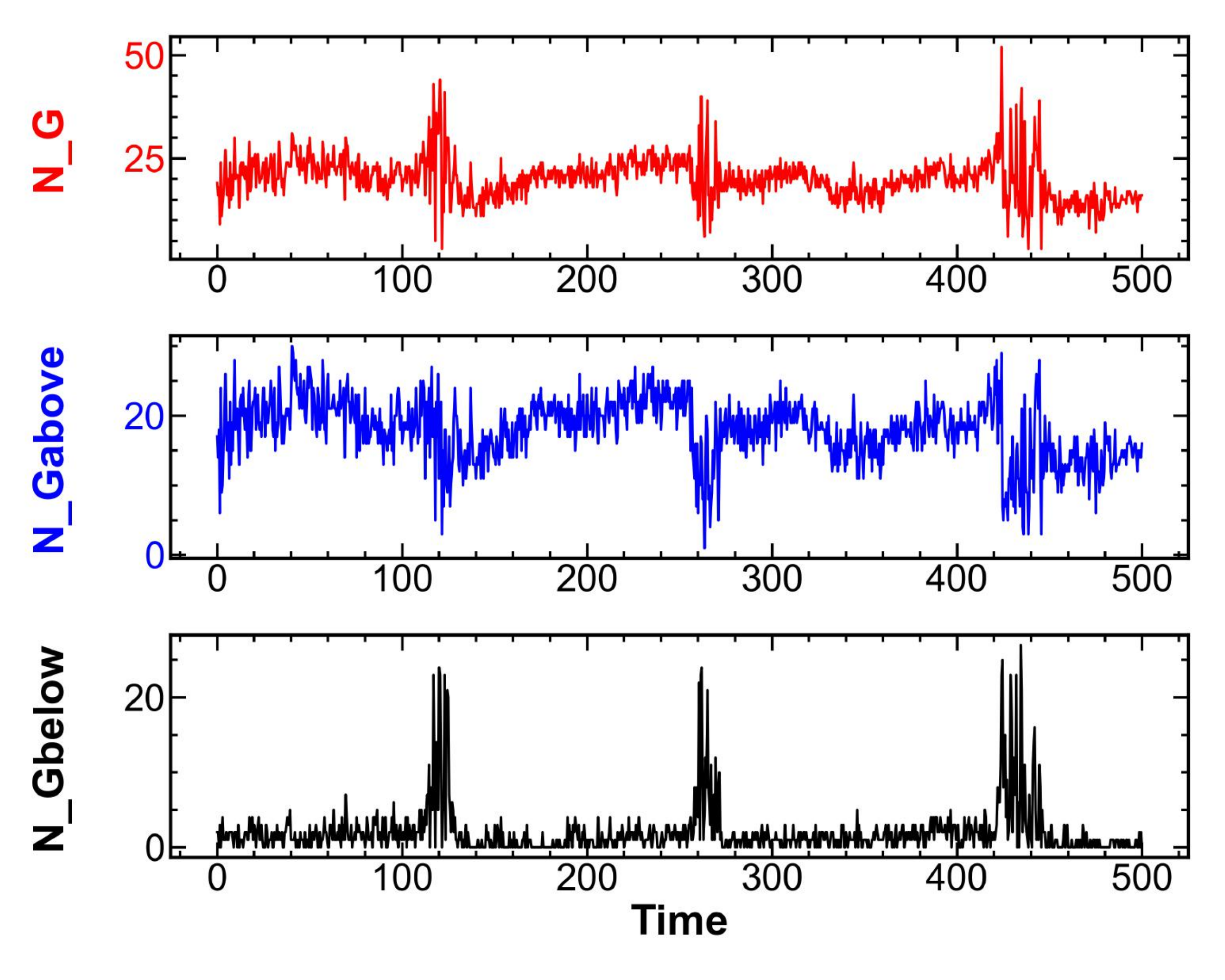}
                 \caption{FP, N$_{\rm G}$.}
             \end{subfigure}
             
    \caption{Pentagons with basal friction, $\beta_0$ (components).  This and the following figures 
    report the results obtained after removing the band of generators next to the diagonal of the PDs, as
    discussed in the text.}
    \label{fig:TP0_pentagon_withbf}
\end{figure}

\begin{figure}[!ht]
    \centering
             \begin{subfigure}[b]{.4\textwidth}
                 \centering
    \includegraphics[width=\linewidth]{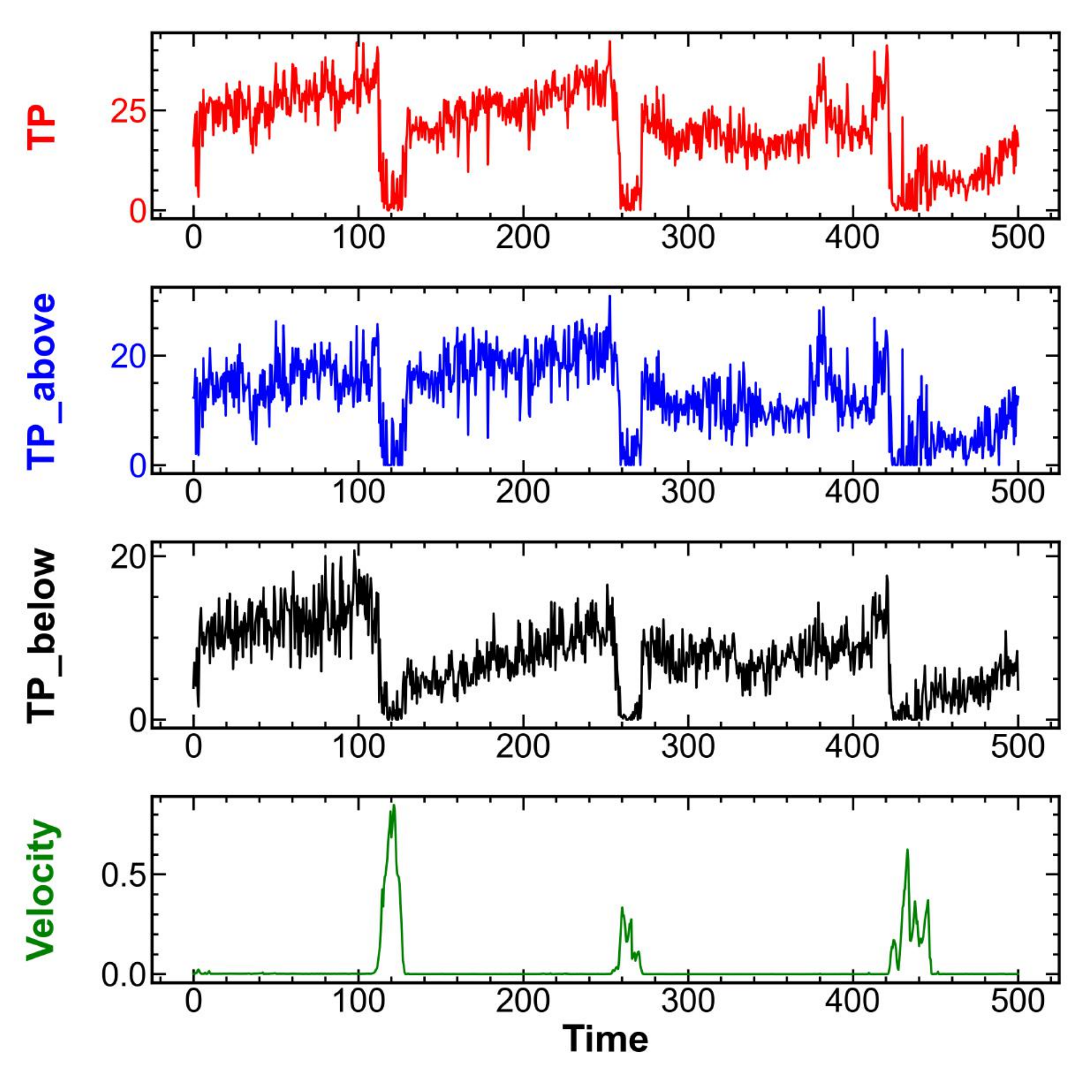}
    \caption{FC, TP.}
    \end{subfigure}
             \hfill
             \begin{subfigure}[b]{.4\textwidth}
                 \centering
    \includegraphics[width = \linewidth]{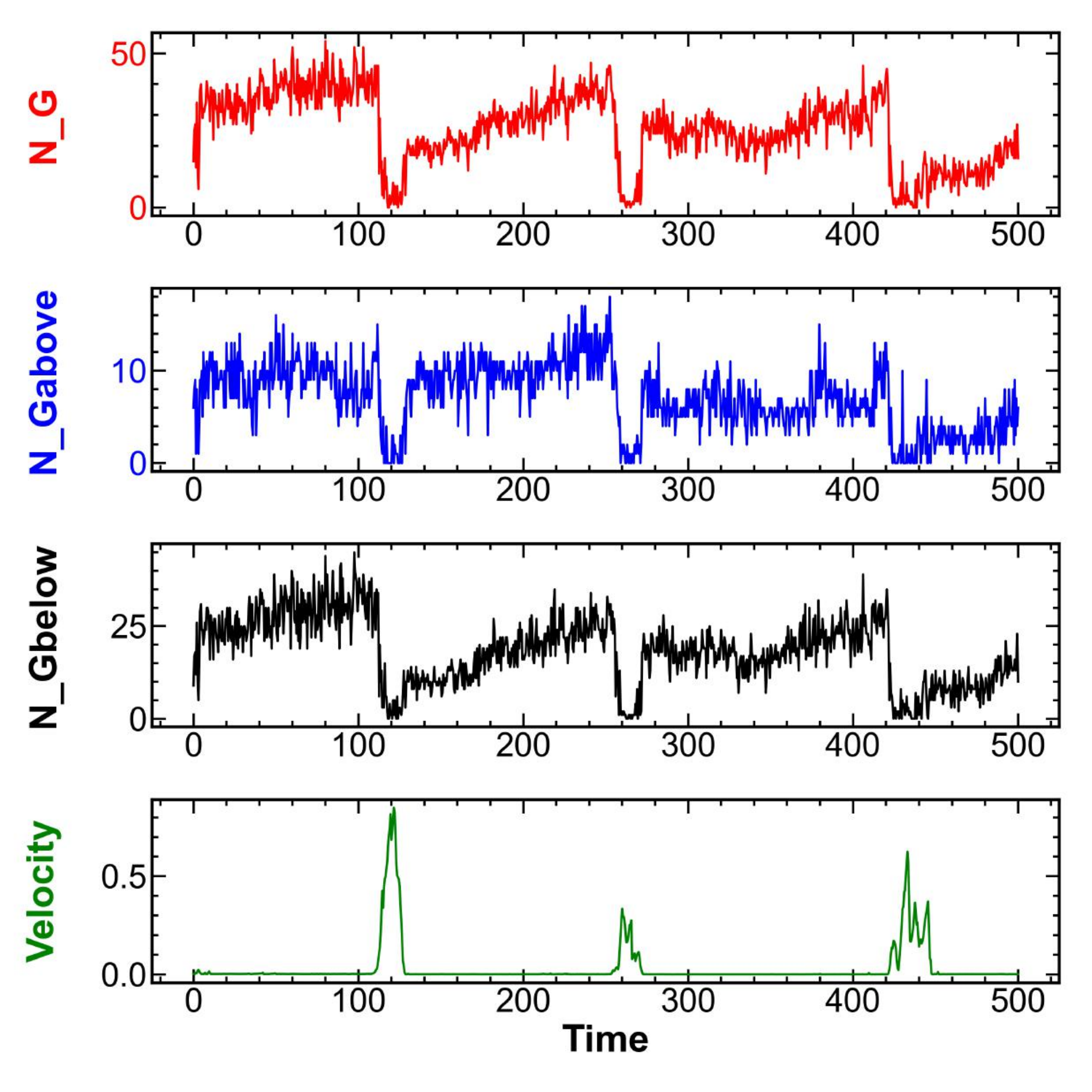}
    \caption{FC, N$_{\rm G}$.}
    \end{subfigure}
             \hfill
             \begin{subfigure}[b]{.4\textwidth}
                 \centering
    \includegraphics[width=\linewidth]{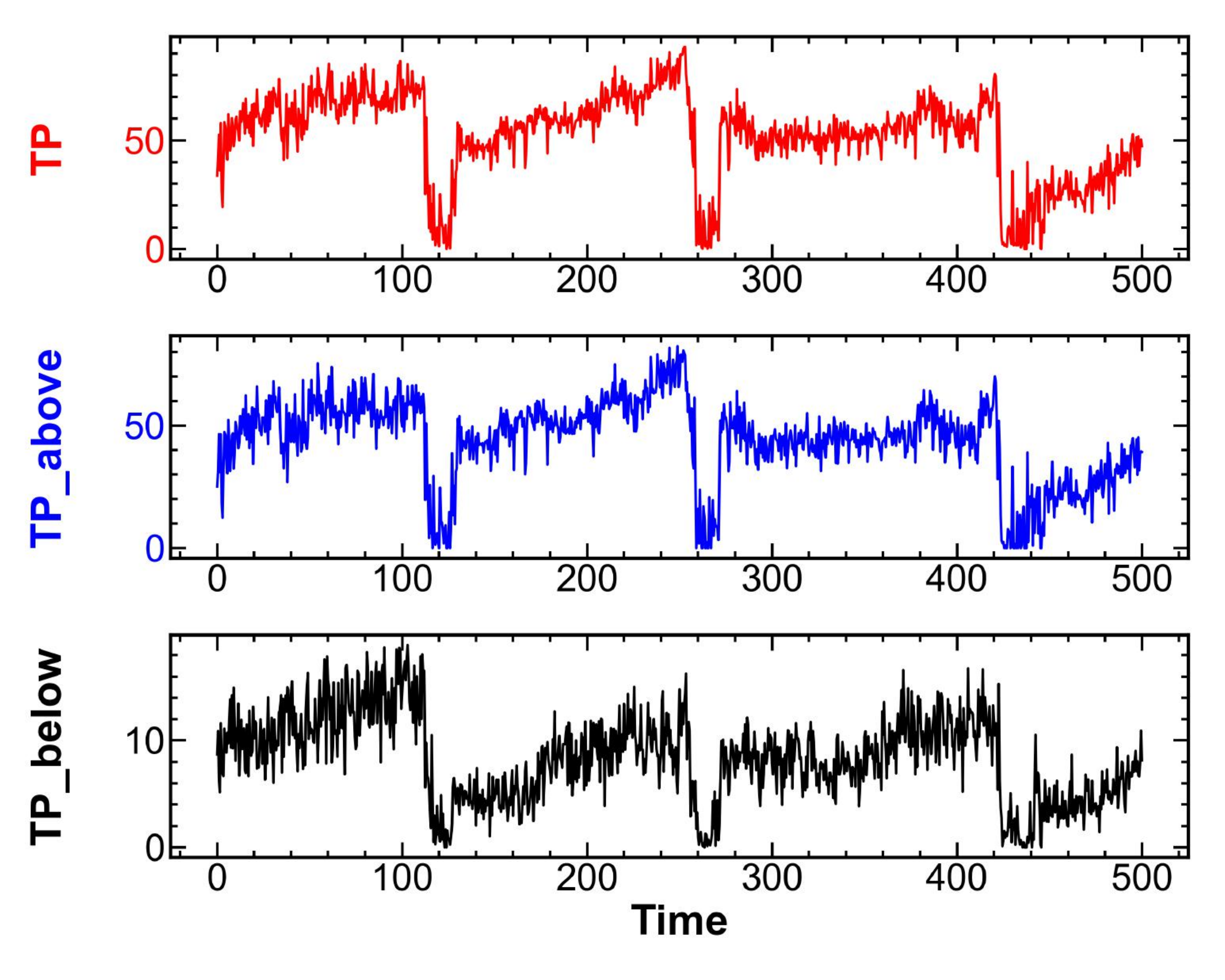}
    \caption{FP, TP.}
    \end{subfigure}
             \hfill
             \begin{subfigure}[b]{.4\textwidth}
                 \centering
                 \includegraphics[width = \linewidth]{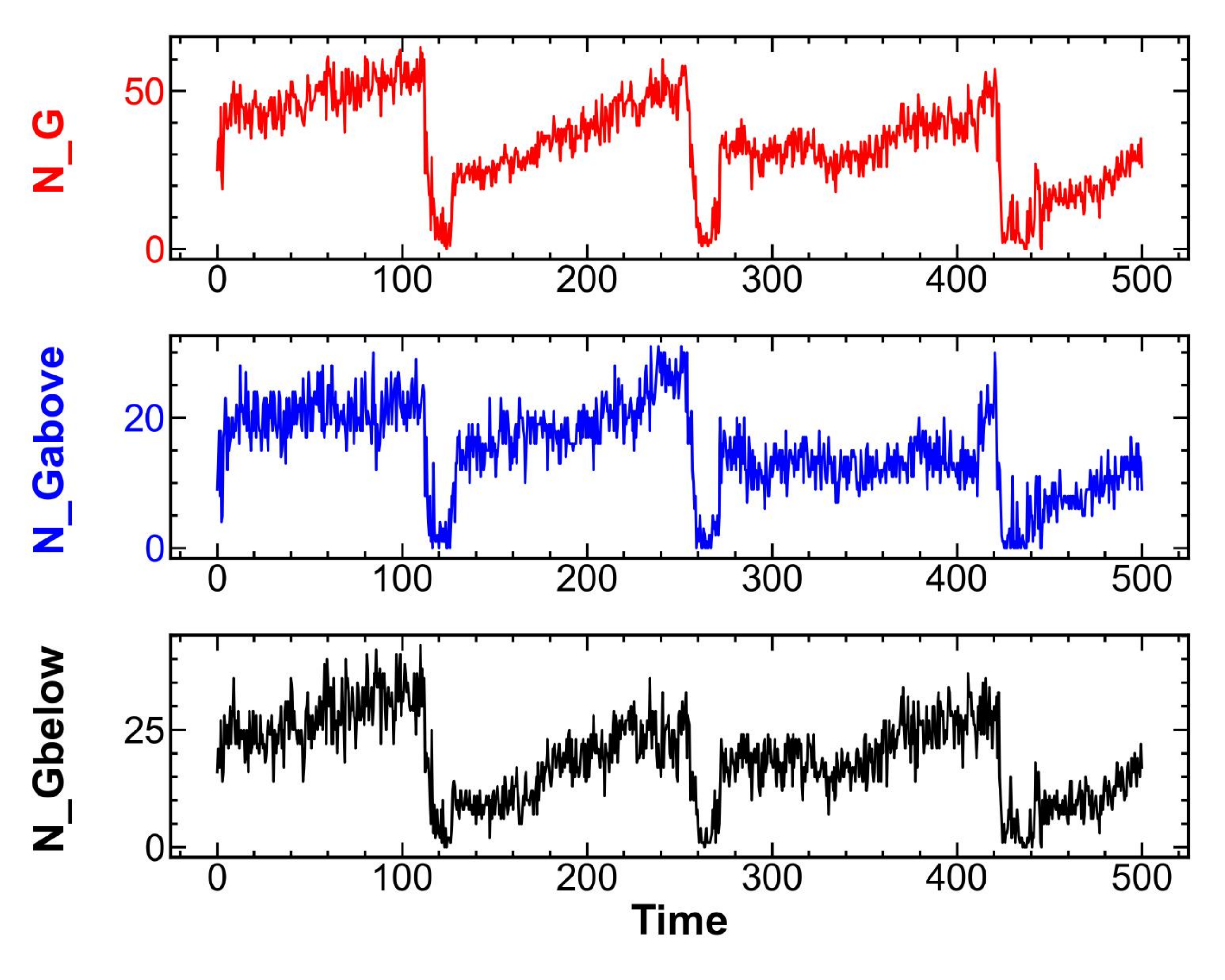}
                 \caption{FP, N$_{\rm G}$.}
             \end{subfigure}
             
    \caption{Pentagons with basal friction, $\beta_1$ (loops).}
    \label{fig:TP1_pentagon_withbf}
\end{figure}

Figures~\ref{fig:TP0_pentagon_withbf} and~\ref{fig:TP1_pentagon_withbf} show 
the results for pentagons.  The comparison with the results for disks show that 
the measures that we consider (TP and N$_{\rm G}$) are considerably different 
between the two systems, despite the fact that both system lead to the stick-slip
type of dynamics.  For pentagons we observe that the results for TP and N$_{\rm G}$ 
are much less noisy,  with clearly defined slip events, and not many changes in TP during 
the stick phases.  The comparison of the results for the loops, 
Figs.~\ref{fig:TP1_disk_withbf_removeband} and~\ref{fig:TP1_pentagon_withbf} is 
interesting as well.  The pentagons show a significantly smaller 
number of loops.  This feature of the results will be discussed in more details in our future work.  
For present purposes, the main question is whether the correlation between 
the FC and FP networks is still as good as found for the disks.  
Table~\ref{table:pentagon_bf} show that this is indeed the case: the correlation 
between two measures that we consider is still excellent.

Finally, we comment briefly on the results obtained without basal friction.  In such
systems, one finds clogging type of dynamics, leading to PH results that are much more noisy 
for both disks and pentagons (not shown for brevity). However, despite the noisy behavior, 
the correlations between the considered measures, shown in Tables~\ref{table:disk_nobf}
and~\ref{table:pentagon_nobf} are still excellent.  This result suggest that even for 
dynamic systems, one can still obtain an excellent understanding regarding the evolving 
force networks even if only total force on the particles is known.  Of course, we 
have considered only one particular setup, and only relatively crude measures 
for the purpose of quantifying the considered networks: further research should
consider other type of dynamics, as well as more detailed measures for analysis 
of the considered networks and associated persistence diagrams. 

\begin{table}
\caption{Correlation between the topological measures for pentagons with basal friction shown in 
Figs.~\ref{fig:TP0_pentagon_withbf} and~\ref{fig:TP1_pentagon_withbf}.  This and the following tables 
report the results obtained after removing the band of generators next to the diagonal of the PDs, as discussed in the text.}
\label{table:pentagon_bf}
\centering
\small
\renewcommand{\arraystretch}{1.25}
\begin{tabular}{lll}
\hline\hline
\multicolumn{1}{c}{} &
\multicolumn{1}{c}{$\beta_0$} &
\multicolumn{1}{c}{$\beta_1$ } \\
\hline
TP& 0.86 & 0.92 \\
TP$_{\rm above}$ & 0.86 & 0.85 \\
TP$_{\rm below}$ & 0.90 & 0.85  \\
N$_{\rm G}$ &  0.84 & 0.97\\
N$_{\rm Gabove}$ &  0.81 & 0.84 \\
N$_{\rm Gbelow}$ & 0.86 & 0.91  \\
\hline\hline
\end{tabular}
\normalsize
\end{table}

\begin{table}
\caption{Correlation between the topological measures for disks without basal friction.
}
\label{table:disk_nobf}
\centering
\small
\renewcommand{\arraystretch}{1.25}
\begin{tabular}{lll}
\hline\hline
\multicolumn{1}{c}{} &
\multicolumn{1}{c}{$\beta_0$ } &
\multicolumn{1}{c}{$\beta_1$ } \\
\hline
TP& 0.84 & 0.95 \\
TP$_{\rm above}$ & 0.84 & 0.94 \\
TP$_{\rm below}$ & 0.84 & 0.97  \\
N$_{\rm G}$ &  0.95 & 0.98  \\
N$_{\rm Gabove}$ &  0.94 & 0.96 \\
N$_{\rm Gbelow}$ & 0.86 & 0.98  \\
\hline\hline
\end{tabular}
\normalsize
\end{table}

\begin{table}
\caption{Correlation between the topological measures for pentagons without basal friction.}
\label{table:pentagon_nobf}
\centering
\small
\renewcommand{\arraystretch}{1.25}
\begin{tabular}{lll}
\hline\hline
% \hline\hline
\multicolumn{1}{c}{} &
\multicolumn{1}{c}{$\beta_0$ } &
\multicolumn{1}{c}{$\beta_1$ } \\
\hline
TP& 0.78 & 0.87 \\
TP$_{\rm above}$ & 0.78 & 0.74 \\
TP$_{\rm below}$ & 0.80 & 0.83  \\
N$_{\rm G}$ &  0.85 & 0.92\\
N$_{\rm Gabove}$ &  0.85 & 0.75 \\
N$_{\rm Gbelow}$ & 0.85 & 0.88 \\
\hline\hline
\end{tabular}
\normalsize
\end{table}

\section{Conclusions}

While force networks and their static and dynamic properties are known to be a crucial factor
determining macroscopic behavior of particulate-based systems, they are difficult to extract 
from experiments.  In this work, we have shown that even if only incomplete information is available, 
a still very good understanding of the main features of the force networks can be extracted.  
This result is found to hold for the particles independently of their shape (disks and pentagons
have been considered) and both for stick-slip and clogging type of dynamics.   In reaching 
this conclusion, we were helped greatly by the tools of persistent homology, which allow for
extraction of simple but objective  measures describing complicated weighted networks.  

While in this work we focus on two dimensional (2D) systems, the methods used are not limited 
to 2D - they are applicable equally well in 3D.  Our results therefore set a stage for more
in-depth analysis of the properties of force networks in 3D, where the interparticle forces
are even more difficult to obtain.   

Furthermore, the presented results set the stage for comparing the results of simulations (for which we have
complete data about interparticle forces available) and of experiments, for which only partial
information may be available.  Such a comparison will be the subject of future work.   

%\section{ Appendix(es)}
\section{ Data Availability Statement}  All the data used in this study is available from the 
authors upon request. 
\section{ Acknowledgments} This study was supported by the US Army Research Office Grant No.~W911NF1810184.  
L.A.P. and C.M.C. acknowledge support by Universidad Tecnológica Nacional through Grants No. PID- MAUTNLP0004415 and 
No. PID-MAIFIBA0004434TC and CONICET through Grant No. RES-1225-17 and PUE 2018 229 20180100010 CO.
%\section{Disclaimers}
%\section{ Notation list}
\section{ Supplemental Materials} The animations of the force networks shown in 
Fig.~\ref{fig:sim} and corresponding persistence diagrams, Fig.~\ref{fig:sim_pd} are
available.  

\bibliography{granulates}

\end{document}